\documentclass[a4paper,11pt]{article}
\pdfoutput=1 

\usepackage{format} 

\usepackage[T1]{fontenc} 
\usepackage{longtable}
\usepackage[utf8]{inputenc} 

\usepackage{amssymb}
\usepackage{amsmath}
\usepackage{yfonts}
\usepackage{dsfont} 
\usepackage[bottom]{footmisc} 

\usepackage{titlesec}

\titleformat*{\section}{\Large\bfseries}

\usepackage[normalem]{ulem}

\usepackage{tikz}
\usetikzlibrary{positioning,arrows}
\usetikzlibrary{decorations.pathmorphing}
\usetikzlibrary{decorations.markings}
\usepackage[compat=1.1.0]{tikz-feynman}

\usepackage{subcaption}

\usepackage{placeins}

\usepackage{enumitem}
\setlist{nosep}

\newcommand{\beq}{\begin{equation}}
\newcommand{\eeq}{\end{equation}}
\newcommand{\bea}{\begin{eqnarray}}
\newcommand{\eea}{\end{eqnarray}}
\newcommand{\barr}{\begin{array}}
\newcommand{\earr}{\end{array}}
\newcommand{\dis}{\displaystyle}

\newcommand{\nn}{\nonumber}

\newcommand{\lag}{{\cal L}}

\newcommand{\zLN}{z_{{}_{LN}}}
\newcommand{\zLE}{z_{{}_{LE}}}
\newcommand{\zRE}{z_{{}_{RE}}}

\newcommand{\zLU}{z_{{}_{LU}}}
\newcommand{\zLD}{z_{{}_{LD}}}
\newcommand{\zRU}{z_{{}_{RU}}}
\newcommand{\zRD}{z_{{}_{RD}}}

\newcommand{\ylN}{y_{{}_{lN}}}
\newcommand{\ylE}{y_{{}_{lE}}}
\newcommand{\yeL}{y_{{}_{eL}}}

\newcommand{\yqU}{y_{{}_{qU}}}
\newcommand{\yqD}{y_{{}_{qD}}}
\newcommand{\yuQ}{y_{{}_{uQ}}}
\newcommand{\ydQ}{y_{{}_{dQ}}}

\newcommand{\lfb}{\left(}
\newcommand{\rfb}{\right)}

\newcommand{\ltb}{\left[}
\newcommand{\rtb}{\right]}

\newcommand{\CLR}{C^{LR}}
\newcommand{\CLL}{C^{LL}}
\newcommand{\CRR}{C^{RR}}
\newcommand{\NLR}{N^{LR}}
\newcommand{\NLL}{N^{LL}}
\newcommand{\NRR}{N^{RR}}

\newcommand{\NDL}{N^D_L}

\newcommand{\NSR}{N^S_R}

\newcommand{\ESL}{E^S_L}
\newcommand{\ESR}{E^S_R}

\newcommand{\LDL}{L^D_L}
\newcommand{\LDR}{L^D_R}
\newcommand{\QDL}{Q^D_L}
\newcommand{\QDR}{Q^D_R}

\newcommand{\USL}{U^S_L}
\newcommand{\USR}{U^S_R}
\newcommand{\DSL}{D^S_L}
\newcommand{\DSR}{D^S_R}

\newcommand{\mN}{\widetilde m_N}
\newcommand{\mL}{\widetilde m_L}
\newcommand{\mE}{\widetilde m_E}
\newcommand{\muphi}{\mu_{\Phi}}

\newcommand{\ES}{\widetilde E^S}
\newcommand{\ED}{\widetilde E^D}
\newcommand{\NS}{\widetilde N^S}
\newcommand{\ND}{\widetilde N^D}

\long\def\/*#1*/{}
\usepackage{color}
\usepackage{colortbl} 
 \usepackage[normalem]{ulem}
\definecolor{darkgreen}{cmyk}{1,0,1,0.4}
\definecolor{darkred}{cmyk}{0,1,1,0.4}
\definecolor{pink}{rgb}{0.93,0.23,0.51}

\providecommand{\keywords}[1]
{
  \small	
  \textbf{\textit{Keywords---}} #1
}

\title{\boldmath Neutrino masses, anomalous magnetic moments and dark matter with vector-like fermions and an inert scalar doublet}

\author{Vandana Sahdev}

\affiliation{Department of Physics and Astrophysics, University of Delhi, Delhi-110007, India}

\emailAdd{vandanasahdev20@gmail.com}

\abstract{The beyond-the-standard-model scenario in this work is motivated from the observations of neutrino masses, anomalous magnetic moments of electron and muon, and dark matter in the Universe. We explain these observations by extending the standard model with two generations of vector-like fermions and  an inert scalar doublet, all odd under a $Z_2$ symmetry. The light neutrino masses and mixings are generated radiatively while maintaining consistency with bounds on lepton flavor violation. Loop diagrams with the very same fields also serve to explain the anomalous magnetic moments. Similarly, the correct dark matter relic abundance is reproduced without coming into conflict with direct detection constraints, or those from big bang nucleosynthesis or the cosmic microwave observations. Finally, prospective signatures at the LHC are discussed.}


\begin{document}

\tikzset{
  vector/.style={decorate, decoration={snake,amplitude=.4mm,segment length=2mm,post length=1mm}, draw},
  tes/.style={draw=black,postaction={decorate},
    decoration={snake,markings,mark=at position .55 with {\arrow[draw=black]{>}}}},
  provector/.style={decorate, decoration={snake,amplitude=2.5pt}, draw},
  antivector/.style={decorate, decoration={snake,amplitude=-2.5pt}, draw},
  fermion/.style={draw=black, postaction={decorate},decoration={markings,mark=at position .55 with {\arrow[draw=black]{>}}}},
  fermionbar/.style={draw=black, postaction={decorate},
    decoration={markings,mark=at position .55 with {\arrow[draw=black]{<}}}},
  fermionnoarrow/.style={draw=black},
  scalar/.style={dashed,draw=black, postaction={decorate},decoration={markings,mark=at position .55 with {\arrow[draw=blue]{>}}}},
  scalarbar/.style={dashed,draw=black, postaction={decorate},decoration={marking,mark=at position .55 with {\arrow[draw=black]{<}}}},
  scalarnoarrow/.style={dashed,draw=black},
  electron/.style={draw=black, postaction={decorate},decoration={markings,mark=at position .55 with {\arrow[draw=black]{>}}}},
  bigvector/.style={decorate, decoration={snake,amplitude=4pt}, draw},
  particle/.style={thick,draw=blue, postaction={decorate},
    decoration={markings,mark=at position .5 with {\arrow[blue]{triangle 45}}}},
  gluon/.style={decorate, draw=black,
    decoration={coil,aspect=0.3,segment length=3pt,amplitude=3pt}},
  majorana/.style={draw=black},
  linearrow/.style={draw=black, postaction={decorate},decoration={markings,mark=at position 1.0 with {\arrow[draw=black]{>}}}},
  scalar2/.style={dashed,draw=black}
}

\maketitle

\flushbottom

\section{Introduction}
\label{sec:intro}
\vspace{0.25cm}
 Notwithstanding the remarkable success of the Standard Model (SM), it
 continues to suffer from several lacunae, in particular its inability
 to explain the existence of dark matter (DM) on the one hand and to
 address the flavour problem on the other. Added to this are
 long-standing discrepancies between the data and
 expectations. 
Foremost amongst these are the anomalous magnetic moments of the muon and electron. The discrepancy in the experimental measurement and SM prediction of muon g-2 has been a long-standing puzzle, with the calculation of hadronic vacuum polarization (HVP) accounting for
the major uncertainties. The discrepancy, derived from the comparison of the experimental average \cite{Muong-2:2023cdq} with the theoretical prediction based on the data-driven method \cite{Aoyama:2020ynm}, reached a significance of $\sim 5.1 \sigma$. Recently, however, the cross-section measurement of $e^+e^- \to \pi^+\pi^-$ has created tension between experimental inputs for the data-driven method, while the lattice-QCD calculation of the SM prediction has got more precise. After inclusion of the latter by the Muon g-2 Theory Initiative in their 2025 White Paper \cite{Aliberti:2025beg}, the discrepancy has greatly reduced.
The observation of neutrino
oscillations~\cite{Fukuda:1998mi,Ahmad:2001an,Abe:2011fz,An:2012eh,Ahn:2012nd}
 have long demanded small, but nonzero neutrino
masses. Of the parameters in the neutrino sector, the three mixing
angles as well as one of the two differences in the squares of masses
are quite well-determined~\cite{ParticleDataGroup:2024cfk,Esteban:2018azc}, with
only the magnitude (and not the sign) of the other difference being
known. Of the nontrivial phases that are possible in the mixing
matrix, only one can, in principle, be measured in an oscillation
experiment. Poorly known as of date, recent measurements have
indicated a nonzero value for the
same~\cite{Abe:2019vii,Pascoli:2020swq}.  Although the oscillation
data is only sensitive to the difference in mass-squareds and not the
absolute mass scale, the latter is very-well constrained to $\sum
m(\nu_i) < (0.340-0.715)$ eV~\cite{Zyla:2020zbs}\footnote{More recent 
cosmological analyses, such as those from Planck \cite{Planck:2018vyg},
yield a stronger bound of $\sum m_\nu \lesssim 0.12$ eV.}---where the
$\nu_i$ are the (cosmologically) stable light neutrinos---from
a host of cosmological data, with such bounds comfortably outdoing
those from terrestrial experiments \cite{Aker:2019uuj}.

Neutrino masses and mixings can, of course, be trivially arranged by
introducing right-handed fields alongwith appropriately tiny Yukawa
terms. While the additional hierarchy in the couplings might be
aesthetically repugnant, there are no technical objections to such a
paradigm. However, with the right-handed neutrinos being gauge
singlets, terms such as $\overline{(\nu_{jR})^c} \nu_{kR}$ can have
arbitrarily large coefficients. Beset with such large Majorana masses,
the $\nu_{jR}$ can be integrated out from the low-energy theory,
leaving the SM neutrinos with tiny masses. Evading the need for highly
suppressed Yukawa couplings, this {\em seesaw
  mechanism} has proven to be a very popular one with three large
classes of variants
\cite{Chen:2019nud,Minkowski:1977sc,Yanagida:1979as,Mohapatra:1979ia,Konetschny:1977bn,Ashanujjaman:2021txz,Cheng:1980qt,Lazarides:1980nt,Schechter:1980gr,Magg:1980ut,Mohapatra:1980yp,Foot:1988aq,Ashanujjaman:2021jhi,Ashanujjaman:2021zrh}. 
An attractive possibility is that neutrino masses are generated radiatively,
with the particles running in the loop belonging to a dark sector protected by
a discrete symmetry. A well-known example is the scotogenic model proposed 
independently by Ma \cite{Ma:2006km} and Tao \cite{Tao:1996vb}, 
where neutrino masses arise at one loop with an inert
scalar doublet and singlet fermions running in the loop.

The almost unfettered freedom in arranging the low-energy neutrino
sector parameters implies that, by themselves, these observables are
of little use in delineating the structure of possible physics beyond
the SM.  However, when taken in conjunction with other observables,
the quest for a unified resolution can be profitable. In this spirit,
we now consider the other longstanding issues. One of these, namely
the existence of the DM we have already mentioned. Not only must we
have a suitable candidate (whose stability, typically, owes itself to
a discrete symmetry) but we also need to ensure that the correct relic
abundance, as determined so accurately by the {\sc
  wmap}~\cite{Hinshaw:2012aka} and {\sc planck}~\cite{Ade:2015xua}
observations, is obtained. Furthermore, in doing this, it needs to be
ascertained that other astrophysical and cosmological data such as
those arising from observables relating to big-bang nucleosynthesis,
the cosmic microwave background radiation as well as the comparison of
the gamma-ray or radio-frequency spectra arising from the
pair-annihilation of such DM with those observed by the
Fermi-LAT~\cite{Abdo:2010ex} on one side and
GMRT~\cite{Intema:2016jhx} or VLA~\cite{condon1998} on the other.  As
is well-known, the so-called WIMP miracle can successfully answer all
of these questions, albeit only for a constrained set of parameters,
and often, at the cost of introducing additional `dark-sector'
particles.

Attempts to construct a unified theory that explains more than one of
the aforementioned phenomena suffer, typically, from roadblocks in the
form of introducing unwelcome phenomenological consequences. For
example, any ultraviolet complete theory that seeks to explain the
anomalous magnetic moments would necessarily contain additional
fields.  Were these to play a role in generating neutrino masses and
mixings, there is always the danger that flavour-violating processes
involving the charged leptons themselves would be set up, thereby
running afoul of strong constraints on processes such $\mu \to e
\gamma$ or $\mu \to 3 e$ or even $K_L \to \mu^+ e^-$ etc. In other
words, the introduction of new fields cannot be arbitrary, for not
only must low-energy observables remain consistent with measurements,
but the failure of collider experiments to observe such particles must
be explained. The said non-observation can, of course, be explained by
postulating these to be either SM gauge-singlets or very massive, with
the first choice precluding most roles.

Were such ultra-massive fields to be fermionic (as some of them would
need to be to explain, for example, the anomalous magnetic moments), a
chiral assignment of quantum numbers would necessitate very large
Yukawa couplings. The latter, in turn, would not only introduce
significant corrections to the electroweak precision tests, but also
contribute to one or both of Higgs production and decay.  Vector-like (VL) 
fermions, on the other hand, may have gauge-invariant bare mass terms,
thereby easily evading such restrictions. In addition, the inclusion
of such fermions does not introduce additional
contributions to the chiral anomaly, and is,
therefore, theoretically well-motivated. 

Such vector-like fermions appear in diverse
scenarios \cite{Aguilar-Saavedra:2013qpa,Ellis:2014dza}, a trivial
example being the Higgsinos in the minimal supersymmetric SM.
Additional such fields may arise in the quest of enlarging the gauge
symmetry~\cite{Kang:2007ib,Dermisek:2012ke,Bhattacherjee:2017cxh,EmmanuelCosta:2005nh,Barger:2006fm,Dorsner:2014wva},
as also in extended supersymmetry~\cite{Choi:2010gc}, wherein the
(normally Majorana) gauginos are promoted to Dirac-like particles,
thereby suppressing pair production channels as well as cascade
decays. In an analogous fashion, the Higgs-sector could be extended to
obtain an $R$-symmetric theory~\cite{Choi:2010an}. Such matter can
also help alleviate the tension with the mass of the SM-like
Higgs~\cite{Martin:2009bg,Martin:2010dc} in supersymmetric theories on
the one hand, and the little hierarchy problem~\cite{Graham:2009gy} on
the other.  Similarly, many of the problems faced by gauge-mediated
breaking of supersymmetry may be cured
too~\cite{Moroi:2011aa,Endo:2011xq,Martin:2012dg,Fischler:2013tva}. A
particularly intriguing example was offered in
Refs.~\cite{Babu:1990gp,Babu:1994pd} wherein a supersymmetric theory of
compositeness not only predicts three chiral families of fermion (as
seen in the SM) but is also accompanied by two heavier vector-like generations.

Theories of compositeness, such as models wherein the electroweak
symmetry was broken dynamically through the condensation of top quarks
(or its partners)~\cite{Dobrescu:1997nm,Chivukula:1998wd,He:1999vp}
provide historically interesting examples of non-chiral fermions.  And
while vector-like fermions are ubiquitous in any extra-dimensional
scenario
\cite{Appelquist:2000nn,Cheng:2002ab,Choudhury:2009kz,Servant:2002aq,Appelquist:2001mj,Ghosh:2008ix,Ghosh:2008ji,Bhattacherjee:2010vm,Dobrescu:2001ae,Burdman:2006gy,Choudhury:2016tff,Avnish:2020atn}
wherein the SM fields extend into the bulk they are also present in a
variety of scenarios such as composite
Higgs~\cite{Contino:2006qr,Anastasiou:2009rv,Vignaroli:2012sf,
  DeSimone:2012fs,Delaunay:2013iia,Gillioz:2013pba,Banerjee:2017wmg},
little Higgs
models~\cite{Han:2003wu,Carena:2006jx,Choudhury:2006mp,Choudhury:2006sq,
  Matsumoto:2008fq,Choudhury:2012xc,Berger:2012ec}, or
simple Higgs-portal solutions for obtaining correct DM relic
abundance~\cite{Patt:2006fw,Gopalakrishna:2009yz,Baek:2011aa}.

In this paper, we delve into the potential of vector-like leptons
explaining neutrino masses and mixings, the existence and correct
relic abundance of dark matter as well as the anomalous magnetic
moments of both the muon and the electron. We would demonstrate that a
very simple, and natural, extension, eschewing an inordinate
amount of fine-tuning, solves each one of these problems without
coming into conflict with constraints from flavour-changing neutral
current processes. And while the aforementioned problems do not
require the introduction of vector-like quarks, we show that an
analogous introduction serves to ensure that the SM gauge couplings do
meet at a point under renormalization group flow, thereby raising the
possibility that the scenario could be the low-energy manifestation of
a grand unified theory. Such vector-like quarks have also been shown
to introduce the right amount of mixing in the quark sector that
allows one to explain~\cite{Choudhury:2001hs} the long-standing
$2.9\sigma$ discrepancy in the forward-backward asymmetry in
bottom-quark production as measured at the
$Z$-peak~\cite{ALEPH:2010aa}.
\section{Description of the model}
\label{sec:model}
\vspace{0.25cm}
Maintaining the gauge symmetry to be $SU(3)_c\otimes SU(2)_L\otimes
U(1)_Y$, we augment the SM by the inclusion of a few
vector-like fermion multiplets. While the latter could, in principle,
carry any set of quantum numbers, for the sake of simplicity, we
restrict ourselves to only those combinations as seen within the
SM. To eliminate constraints from flavour-changing neutral currents,
we disallow mixings between the SM fermions and the new ones by
postulating an unbroken $Z_2$, under which the SM fields are charged
$+1$, while all the new fields are charged $-1$. The unbroken $Z_2$
symmetry renders the lightest $Z_2$-odd particle
(L$Z_2$OP) absolutely stable.  Hence, as long as the latter is
  electrically neutral and a color-singlet, it can be a
cosmologically viable candidate for dark matter, provided the
 measured value of
the relic density (RD) is reproduced and the
constraints from all direct and indirect dark matter detection
experiments satisfied.

If neutrino masses or the anomalous magnetic moments for the muon or
the electron are to be explained, the new fermions would need to
interact directly with the SM fermions and not merely through current-current
interactions. To facilitate this, an extra $Z_2$-odd scalar doublet
$\Phi$ is introduced. Since the $Z_2$ needs to remain unbroken, the
parameters of the scalar potential must be so that, unlike the SM
Higgs field $H$, the new one does not acquire a nonzero vacuum
expectation value.

Neutrino oscillation data stipulate that at least two of the light
neutrinos are massive, and this would demand that we must have at
least two generations of the vector-like leptons. This is also
demanded by the resolution of the discrepancies in the anomalous
magnetic moments. While it might seem attractive to have a vector-like
generation accompany each chiral generation, this is neither necessary
nor desirable. For example, a symmetry between quarks and leptons, as
also the desirability of accommodating gauge unification would call
for the inclusion of vector-like quarks as well\footnote{As mentioned
  earlier, such fermions also offer resolutions for certain
  discrepancies in the hadronic sector, such as the long-standing one
  in $A^b_{\rm FB}$ \cite{Choudhury:2001hs} or the recent
  $B$-anomalies by means of generating effective current-current
  interactions~\cite{Choudhury:2017qyt,Choudhury:2017ijp,Bhattacharya:2019eji}. However,
  we shall not delve into this.}. And, as can be appreciated easily,
asymptotic freedom would be lost if there were more than two complete
vector-like generations. Such competing arguments call for the
inclusion of exactly two generations of vector-like fermions, a
construction also favoured by early efforts to explain three chiral
generations in scenarios invoking fermion
compositeness~\cite{Babu:1990gp,Babu:1994pd}.
\begin{table}[!h]
   \centering
  \begin{tabular}{|l| l|| l | l|}
    \hline
    \multicolumn{2} {|c||}{SM Sector} & \multicolumn{2} {|c|}{Exotic Sector}
    \\ \hline $q_{i L}
    \equiv \left(u_{iL} \;\, d_{iL}\right)^T$ &
    $\left(3,2,1/6,+\right)$ & $Q_{\alpha L/R} \equiv
    \left(U^D_{\alpha L/R} \;\, D^{D}_{\alpha L/R}\right)^T$ &
    $\left(3,2,1/6,-\right)$
    \\[1ex]
    $u_{i R}$ &
    $\left(3,1,2/3,+\right)$ & $U^S_{\alpha L/R}$ &
    $\left(3,1,2/3,-\right)$
    \\[1ex]
    $d_{i R}$ &
    $\left(3,1,-1/3,+\right)$ & $D^S_{\alpha L/R}$ &
    $\left(3,1,-1/3,-\right)$ \\[1ex]
    \hline
    $l_{i L} \equiv \left(\nu_{iL}
    \;\, e_{iL}\right)^T$ & $\left(1,2,-1/2,+\right)$ & $L_{\alpha
      L/R} \equiv \left(N^D_{\alpha L/R} \;\, E^D_{\alpha
      L/R}\right)^T$ & $\left(1,2,-1/2,-\right)$ \\[1ex] $e_{iR}$ &
    $(1,1,-1,+)$ & $E^S_{\alpha L/R}$ & $(1,1,-1,-) $ \\[1ex] & &
    $N^S_{\alpha R}$ & $\left(1,1,0,-\right)$ \\[1ex]
    \hline
    $H \equiv
    \left(h^{+} \; \, [h + i\eta] / \sqrt{2} \right)^T$ &
    $\left(1,2,1/2,+\right)$ & $\Phi \equiv \left(\phi^{+} \; \,
         [\phi_S + i\phi_P]/\sqrt{2}\right)^T$
         &$\left(1,2,1/2,-\right)$\\
         \hline
  \end{tabular}
  \caption{Field content of the model along with their quantum numbers
    under $SU(3)_C\otimes SU(2)_L \otimes U(1)_Y \otimes Z_2$.
    Here $i=1\dots3$ and $\alpha=1,2$ are the generation indices for the SM and $Z_2$-odd fermions, respectively.}
     \label{table:fields}
\end{table}
The fields in the model are listed in Table~\ref{table:fields}. It
might seem that the absence of $N^S_{\alpha L}$ violates our stated
principle of introducing vector-like fermions alone. However, the
introduction of such fields would only add a further layer of
complication in the neutrino sector, without bringing in any
qualitative change to the phenomenology, whether in this sector or in
any other.  And since, with the $N^S_{\alpha R}$ being gauge singlets,
the absence of the left-handed counterparts does not introduce any
gauge or chiral anomalies, we omit such $N^S_{\alpha L}$ from
consideration.
  
\vspace{0.25cm}
\subsection{The spin-zero spectrum}
\vspace{0.25cm}
The most general gauge and $Z_2$ invariant renormalizable potential reads
\beq
\barr{rcl}
V(H, \Phi)=&\dis-& \mu_H^2 H^\dagger H + \lambda_H \lfb H^\dagger H\rfb^2
+ {\mu^2_{\Phi}}\lfb \Phi^{\dagger}\Phi \rfb
+ \lambda_{\Phi}\lfb{\Phi}^{\dagger}\Phi\rfb^2\\[1ex]
&+&\dis \lambda_1\lfb{H}^{\dagger}H\rfb\lfb{\Phi}^{\dagger}\Phi\rfb
+ \lambda_2\left|{H^{\dagger}\Phi}\right|^2
+ \ltb {\color{blue}{\lambda_3}} ({H}^{\dagger}\Phi)^2+h.c.\rtb .
\earr
\label{eqn:lag_scalar}
\eeq
Without loss of generality, the coupling $\lambda_3$ can be taken to be real, as any complex phase can be absorbed by a redefinition of the scalar fields. The stability of the vacuum, corresponding to the potential of
Eq.~\ref{eqn:lag_scalar} requires that
\beq
\lambda_{H, \Phi} > 0 ,~~~~~~~
\lambda_{1} > -2 \sqrt{\lambda_{H} \lambda_{\Phi}} ,~~~~~~~
\lambda_{1}+\lambda_{2} \pm 2 |\lambda_{3}| > -2 \sqrt{\lambda_{H} \lambda_{\Phi}}.
\eeq
while the absence of charge-breaking minima implies that
\beq
\lambda_{2} - 2 |\lambda_{3}| < 0.
\eeq
As long as $\mu_{H}^2 > 0$, electroweak symmetry breaking (EWSB)
may be achieved by the neutral
component of $H$ acquiring a nonzero vacuum expectation value (vev),
$v$, {\em viz.,}
\beq
\langle H^\dagger H \rangle = \frac{\mu_H^2}{2 \lambda_H} \equiv \frac{v^2}{2} \ .
\eeq
Since we would not want the $Z_2$ symmetry to be broken, we further need 
\beq
\mu_\Phi^2 + \frac{\mu_H^2}{2 \lambda_H} \lambda_1 > 0 \ , \qquad
\mu_\Phi^2 + \frac{\mu_H^2}{2 \lambda_H} \lfb \lambda_{1}+\lambda_{2} - 2 |\lambda_{3}| \rfb > 0.
\eeq
In the absence of $Z_2$ breaking, there is no mixing between the
components of $H$ and $\Phi$, not only at the tree-level, but to
all orders. And while each of $\mu_H^2$ and $\lambda_H$ would receive
quantum corrections from the Higgs-sector couplings (as also the new
Yukawa couplings that we would see shortly), we shall neglect such
effects in the course of this paper.

After EWSB, the (tree-level) masses of the $Z_2$-odd charged scalar
($\phi^\pm$), neutral scalar ($\phi_S$) and pseudo-scalar ($\phi_P$)
are given by
\beq
m^2_{\phi^+}~=~{\mu_{\Phi}^2+\frac{v^2}{2} \lambda_1},~~~~ m^2_{\phi_S} = {\mu_{\Phi}^2+\frac{v^2}{2} \lfb \lambda_1+\lambda_2{+}2\lambda_3 \rfb},~~~~ m^2_{\phi_P} = {\mu_{\Phi}^2+\frac{v^2}{2} \lfb \lambda_1+\lambda_2{-}2\lambda_3 \rfb}.
\label{eqn:scalar_mass}
\eeq
The non-observation of charged scalars at the LHC implies $m_{\phi^+} >
80$ GeV \cite{Zyla:2020zbs}. It
  would turn out, though, that an analytic understanding of several
  constraints is easier if the mass splitting between the neutral
  $Z_2$-odd scalar and the pseudo-scalar is relatively small. This is
  most easily arranged for when $\mu_\Phi \gg v$ with only
  perturbativity limits being imposed on $\lambda_3$. The consequent
  splitting is easily seen to be $m_{\phi_S}-m_{\phi_P} \approx
  \lambda_3 v^2/ \mu_\Phi$. 
  
\vspace{0.25cm}
\subsection{The $Z_2$-odd fermions}
\vspace{0.25cm}
The rich structure of the theory, with the existence of both $Z_2$-odd
fermions and a $Z_2$-odd scalar doublet, permits a variety of new mass
and interaction terms. As these are crucial for the phenomenology, we
present here a brief discussion of the same.

\vspace{0.25cm}
\subsubsection{Direct mass terms}
\vspace{0.25cm}
The vector-like nature permits bare terms ({\em i.e.}, sans a Higgs) for the new fermions.
These can be both Dirac-like and Majorana-like, namely
\beq
\barr{rcl}
\lag_{\rm Mass}&\supset& \dis
m_Q^{\alpha\beta}\, \overline{Q_{\alpha L}}Q_{\beta R}
+ m_U^{\alpha\beta} \, \overline{U^S_{\alpha L}}U^S_{\beta R}+
m_D^{\alpha\beta}\, \overline{D^S_{\alpha L}}D^S_{\beta R}
\\[1ex]
& + & \dis
m_L^{\alpha\beta}\, \overline{L_{\alpha L}}L_{\beta R}~+~
m_E^{\alpha\beta}\, \overline{E^S_{\alpha L}}E^S_{\beta R}
+\frac{1}{2} \, m_N^{\alpha\beta}\, \overline{\left(N^S_{\alpha R}\right)^c} \, N^S_{\beta R} + {\rm H.c.},
  \earr
  \label{eqn:lag_mass}
\eeq
where $m_{Q, U, D, L, E, N}$ are, in general, complex $2\times 2$
matrices with $m_N$ being symmetric.  Without any loss of generality,
though, we may consider these to be diagonal.

Unlike the Higgs-mediated mass terms, the largest of which cannot far
exceed the EWSB scale, the eigenvalues of the aforementioned matrices
could, in principle, assume any value, constrained only by the cutoff
scale of the theory, if any. We exploit this freedom to invoke
vector-lepton masses at the TeV scale (motivated by the need to
address leptonic observables), while allowing the quarks to be very
heavy. 
The latter choice not only allows us to evade the
constraints from the LHC but
also (as we would see later) facilitates unification of gauge couplings. And while such a mass-splitting may seem arbitrary, it
is technically natural.  
\vspace{0.25cm}
\subsubsection{The Yukawa lagrangian}
\vspace{0.25cm}
Apart from the usual Yukawa terms involving the SM fermions alone, we
now have a whole set of new ones. These can be divided into classes,
those involving the SM Higgs $H$ and those involving $\Phi$. The
former can be represented (with $\tilde H \equiv i\sigma_2 H^*$) as
\beq
\barr{rcl}
\lag_{\rm Yukawa}^H &\supset& \dis
\zLU^{\alpha\beta}\, \overline{Q_{\alpha L}}\tilde H U^S_{\beta R}
+\zRU^{\alpha\beta} \, \overline{Q_{\alpha R}}\tilde H U^S_{\beta L}
+ \zLD^{\alpha\beta}\, \overline{Q_{\alpha L}}H D^S_{\beta R}
+ \zRD^{\alpha\beta}\, \overline{Q_{\alpha R}}H D^S_{\beta L} \\[1ex]
&+& \dis \zLN^{\alpha\beta}\, \overline{L_{\alpha L}}\tilde H N^S_{\beta R}
+ \zLE^{\alpha\beta} \, \overline{L_{\alpha L}}H E^S_{\beta R}
+ \, \zRE^{\alpha\beta}\, \overline{L_{\alpha R}}H E^S_{\beta L} + {\rm H.c.}\ ,
\earr
\label{eqn:lag_yuk_H}
\eeq
where $\zLN,~\zLE,~\zRE,~\zLU,~\zLD,~\zRU~{\rm and}~\zRD$ are
$2\times2$ complex matrices. Only some of the phases can be reabsorbed
by phase redefinitions of the vector-like fermion fields. For example,
in the presence of nonzero Majorana mass terms ($m_N$), phase
redefinitions of $N_{\beta R}^S$'s are not possible (without
introducing phases in $m_N$).  However, two phases
of $\zLN$ can be absorbed in $L_{\alpha L}$. Similarly, $E_{\beta
  R}^S$ can absorb two phases of $\zLE$. Once
$L_{\alpha L}$ and $E_{\beta R}^S$ are phase redefined, no further
redefinitions of $L_{\alpha R}$ and $E_{\beta L}^S$ are
possible without introducing additional phases in $m_L$ and $m_E$,
respectively. Therefore, in the basis where $m_L,~m_E~{\rm and}~m_N$
are diagonal with real positive diagonal elements, $\zLN$ and $\zLE$
are defined by six real parameters while $\zRE$ needs
 eight. Analogous arguments follow for the quark
sector too.

While, after EWSB, the terms in Eq.~\ref{eqn:lag_yuk_H} would contribute
to the masses of the vector-like fermions, these contributions would
be expected to be small compared to the direct terms as in
Eq.~\ref{eqn:lag_mass}.  Consequently, their major role would be to
introduce small mass splittings and mixings. The splittings,
especially between the leptonic states would turn out to be crucial in
deciding the DM relic density. 

Turning to Yukawa interactions terms involving $\Phi$, these can be
represented (with $\tilde \Phi \equiv i\sigma_2 \Phi^*$) as
\beq
\barr{rcl}
\lag_{\rm Yukawa}^{\Phi}&=& \dis
\yqU^{i\alpha} \, \overline{q_{i L}}\tilde \Phi U^S_{\alpha R} +
\yuQ^{i\alpha} \, \overline{u_{i R}}{\tilde \Phi}^\dagger Q_{\alpha L} +
\yqD^{i\alpha} \, \overline{q_{i L}}\Phi D^S_{\alpha R} +
\ydQ^{i\alpha} \, \overline{d_{i R}}{\Phi}^\dagger Q_{\alpha L}
\\[1.5ex]
& + & \dis
\ylN^{i\alpha} \, \overline{l_{i L}}\tilde \Phi N^S_{\alpha R} +
\ylE^{i\alpha} \, \overline{l_{i L}}\Phi E^S_{\alpha R} +
\yeL^{i\alpha} \, \overline{e_{i R}}\Phi^{\dagger} L_{\alpha L}
+ {\rm H.c.},
\earr
\label{eqn:lag_yuk_phi}
\eeq
where $\ylN,~\ylE,~\yeL,~\yqU,~\yqD,~\yuQ~{\rm and}~\ydQ$ are $3\times
2$ complex matrices with each being defined by  six
complex (equivalently,  twelve real) parameters.
Again, some of these are unphysical.  For example, concentrating on
the leptonic sector, the fields $l_{iL}$ can be redefined to absorb
 three phases from $\ylN$.  Note that, once this choice
is made, no other phase can be absorbed. Analogous arguments are
applicable to the quark sector as well.
  
\vspace{0.25cm}
\subsubsection{The $Z_2$-odd fermion spectrum}
\label{sec:ana_mixing}
\vspace{0.25cm}
While the bulk of the masses for the $Z_2$-odd fermions 
are expected to arise from Eq.~\ref{eqn:lag_mass}, the
post-EWSB contribution due to Eq.~\ref{eqn:lag_yuk_H} can be
non-negligible, thereby introducing substantial mixing in the
$Z_2$-odd fermion sector. In the gauge basis, the resultant mass
matrices can be expressed through
\beq
\lag_{\rm Mass}= \frac{1}{2} \overline{(\Psi^N_R)^c} \, {\cal M}_{N} \,\Psi^N_R
+ \overline{\Psi^E_{L}} \, {\cal M}_{E} \,\Psi^E_{R}~+~
\overline{\Psi_{L}^U} \, {\cal M}_{U} \, \Psi_{R}^U~+~
\overline{\Psi_{L}^D} \, {\cal M}_{D} \, \Psi_{R}^D~+~{\rm h.c.},
\label{eqn:lag_mix}
\eeq
where
\beq
\Psi^N_R~=~\lfb {\begin{array}{c}
   (N^D_L)^c\\
   {N^D_R}\\
   {N^S_R}\\
\end{array} } \rfb,~
\Psi^E_{L(R)}~=~\lfb {\begin{array}{c}
   E^D\\
   E^S\\
\end{array} } \rfb_{L(R)},~
\Psi^U_{L(R)}~=~\lfb {\begin{array}{c}
   U^D\\
   U^S\\
\end{array} } \rfb_{L(R)},~
\Psi^D_{L(R)}~=~\lfb {\begin{array}{c}
   D^D\\
   D^S\\
\end{array} } \rfb_{L(R)},~\nonumber
\eeq
where the generation indices ($\alpha = 1,2$ for each field type) have been
subsumed. The matrices in
Eq.~\ref{eqn:lag_mix} can be obtained from Eqns.~\ref{eqn:lag_mass} and
\ref{eqn:lag_yuk_H} and are given by
  \beq \dis
{\cal M}_N~=~
  \lfb {\begin{array}{ccc}
   {0} & {m_L} & \dis \frac{v}{\sqrt 2}\zLN \\[1ex]
   {m_L^T} & {0} & {0} \\[1ex]
   \dis \frac{v}{\sqrt 2}\zLN^T & {0} & {m_N} \\
  \end{array} } \rfb, \qquad \quad
{\cal M}_E~=~
  \lfb {\begin{array}{cc}
   m_L &  \dis \frac{v}{\sqrt 2}\zLE\\
  \dis \frac{v}{\sqrt 2}\zRE^\dagger & m_E\\
  \end{array} } \rfb.
  \label{eqn:mass_matrices}
  \eeq
The structures of ${\cal M}_{U(D)}$ are similar to that for ${\cal M}_E$ with
$m_L,~m_E,~\zLE~{\rm and}~\zRE$ replaced by
$m_Q,~m_{U(D)},~\zLU(\zLD)$ and $\zRU(\zRD)$, respectively.  ${\cal M}_N$,
being a symmetric matrix, can be diagonalized by a unitary matrix
$U_N$, namely ${\cal M}_N^{\rm diag}~=~U_N^T {\cal M}_N U_N$. 
Being arbitrary complex matrices, the diagonalization of ${\cal M}_E({\cal M}_{U(D)})$
can only proceed through a biunitary transformation (allowed since the
left- and right-handed fields can be rotated independently), namely
${\cal M}_{E(U)}^{\rm diag}~=~U_L^{E(U)^\dagger}{\cal M}_{E(U)}U_R^{E(U)}$. These rotation matrices
relate the mass and  gauge eigenstates through
\beq
\widetilde \Psi_R^N \equiv \lfb {\begin{array}{c}
   {{\ND}_X}\\
   {{\ND}_Y}\\
   {\widetilde N^S}\\
\end{array}} \rfb_{\hskip -5pt R}~=~U_N^\dagger \lfb {\begin{array}{c}
   {(N^D_L)^c}\\
   {N^D_R}\\
   {N^S_R}\\
\end{array} } \rfb,  \qquad
\widetilde \Psi^E_{L(R)} \equiv \lfb {\begin{array}{c}
   \widetilde E^D\\
   \widetilde E^S\\
\end{array} } \rfb_{\hskip -5pt L(R)}~=~U_{L(R)}^{E^\dagger}\lfb {\begin{array}{c}
   E^D\\
   E^S\\
\end{array} } \rfb_{\hskip -5pt L(R)}.
\label{eqn:mass_basis}
\eeq
Here, $\widetilde N^D_{a,b}$ denote the mass eigenstates that are
dominated by the left(right)-handed doublet fields while $\widetilde
N^S$ are dominated by the gauge-singlets.  While such a nomenclature
might seem strange, it would turn out to be useful in understanding
 the loop-mediated effects.

While the exact diagonalization can be achieved numerically, it is
useful to obtain analytic results, even approximate ones, if only as an aid to understanding
the dependence of different low energy observables (to be undertaken
in the next section) on the different parameters of this sector. This
is particularly straightforward when the EWSB-generated mass terms are much
smaller than the direct ones. Starting with ${\cal M}_N$, at the first
step, it can be approximately block-diagonalized
using a unitary matrix $U_N^0$, {\em viz.,}
\beq
{U_N^0}^T {\cal M}_N U_N^0 \approx {\cal M}_N^{(1)} = 
  \lfb {\begin{array}{ccc}
\dis   -\textfrak{N}_1 & {m_L} & 0 \\
   {m_L^T} & {0} & {0} \\
   0 & {0} & {m_N} \\
  \end{array} } \rfb ,
  \label{eqn:step1}
  \eeq
  where $\textfrak{N}_1 = \frac{v^2}{2}\zLN m_N^{-1}\zLN^T~~$ and it has
   been assumed that the eigenvalues of $v^2 \zLN \zLN^T$
  are much smaller than those of $m_N^2$. Indeed, a correction $\sim
  v^2 \zLN^T m_N^{-1} \zLN$ to the ``33'' block submatrix $m_N$ has
  been omitted in the expression above.  The mixing induced Majorana
  mass term ( ``11'' element of the matrix on the right side of
  Eq.~\ref{eqn:step1}) $-\frac{v^2}{2}\zLN m_N^{-1}\zLN^T$, while small
  compared to $m_{L,R}$ is, however, phenomenologically important and,
  hence, retained. Working in the basis where $m_{L(N)}$ is diagonal, following Ref.~\cite{Grimus:2000vj},
  the unitary matrix $U_N^0$ can be written as
\beq U_N^0~=~ \lfb {\begin{array}{cc} 1-\frac{1}{2}B B^\dagger &
     B \\ -B^\dagger & 1-\frac{1}{2}B^\dagger B \\
\end{array} } \rfb.  
\eeq
where
  \beq
 B \equiv 
 \frac{v}{\sqrt 2}
 \lfb {\begin{array}{c}
   B_1\\
   B_2\\
  \end{array} } \rfb m_N^{-1} \, \approx \, \frac{v}{\sqrt 2}
 \lfb {\begin{array}{c}
   \zLN^*+m_L^2\zLN^*m_N^{-2}+m_L^4\zLN^*m_N^{-4}+....\\
    m_L\zLN m_N^{-1} + m_L^3\zLN m_N^{-3}+m_L^5\zLN m_N^{-5}+....\\
 \end{array} } \rfb m_N^{-1},  \nonumber
 \eeq
 with the approximation being valid for the case of the Majorana
 masses being substantially larger than Dirac
 masses\footnote{Analogous expansions can be obtained for Dirac
   masses larger than the Majorana masses.}. Here, $B_1$ is a $2
 \times 2$ matrix satisfying the Sylvester equation, namely,
 $m_L^2B_1-B_1m_N^2~=~-\zLN^*m_N^2$ whereas
 $B_2~=~m_LB_1^*m_N^{-1}$. For
 diagonal $m_N$ and $m_L$, the elements of $B_1$ are given by
\beq
 B_1^{\alpha\beta}~=~ \frac{\lfb m_N^{\beta\beta}\rfb^2}{\lfb m_N^{\beta\beta}\rfb^2-\lfb m_L^{\alpha\alpha}\rfb^2}\lfb{\zLN^*}\rfb^{\alpha\beta}.
\eeq
 If we make a
 simplifying assumption of quasi-universal masses for the heavy
 sector, namely $m_L \approx {\rm diag}(\mL, \mL)$ and $m_N \approx 
     {\rm diag}(\mN, \mN)$, the matrices
     $B_{1,2}$ can be written a compact form given by $B_1~=~\lfb
   1-\epsilon_N^2\rfb^{-1}\zLN^*$ and $B_2~=~\lfb 1-\epsilon_N^2
   \rfb^{-1} \epsilon_N \zLN$ where $\epsilon_N = \widetilde
   m_L/\mN$. In this {\em simplified scenario}, $U_N^0$ can
   be written as
\beq
 U_N^0~=~
 \lfb {\begin{array}{ccc}
   1-\chi_N^2\zLN^* \zLN^T ~~& -\epsilon_N\chi_N^2\zLN^* \zLN^\dagger ~~& \sqrt 2 \chi_N \zLN^*\\
   -\epsilon_N\chi_N^2 \zLN \zLN^T  & 1-\epsilon_N^2\chi_N^2\zLN \zLN^\dagger & \sqrt 2 \epsilon_N \chi_N \zLN\\
   -\sqrt 2 \chi_N \zLN^T & -\sqrt 2 \epsilon_N \chi_N \zLN^\dagger & 1-\chi_N^2\lfb \zLN^T \zLN^*+\epsilon^2_N \zLN^\dagger \zLN \rfb\\
 \end{array} } \rfb,
 \label{eqn:UN0_1}
\eeq
 where $\chi_N = v / [2 \lfb 1-\epsilon_N^2 \rfb 
\mN] ~\sim~0.1$ for TeV scale $\mN$ and
 $\mL$.  It can be easily checked that the same $U_N^0$ is
 obtained by repeating the steps for the case of ${\widetilde
   m_L}>{\mN}$. With a little rearrangement, the matrix can be expressed as: 
\beq
 U_N^0 ~=~
 \lfb {\begin{array}{ccc}
   1-\epsilon_L^2 \chi_L^2 \zLN^* \zLN^T ~~& -\epsilon_L \chi_L^2\zLN^* \zLN^{\dagger} ~~& -\sqrt 2 \epsilon_L \chi_L \zLN^*\\
   -\epsilon_L\chi_L^2 \zLN \zLN^T  & 1-\chi_L^2 \zLN \zLN^{\dagger} & -\sqrt{2} \chi_L \zLN\\
   \sqrt 2 \epsilon_L \chi_L \zLN^T & \sqrt 2 \chi_L \zLN^{\dagger} & 1-\chi_L^2 \lfb \epsilon_L^2  \zLN^T \zLN^* + \zLN^{\dagger} \zLN \rfb \\
 \end{array} } \rfb, 
 \label{eqn:UN0_2}
\eeq
where
\beq
\epsilon_L~=~\frac{\mN}{\mL}, ~~\chi_L~=~\frac{1}{2\lfb 1-\epsilon_L^2 \rfb} \frac{v}{\mL}. \nonumber
\eeq
\vskip 5pt
 \noindent{\em Step 2:} Diagonalize the $2\times 2$ non-diagonal block
 of the matrix on the right hand side of Eq.~\ref{eqn:step1}. For Dirac
 masses much larger than the elements of $\textfrak{N}_1$, an approximate 
 diagonalization can be achieved through another matrix $U_N^1$ namely,
\beq
{U_N^1}^T {\cal M}_N^{(1)} U_N^1~=~\lfb {\begin{array}{ccc}
     m_L-\, \frac{1}{2} \textfrak{N}_1 & 0 & 0\\
     0 & -m_L -\, \frac{1}{2} \textfrak{N}_1 & 0\\
     0 & 0 & m_N\\
     \end{array} }\rfb,
\eeq
where
\beq
 {U_N^1} ~=~ \frac{1}{\sqrt 2} \lfb {\begin{array}{ccc}
    1 & -1 & 0 \\
   1 & 1 & 0 \\
   0 & 0 &\sqrt 2 \\
\end{array} } \rfb.
\label{eqn:UN1}
\eeq
 The mixing matrix $U_N$ introduced in Eq.~\ref{eqn:mass_basis} is,
 thus, approximated by $U_N~=~U_N^0 U_N^1$. It is important to note
 that the mixings between doublet and singlet neutral fermions
 introduce a small mass splitting (${\cal O}\lfb v^2\zLN^2/m_N\rfb$)
 between the predominantly doublet states and, hence, give rise to a pseudo-Dirac
 pair. As we shall see
 later, this has profound consequences in the context of dark matter
 phenomenology (especially in the context of direct detection).

Diagonalising the charged lepton mass matrix $M_E$ needs a bi-unitary
transformation on account of its being non-hermitian in general. If
$U_L$ and $U_R$ diagonalise the matrices $M_EM_E^{\dagger}$ and 
$M_E^{\dagger}M_E$ respectively, then
\beq
M_E^D \equiv U_L^{\dagger}M_EU_R
\eeq
is diagonal.  While such a diagonalization can be carried analogously
to that for $M_N$, in the limit of quasi-universal masses for the
heavy sector, {\em viz.,}  $m_E~=~\mE\times I_{2\times 2}$
and $m_L~=~\mL\times I_{2\times 2}$, the form of $U_{L,R}$
is simplified considerably yielding
\beq
 {U_L} \approx \lfb {\begin{array}{cc}
    I_{2\times 2} & \chi_E \left(\zLE + \epsilon_E \zRE \right) \\
    -\chi_E \left(\zLE^{\dagger}+\epsilon_E \zRE^{\dagger} \right)  & I_{2\times 2}, \\
 \end{array} } \rfb,
\eeq
and,
\beq
 {U_R} ~=~\lfb {\begin{array}{cc}
    I_{2\times 2} & \chi_E \left(\epsilon_E \zLE + \zRE \right) \\
    -\chi_E \left(\epsilon_E \zLE^{\dagger} + \zRE^{\dagger} \right)  & I_{2\times 2} \\
 \end{array} } \rfb,
\eeq
where $\chi_{E}= v/\lfb \sqrt{2} \mE~(1-\epsilon_E^2)\rfb$ with
$\epsilon_E \equiv \mL / \mE$.
Finally,
\beq
 {U_L^{\dagger}} \lfb {\begin{array}{cc}
   m_L &  \frac{v}{\sqrt 2}\zLE\\
   \frac{v}{\sqrt 2}\zRE^\dagger & m_E\\
  \end{array} } \rfb {U_R} \sim \lfb {\begin{array}{cc}
   \mL I_{2\times 2} - \textfrak{N}_2 & 0 \\
   0 & \mE I_{2\times 2} + \textfrak{N}_3,\\
  \end{array} } \rfb,
\eeq
where $\textfrak{N}_2=-\frac{v^2}{4 \mL} \epsilon_E^2 \lfb \zRE \zRE^T +\zLE \zRE^T + \zRE \zLE^T \rfb $ and $\textfrak{N}_3=\frac{v^2}{4 \mE} \lfb \zLE^T \zLE + \zRE^T \zRE \rfb + \frac{v^2}{4 \mE} \epsilon_E^2 \lfb \zRE^T \zRE + \zLE^T \zRE +  \zRE^T \zLE \rfb $, considering real $\zRE(\zLE)$.
\section{Unification of gauge couplings}
\label{sec:unification}
\vspace{0.25cm}
The presence of the additional fermions and scalars affects the
running of the gauge couplings above the TeV
scale. The exotic particles contribute to SM $\beta$-functions 
in the same way as the SM fermions and the Higgs doublet, 
with the exception that for the
vector-like counterparts to the SM fermions, each contribution gets
doubled with both left- and right-handed components contributing equally. The expressions are given in Appendix~\ref{app:beta}.

To this order, the 
renormalization group equations (RGEs) 
for the couplings ($g_i$) can be expressed as
\beq
\frac{dg_i}{d \ln Q}=\beta_i {(g_i)},
\label{eqn:running}
\eeq
where 
$\beta_i {(g_i)}$ are the $\beta$-functions and
$Q$ denotes the scale at which the couplings are being considered,
  with the boundary values fixed experimentally at $Q = m_Z$.
  For our purpose, it suffices to consider the $\beta$-functions upto
    two-loops.
 At this order of sophistication, one
  could neglect threshold effects and include the contributions of a
  new species $J$ only for $Q > m_J$. Within this approximation, thus, the
  $\beta$-functions would be composed of a series of step functions.
  
For\footnote{$g_1=\sqrt{5/3}~g_Y$} $i=1,2,3$, $g_i$ denotes the coupling for $U(1)_Y$, $SU(2)_L$ and $SU(3)_C$, respectively. In order to solve Eq.~\ref{eqn:running}, 
we also take into account the contribution of Top-Yukawa coupling, $Y_t$ and the scalar quartic coupling, $\lambda_H$ as well, represented as $g_4$ and $g_5$, respectively. Together with the equations for $g_1,g_2$ and $g_3$, this gives us five coupled differential equations. Representing the scale by $t=\ln Q$ 
such that $t_0=\ln m_Z$ denotes the scale for SM, $t_1$ denotes the next higher energy scale at which one or more BSM particles are introduced and so on,  
the equations can be solved numerically in each region $(t_{n-1},t_n)$ to obtain the solutions $g_i^{(n)} (t)$, using the boundary conditions $g_i^{(n)} (t_{n-1}) = g_i^{(n-1)} (t_{n-1})$.

We assume the scale for $Z_2$-odd leptons and the scalar $\Phi$ as $1$ TeV while $Z_2$-odd quarks are assumed to be much heavier such that there is a unification of the gauge couplings at a next higher scale ($t_G$), given by the condition
\beq
g_1 (t_G) = g_2 (t_G) = g_3 (t_G).
\label{eqn:unif}
\eeq
\begin{figure}[h!]
\begin{center}
\includegraphics[width=7cm,height=5.5cm]{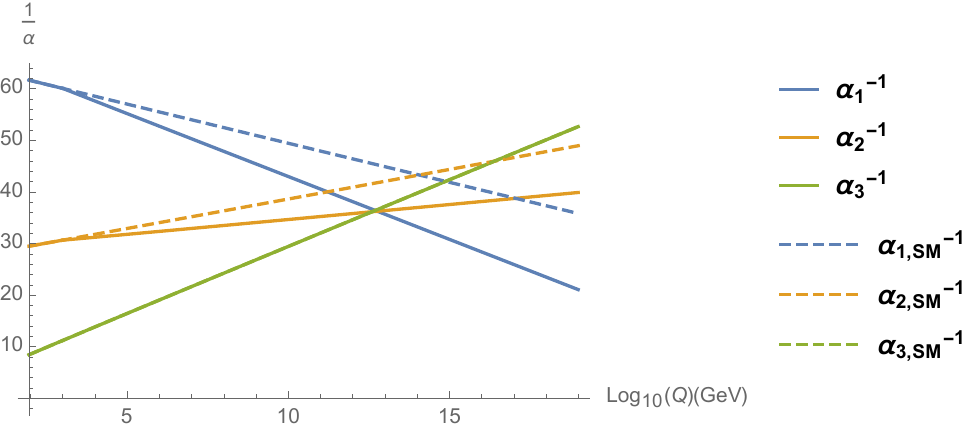}
\caption{Running of couplings $g_1, g_2$ and $g_3$ and their unification at 2-loop level, in terms of $\alpha^{-1}$, in SM (dashed lines) $vs$ BSM (solid lines). Blue, amber and green represent $g_1$, $g_2$ and $g_3$, respectively. The BSM consists of inert scalar doublet and two generations of $Z_2$-odd vector-like leptons, with masses of ${\cal{O}}(1 ~\text{TeV})$.}
\label{fig:unif1}
\end{center}
\end{figure}
In Fig.~\ref{fig:unif1}, we first illustrate the unification of gauge couplings, considering only the inert scalar doublet and two generations of $Z_2$-odd vector-like leptons as the BSM content, all with masses of order ${\cal O}(1~\text{TeV})$. Without any BSM quarks, the slope of $\alpha_3^{-1}$ is unchanged, in comparison to the SM. On the other hand, the slopes of $\alpha_1^{-1}$ and $\alpha_2^{-1}$ are increased as both get positive contribution from the BSM. This brings down the scale of unification.
We also observe that the presence of new particles raises the value of the unified gauge coupling. Embedding the model into grand unified theories may predict proton decay which has not been observed experimentally. 
The most-stringent limit on proton lifetime comes from  Super-Kamiokande \cite{Proceedings:2012ulb} {\em viz.,}  $\tau (p \to \pi^0 e^+) > 2.4 \times 10^{34}$ years. Although the proton lifetime is model-dependent, a naive estimate \cite{Nath:2006ut} is $\tau \sim M_G^4/(\alpha_G^2m_p^5)$ where $M_G$ is the unification scale, $\alpha_G=\frac{g_G^2}{4\pi}$, $g_G$ being the coupling value at $M_G$ and $m_p$ is the proton mass. 
For $1/\alpha_G \sim 30$, we require $M_G \gtrsim 6 \times 10^{15}$ GeV.

\begin{figure}[h!]
\begin{center}
\includegraphics[width=7cm,height=5.5cm]{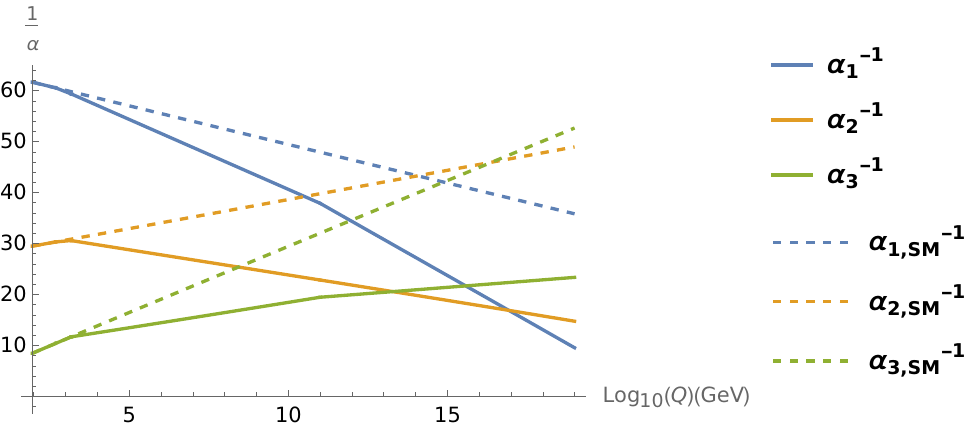}
\includegraphics[width=7cm,height=5.5cm]{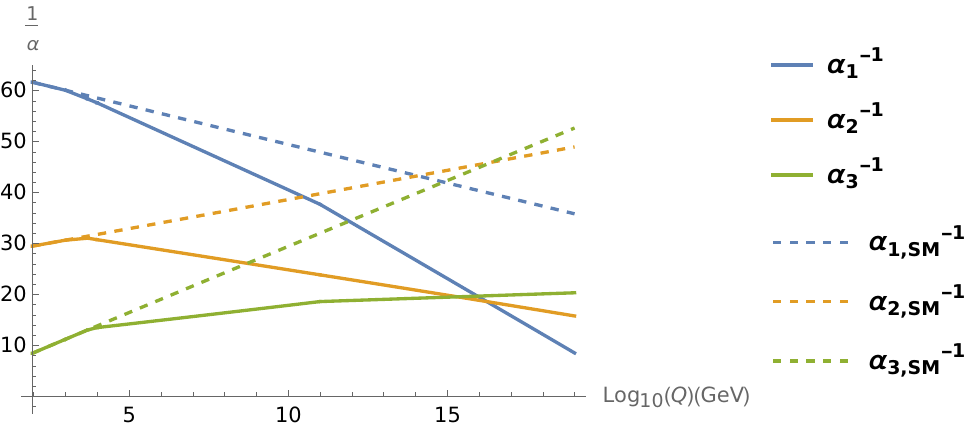}
\caption{Running of couplings $g_1, g_2$ and $g_3$ and their unification at 2-loop level, in terms of $\alpha^{-1}$, in SM (dashed lines) $vs$ the new model (solid lines). Blue, amber and green represent $g_1$, $g_2$ and $g_3$, respectively. The two figures correspond to following mass scales and no. of generations for the $Z_2$-odd quarks: Left: $2 \cdot 10^3$ GeV ($Q_{L/R} \times 2$ and $D^S_{L/R} \times 2$) and $10^{11}$ GeV ($U^S_{L/R} \times 2$); Right: $5 \cdot 10^3$ GeV ($Q_{L/R} \times 2$), $10^4$ GeV ($D^S_{L/R} \times 3$) and $10^{11}$ GeV ($U^S_{L/R} \times 2$).}
\label{fig:unif}
\end{center}
\end{figure}

With more number of scalars or vector-like leptons, the unification scale would be further decreased. The only plausible way to achieve $M_G \gtrsim 10^{16}$ GeV seems to be addition of quarks. In order to estimate the scale of the exotic quarks for unification, we utilise the RGEs upto 1-loop, 
which are a set of three decoupled equations corresponding to the gauge couplings $g_1$, $g_2$ and $g_3$.
The calculation is shown in Appendix~\ref{app:rge_1loop}. 
Using these scales for defining the $\beta$-functions, we numerically solve the RGEs upto two-loops {\em i.e.,} Eq.~\ref{eqn:running} for the couplings $g_i$. Out of the many solutions, two are plotted in Fig.~\ref{fig:unif}.

In Fig.~\ref{fig:unif} (Left), we have considered two generations of each of the exotic quarks and find out that a unification scale above $10^{15}$ GeV requires the exotic singlet up-type quarks to be at a very high scale, just a few orders shy of the unification scale while the exotic doublet quarks as well as the singlet down-type quarks must be close to the other $Z_2$-odd particles \footnote{
Due to its high hypercharge, $U^S_{L/R}$ greatly affects the slope of $\alpha_1^{-1}$, bringing down the unification scale while $Q_{L/R}$ and $D^S_{L/R}$ mainly affect the slopes of $\alpha_2^{-1}$ and $\alpha_3^{-1}$
.}. 
However, 
it is difficult to push the unification scale to $10^{16}$ GeV. Observing that the singlet down-type quark, $D^S_{L/R}$, considerably affects the slope of $\alpha_3^{-1}$ while its effect on the slope of $\alpha_1^{-1}$ is negligible, we are able to achieve unification at $m_G \sim 10^{16}$ GeV\footnote{We note that a unification scale of order $10^{16}\,\mathrm{GeV}$ is close to the lower limit required to satisfy current bounds on proton decay from dimension-6 operators. The precise constraint on the unification scale depends on the value of the unified gauge coupling as well as possible threshold corrections and model-dependent effects.
Therefore, the value obtained here should be regarded as marginally consistent with present proton decay limits.} by considering an extra generation of $D^S_{L/R}$. This is shown in Fig.~\ref{fig:unif} (Right). Similar results can be achieved by instead considering an extra generation of the doublet VL quarks ($Q_{L/R}$) at somewhat different mass scales for the VL quarks. The interpretation of these results remains open, leaving room for curiosity.
\section{Radiative neutrino mass generation}
\label{sec:numass}
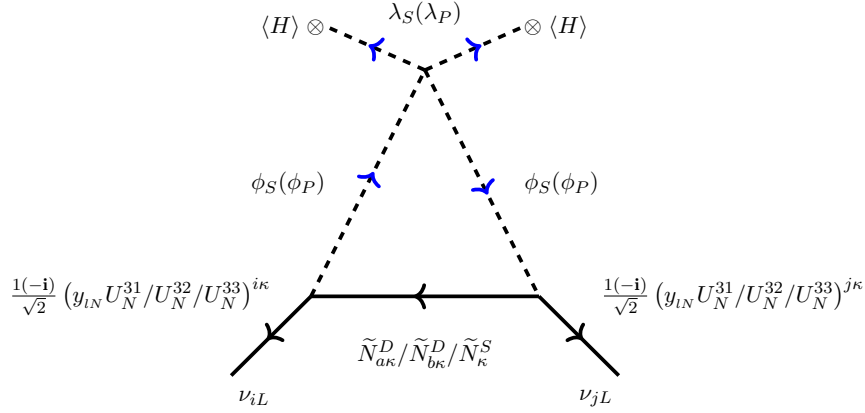
\begin{figure}[!t]
  \centering
  \begin{minipage}{0.8\textwidth}
  \hspace{1cm}
    \begin{tikzpicture}[line width=1.4 pt, scale=1.5,every node/.style={scale=1}]
      \draw[fermion,black] (0.0,0.0)  --(-0.7,-0.7);
      \draw[fermion,black] (2,0) --(0,0);
      \draw[fermion,black] (2,0) --(2.7,-0.7);

      \draw[scalar,black] (0,0)  --(1,2.0);
      \draw[scalar,black] (1,2)  --(2,0);
      \draw[scalar,black] (1,2)  --(0.1,2.4);
      \draw[scalar,black] (1,2)  --(1.9,2.4);

      \node[scale=0.8] at (1,-0.5) {$\widetilde N^D_{a\kappa}/\widetilde N^D_{b\kappa}/\widetilde N^S_{\kappa}$};
      \node[scale=0.8] at (-0.2,1.0) {$\phi_S(\phi_P)$};
      \node[scale=0.8] at (2.2,1.0) {$\phi_S(\phi_P)$};
      \node[scale=0.8] at (1,2.5) {$\lambda_S(\lambda_P)$};
      \node[scale=0.8] at (-0.16,2.38) {$\langle H\rangle \; \otimes$};
      \node[scale=0.8] at (2.15,2.38) {$\otimes \; \langle H\rangle$};
      \node[scale=0.8] at (2.5,-0.9) {$\nu_{jL}$};
      \node[scale=0.8] at (-0.5,-0.9) {$\nu_{iL}$};
      \node[scale=0.8] at (-1.5,-0) {$ \frac{1(-\mathbf{i})}{\sqrt 2}\left(\ylN U_N^{31}/U_N^{32}/U_N^{33}\right)^{i\kappa}$};
      \node[scale=0.8] at (3.7,-0) {$\frac{1(-\mathbf{i})}{\sqrt 2}\left(\ylN U_N^{31}/U_N^{32}/U_N^{33}\right)^{j\kappa}$};
    \end{tikzpicture}
  \end{minipage}
  \caption{Feynman diagrams generating light neutrino masses at the
    1-loop level. There are two diagrams arising from $\phi_S$ and
    $\phi_P$ in the loop, respectively. Expressions corresponding to
    $\phi_P$ are indicated in parentheses. The coupling combinations
    $\lambda_{S,P} \equiv -\frac{1}{4} \left(\lambda_1 + \lambda_2 \pm 2\lambda_3\right)$. 
    Indices
    $\kappa=1,2$ denotes the generation index for the heavy neutrinos
    ${\ND}_X$, ${\ND}_Y$, and $\widetilde N^S$
    each while $i,j=1-3$ denote the SM generation indices.}
\label{fig:nu_mass}
\end{figure}

\vspace{0.25cm}
With the $Z_2$ remaining unbroken, there are no mass terms connecting
the SM neutrinos to the new fields and, thus, the
  former remain exactly massless at the tree level. However, at the
one-loop level, Weinberg operators are generated; the lepton
number violation inherent to such operators is
induced by the Majorana mass terms for the singlet
fermions $N_{\alpha R}^S$. 
The generic Feynman diagram illustrating the
generation of neutrino masses at the one-loop level is shown in Fig.~\ref{fig:nu_mass}, 
with all $Z_2$-odd particles expressed in the
mass basis. 
This kind of radiative generation of neutrino masses is reminiscent 
of the Scotogenic mechanism proposed in Ref.~\cite{Ma:2006km,Tao:1996vb}. 
The resultant contribution to the effective ($3\times 3$)
mass matrix for the light neutrino sector is given by
\beq
\frac{m_\nu}{\langle H \rangle^2}~=~\frac{\lambda_3}{16\pi^2} \,\, \ylN M_\Delta^{-1} \ylN^T,
\label{eqn:nu_mass_main}
\eeq
where $M_\Delta^{-1}$ is a $2\times 2$
  symmetric complex matrix and to the leading order\footnote{The mass
  splitting between the $Z_2$-odd scalar and pseudoscalar is given by
  $m_{\phi_S}-m_{\phi_P} = \lambda_3 \, v^2 / \mu_\Phi$ and hence, is
  estimated to be small (of the order of GeV) for $\mu_\Phi\sim$ TeV
  and $\lambda_3\sim0.1$. Therefore, the loop-integrals resulting
  from the scalar and pseudoscalar loops are nearly the same with 
  differences being of the order of $\lambda_3 \, v^2/\mu_\Phi^2$.}
   in  $\lambda_3 \, v^2/\mu_\Phi^2$, is
given by
\beq
\lfb M_\Delta^{-1} \rfb^{\alpha\beta}~=~\sum_{\gamma}\ltb \frac{U_N^{\alpha^\prime \gamma}\, I\lfb x_{\widetilde N_{\gamma}}\rfb\,U_N^{\beta^\prime \gamma}}{m_{\widetilde N_{\gamma}}}\rtb.
\label{eqn:m_delta}
\eeq
In the above equation, indices $\alpha,\beta=1,2$, $\alpha'=4+\alpha$ 
and $\beta'=4+\beta$ pertain to gauge basis whereas $\gamma=1-6$ represents 
the index for mass basis. $I\lfb x_{\widetilde N_{\gamma}}\rfb$ is the loop
factor with $x_{\widetilde N_{\gamma}}={m_{\phi_S}^2}/{m_{\widetilde
    N_{\gamma}}^2}$ and given by, $I(x)~=~-\lfb{\rm ln}\,x+1-x\rfb\lfb
x-1 \rfb^{-2}$. Please note that the analytical forms of the mixing
matrices, as derived in Sec.~\ref{sec:ana_mixing}, are presented in
the form of $2 \times 2$ block matrices, utilising the $2 \times 2$
Yukawa couplings outlined in Eq.~\ref{eqn:lag_yuk_H}. Consequently, it
would be advantageous to compute the Feynman diagrams for radiative
neutrino mass generation in this section and for lepton flavor
violation and charged lepton $g-2$ in the subsequent section,
employing the $2 \times 2$ blocks of the mixing matrix. This approach
will facilitate obtaining of analytical results,
expressed in terms of the Yukawa couplings. Thus, expanding the mass
basis as $\widetilde N$=\{${\ND}_X$, ${\ND}_Y$, $\widetilde N^S$\} 
and writing out these contributions more explicitly,
\beq
\lfb M_\Delta^{-1} \rfb^{\alpha\beta}~=\left[ U_N^{31}{\cal D}_{\widetilde N_{a}^D}\left(U_N^{31}\right)^T + U_N^{32}{\cal D}_{\widetilde N_{b}^D}\left(U_N^{31}\right)^T + U_N^{33}{\cal D}_{\widetilde N^S}\left(U_N^{33}\right)^T\right]^{\alpha \beta},
\label{eqn:m_delta_2}
\eeq
where we have denoted $2 \times 2$ blocks for mixing of $\NSR$ with each of 
${\ND}_{a}$, ${\ND}_Y$, and $\widetilde N^S$ as $U_N^{ij}$=`$ij^{th}$' $2 \times 2$ block of the $6 \times 6$ mixing matrix $U_N$ ($=U_N^0 U_N^1$) for the exotic neutral fermion sector (see equations~\ref{eqn:mass_basis}, \ref{eqn:UN0_1}, \ref{eqn:UN0_2} and \ref{eqn:UN1}). Factor ${I\lfb x_{\widetilde N_{\gamma}}\rfb}/{m_{\widetilde N_{\gamma}}}$ is encapsulated into $2 \times 2$ diagonal matrices ${\cal D}_Y={\rm Diag}\left[\frac{I\lfb x_{Y_1}\rfb}{m_{Y_1}}, \frac{I\lfb x_{Y_2}\rfb}{m_{Y_2}}\right]$. For TeV scale
masses of the $Z_2$-odd (pseudo)scalars and Majorana neutrinos {\em
  i.e.,} $x~\sim~{\cal O}\lfb 1 \rfb$, the loop factor $I(x)$ can be
estimated to be of ~${\cal O}\lfb 1 \rfb$\footnote{In
  particular, $$\lim_{x \to 1} I(x)~=~\frac{1}{2}$$.} and hence, $\lfb
M_\Delta^{-1} \rfb^{\alpha\beta}~\sim~{\cal O}\lfb {\rm TeV}^{-1}
\rfb$. Therefore, naively, one can estimate from
Eq.~\ref{eqn:nu_mass_main} that $\ylN~\sim~{\cal O}\lfb 10^{-5} \rfb$
and $\lambda_3~\sim~0.1$ result in neutrino masses of ${\cal O}\lfb
0.1~{\rm eV} \rfb$. It is important to note that $\ylN~\sim~{\cal
  O}\lfb 10^{-1} \rfb$ is also allowed, provided $\lambda_3~\sim~{\cal O}\lfb
10^{-9} \rfb$ 
.

In view of the relatively complicated structure of neutrino mass matrix, $m_\nu$, in Eq.~\ref{eqn:nu_mass_main}, it is instructive to look for an approximate simplified expression for $m_\nu$ in the {\em simplified scenario} ({\em i.e.,} $m_L~=~\mL\times I_{2\times 2}$ and $m_N~=~\mN\times I_{2\times 2}$) discussed in the previous section. In the limit, $\lfb M_{\Delta}^{-1}\rfb^{\alpha \beta}~\sim~ {\mN}^{-1} I\lfb m_{\phi_S}^2/{\mN}^2\rfb \, \delta^{\alpha\beta}$ (upto the leading order in $\zLN$), the neutrino mass matrix in  Eq.~\ref{eqn:nu_mass_main} can be approximated as,
\beq
\frac{m_\nu}{\langle H \rangle^2}~\sim~\frac{\lambda_3}{16\pi^2\mN} I\lfb \frac{m_{\phi_S}^2}{{\mN}^2}\rfb \, \ylN \ylN^T.
\label{eqn:nu_mass_main_sim}
\eeq
\vspace{0.25cm}
\subsection{Constraints from neutrino oscillation data}
\vspace{0.25cm}
In the low energy effective theory, the neutrino mass matrix is determined by 9 parameters (3 neutrino masses, 3 mixing angles, and 3 phases). Decomposing into mixings and masses, the $m_\nu$ on left-hand side of Eq.~\ref{eqn:nu_mass_main} can be written as $U^*_{MNS}D_\nu U_{MNS}^\dagger$ where $D_\nu~=~{\rm diag}\left(m_1,m_2,m_3\right)$ with $m_1,~m_2~{\rm and}~m_3$ being the masses of the SM neutrinos and $U_{MNS}$ is the Pontecorvo--Maki--Nakagawa--Sakata matrix \cite{Maki:1962mu,Pontecorvo:1967fh}. Therefore, Eq.~\ref{eqn:nu_mass_main} can be expressed as
\beq
\frac{v^2\lambda_3}{16\pi^2} \,\, \ylN M_\Delta^{-1} \ylN^T~=~U^*_{MNS}D_\nu U_{MNS}^\dagger.
\label{eqn:numass_NO}
\eeq
The matrix $U_{MNS}$ consists of 3-angles and 3-phases (one Dirac phase and two Majorana phases), in general. However, with two generations of $\NSR$, matrix $\ylN$ is $3 \times 2$ and $M_\Delta^{-1}$ which encodes the heavy neutrino mass parameter $m_N$ is $2 \times 2$. Thus, matrix structure on LHS is $(3 \times 2)(2 \times 2)(2 \times 3)$ which renders one light neutrino massless and one Majorana phase in $U_{MNS}$ unphysical, on the RHS. Thus, the low energy theory can provide information on 7 parameters, in principle. On the other hand, the $3 \times 2$ complex matrix $\ylN$ contains 12 real parameters, of which 3 phases can be eliminated by a redefinition of the SM lepton doublet $l_L$, in a basis where the SM charged lepton mass matrix and $m_N$ are diagonal. Thus, $\ylN$ contains 9 independent real parameters. Mass parameter $m_N$\footnote{$m_L$ and $\zLN$ do not appear in $m_\nu$ (in Eq.~\ref{eqn:nu_mass_main}) at the leading order in $\zLN$.} is determined by $2$ additional parameters. However, these are not independent of the 9 parameters describing $\ylN$ in view of the scaling symmetry\footnote{At the leading order in $\zLN$, $M_\Delta^{-1}$ is diagonal and given by $\lfb M_\Delta^{-1} \rfb^{\alpha\beta}~=~m_{N^S_\alpha}^{-1}\,I\lfb x_{\widetilde N^S_\alpha}\rfb \, \delta^{\alpha\beta}$. Note that Eq.~\ref{eqn:nu_mass_main} is invariant under the simultaneous rescaling of $\ylN^{i\alpha}$, $i=1,2,3$ with $\Lambda_\alpha$ and $\lfb M_\Delta^{-1} \rfb^{\alpha\beta}$ with $\Lambda_\alpha^{-2}$.} of Eq.~\ref{eqn:nu_mass_main}. Therefore, there are $9$ parameters at the high-scale that determine the leading order light neutrino mass matrix, $m_\nu$ at the low-scale via radiative seesaw mechanism. Since the number of parameters in the high-energy theory is larger than the number of parameters describing the low-energy neutrino phenomenology, proper parameterization \cite{Casas:2006hf,Casas:2001sr,Cordero-Carrion:2019qtu} of Yukawa matrix, $\ylN$, is required to ensure consistency of the model with the available results from the neutrino oscillation experiments. In this regard, Eq.~\ref{eqn:numass_NO} can be re-written as, 
\beq
\ltb \frac{v \sqrt{\lambda_3}}{4\pi} \, D_{\sqrt\nu}^{-1}\, U^T_{MNS}\, \ylN\, M_{\sqrt\Delta}^{-1}\rtb \, \ltb \frac{v \sqrt{\lambda_3}}{4\pi}\, \lfb M_{\sqrt\Delta}^{-1}\rfb^T\,  \ylN^T \,U_{MNS} \,D_{\sqrt\nu}^{-1}\rtb~=~{\bf I}_{3\times 3},
\label{eqn:ylN_R}
\eeq
where, $D_{\sqrt \nu}^{-1}~=~{\rm diag}\left(m_1^{-\frac{1}{2}},m_2^{-\frac{1}{2}},m_3^{-\frac{1}{2}}\right)$. $M_{\Delta}^{-1}$, being a complex symmetric matrix, can be diagonalized by a unitary matrix $U_{\Delta}$ as $M_{\Delta,d}^{-1}~=~U_\Delta^T M_\Delta^{-1} U_\Delta$, where, $M_{\Delta,d}^{-1}$ is a diagonal $2\times 2$ matrix. Therefore, $M_{\Delta}^{-1}~=~U_\Delta^* M_{\Delta,d}^{-1} U_\Delta^{\dagger}~=~M_{\sqrt \Delta}^{-1} \lfb M_{\sqrt\Delta}^{-1}\rfb^{T}$, where, $M_{\sqrt \Delta}^{-1}$~=~$U_\Delta^{*} M_{\sqrt{\Delta,d}}^{-1}$. The most general Yukawa matrix $\ylN$ which is consistent with the physical, low-energy neutrino parameters {\em viz.,} the three light neutrino masses ($m_1$, $m_2$ and $m_3$) and the mixing angles as well as phases (contained in $U_{MNS}$) is,
\begin{alignat*}{5}
  &\ylN&~&=&~\lfb\frac{4\pi}{v}\,\lambda^{-\frac{1}{2}}_3\rfb\,&U^*_{MNS}~&\,&D_{\sqrt\nu}&~\,&R~\,M_{\sqrt \Delta},\\
  &\uparrow&&&&\uparrow&&\uparrow&&\uparrow\\
  {\rm \#~of~Parameters:~~~~} 12&-3& &=& &~5& &~2& ~&2~
\end{alignat*}
where, $R$ is a complex $3\times 2$ matrix subjected to the condition $R\,R^T~=~{\mathds 1}_{3\times 3}$. It is important to note that with one massless neutrino, the RHS of Eq.~\ref{eqn:ylN_R} has only two non-zero diagonal elements. Thus, $RR^T~=~{\rm diag}\lfb 0,1,1\rfb$ for NH and ${\rm diag}\lfb 1,1,0\rfb$ for IH. Consequently, $R$ is described by $2$ independent real parameters that bridge the gap between the low- and the high-energy theories describing the light neutrino masses. In the {\em simplified scenario},
\beq
M_{\sqrt \Delta}~=~{\ltb \frac{\mN} {I\lfb \frac{m_{\phi_S}^2}{\mN^2}\rfb}\rtb^{\frac{1}{2}}}\times {\bf \rm I}_{n\times n} ~~{\rm and}~~\ylN~=~{\ltb \frac{16\, \pi^2\,\mN} {\lambda_3\,\,v^2\,\,I\lfb \frac{m_{\phi_S}^2}{\mN^2}\rfb}\rtb^{\frac{1}{2}}}   U^*_{MNS}~\,D_{\sqrt\nu}~\,R. 
\label{eqn:ylN_nu}
\eeq
It may be noted that amongst the 9 parameters contained in $\ylN$, the 3 imaginary DoFs arise from the phases in $U_{MNS}$ (1 CP-violating Dirac phase and 1 Majorana phase) and the phase in R. In view of the fact that experimental data on Majorana phases is not available and that the CP violating Dirac phase could be zero (check Sec.~\ref{sec:numerical}), we can further assume the phase in R to be zero so that $\ylN$ is rendered completely real, with 6 independent parameters.
\section{Anomalous magnetic moments and cLFV decays}
\label{sec:lfv}
\vspace{0.25cm}
The combination of heavy neutral and charged fermions and scalars in the model gives rise to charged lepton flavor violation (cLFV) processes as well as additional contributions to $g-2$ for the SM charged leptons. The effective operator resulting in cLFV decays $l_j \to l_i \gamma$ and also contribution to the $g-2$ of the SM charged leptons can be expressed as
\beq
\lag_{cLFV}^{ij}~=~\overline{l_i}\sigma_{\mu\nu} \lfb {A^L_{ij} P_L + A^R_{ij} P_R} \rfb l_j F^{\mu\nu},
\label{eqn:lfv_vertex}
\eeq
where $P_L$ and $P_R$ are the chirality projection operators. This leads to
\beq
\Gamma (l_j \to l_i \gamma)=\frac{\alpha_e}{4} m_{l_j}^3 \lfb |A^L_{ij}|^2 + |A^R_{ij}|^2 \rfb,
\label{eqn:lfv_decay}
\eeq
where kinematically allowed and to
\beq
a_{l_i}=\frac{1}{2}(g_{l_i}-2)=-2~m_{l_i}~{\rm Re}\lfb A^L_{ii}\rfb.
\label{eqn:g-2}
\eeq
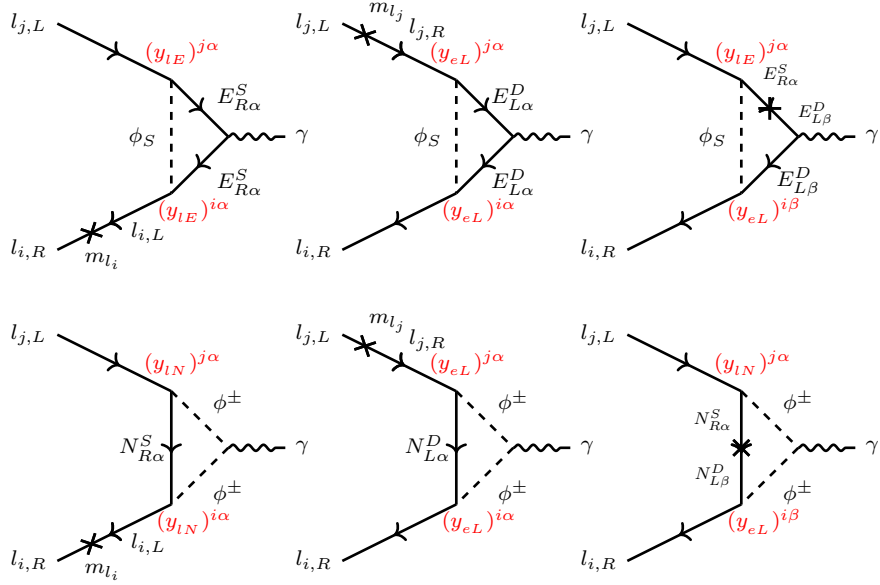
\begin{figure}[!t]
  \centering
  \begin{minipage}{\textwidth}
    \begin{tikzpicture}[line width=1 pt, scale=1.5,every node/.style={scale=1.0}]
\begin{feynman}

      \hspace{1.5cm}

      \draw[fermion,black] (-3,1.25) --(-2,0.75);
      \draw[fermion,black] (-2,0.75)  --(-1.5,0.25);
      \draw[fermion,black] (-1.5,0.25)  --(-2,-0.25);
      \draw[fermion,insertion=0.7,black] (-2,-0.25) --(-3,-0.75);
      \draw[scalar2,black] (-2,0.75)  --(-2,-0.25);
      \draw[vector,black] (-1.5,0.25)  --(-1,0.25);

      \draw[fermion,insertion=0.2,black] (-0.5,1.25) --(0.5,0.75);
      \draw[fermion,black] (0.5,0.75)  --(1,0.25);
      \draw[fermion,black] (1,0.25)  --(0.5,-0.25);
      \draw[fermion,black] (0.5,-0.25) --(-0.5,-0.75);
      \draw[scalar2,black] (0.5,0.75)  --(0.5,-0.25);
      \draw[vector,black] (1,0.25)  --(1.5,0.25);

      \draw[fermion,black] (2,1.25) --(3,0.75);
      \draw[fermion,insertion=0.5,black] (3,0.75)  --(3.5,0.25);
      \draw[fermion,black] (3.5,0.25)  --(3,-0.25);
      \draw[fermion,black] (3,-0.25) --(2,-0.75);
      \draw[scalar2,black] (3,0.75)  --(3,-0.25);
      \draw[vector,black] (3.5,0.25)  --(4,0.25);

      \draw[fermion,black] (-3,-1.5) --(-2,-2);
      \draw[scalar2,black] (-2,-2)  --(-1.5,-2.5);
      \draw[scalar2,black] (-1.5,-2.5)  --(-2,-3);
      \draw[fermion,insertion=0.7,black] (-2,-3) --(-3,-3.5);
      \draw[fermion,black] (-2,-2)  --(-2,-3);
      \draw[vector,black] (-1.5,-2.5)  --(-1,-2.5);

      \draw[fermion,insertion=0.2,black] (-0.5,-1.5) --(0.5,-2);
      \draw[scalar2,black] (0.5,-2)  --(1,-2.5);
      \draw[scalar2,black] (1,-2.5)  --(0.5,-3);
      \draw[fermion,black] (0.5,-3) --(-0.5,-3.5);
      \draw[fermion,black] (0.5,-2)  --(0.5,-3);
      \draw[vector,black] (1,-2.5)  --(1.5,-2.5);

      \draw[fermion,black] (2,-1.5) --(3,-2);
      \draw[scalar2,black] (3,-2)  --(3.5,-2.5);
      \draw[scalar2,black] (3.5,-2.5)  --(3,-3);
      \draw[fermion,black] (3,-3) --(2,-3.5);
      \draw[fermion,insertion=0.5,black] (3,-2)  --(3,-3);
      \draw[vector,black] (3.5,-2.5)  --(4,-2.5);

      \node at (-3.25,1.25) {\scriptsize $l_{j,L}$};
      \node at (-3.25,-0.75) {\scriptsize $l_{i,R}$};
      \node at (-0.85,0.25) {\scriptsize$\gamma$};
      \node at (-1.4,0.6) {\scriptsize$E^S_{R \alpha}$};
      \node at (-1.4,-0.15) {\scriptsize$E^S_{R \alpha}$};
      \node at (-2.25,0.25) {\scriptsize$\phi_S$};
      \node at (-2.2,-0.6) {\scriptsize$l_{i,L}$};
      \node at (-2.6,-0.85) {\scriptsize$m_{l_i}$};
      \node[text=blue] at (-1.9,1) {\scriptsize$({\ylE})^{j\alpha}$};
      \node[text=blue] at (-1.8,-0.4) {\scriptsize$({\ylE})^{i\alpha}$};

      \node at (-0.75,1.25) {\scriptsize$l_{j,L}$};
      \node at (-0.75,-0.75) {\scriptsize$l_{i,R}$};
      \node at (1.65,0.25) {\scriptsize$\gamma$};
      \node at (1,0.6) {\scriptsize$E^D_{L \alpha}$};
      \node at (1,-0.15) {\scriptsize$E^D_{L \alpha}$};
      \node at (0.25,0.25) {\scriptsize$\phi_S$};
      \node at (-0.1,1.35) {\scriptsize$m_{l_j}$};
      \node at (0.25,1.2) {\scriptsize$l_{j,R}$};
      \node[text=blue] at (0.6,1) {\scriptsize$({\yeL})^{j\alpha}$};
      \node[text=blue] at (0.7,-0.4) {\scriptsize$({\yeL})^{i\alpha}$};

      \node at (1.75,1.25) {\scriptsize$l_{j,L}$};
      \node at (1.75,-0.75) {\scriptsize$l_{i,R}$};
      \node at (4.15,0.25) {\scriptsize$\gamma$};
      \node[scale=0.8] at (3.35,0.8) {\scriptsize$E^S_{R \alpha}$};
      \node[scale=0.8] at (3.65,0.45) {\scriptsize$E^D_{L \beta}$};
      \node at (3.5,-0.15) {\scriptsize$E^D_{L \beta}$};
      \node at (2.75,0.25) {\scriptsize$\phi_S$};
      \node[text=blue] at (3.1,1) {\scriptsize$({\ylE})^{j\alpha}$};
      \node[text=blue] at (3.2,-0.4) {\scriptsize$({\yeL})^{i\beta}$};
      
      ----------------------------------------------------------------------------------

      \node at (-3.25,-1.5) {\scriptsize$l_{j,L}$};
      \node at (-3.25,-3.5) {\scriptsize$l_{i,R}$};
      \node at (-0.85,-2.5) {\scriptsize$\gamma$};
      \node at (-1.5,-2.1) {\scriptsize$\phi^{\pm}$};
      \node at (-1.5,-2.9) {\scriptsize$\phi^{\pm}$};
      \node at (-2.25,-2.5) {\scriptsize$N^S_{R \alpha}$};
      \node at (-2.2,-3.35) {\scriptsize$l_{i,L}$};
      \node at (-2.6,-3.6) {\scriptsize$m_{l_i}$};
      \node[text=blue] at (-1.9,-1.75) {\scriptsize$({\ylN})^{j\alpha}$};
      \node[text=blue] at (-1.8,-3.15) {\scriptsize$({\ylN})^{i\alpha}$};

      \node at (-0.75,-1.5) {\scriptsize$l_{j,L}$};
      \node at (-0.75,-3.5) {\scriptsize$l_{i,R}$};
      \node at (1.65,-2.5) {\scriptsize$\gamma$};
      \node at (1,-2.1) {\scriptsize$\phi^{\pm}$};
      \node at (1,-2.9) {\scriptsize$\phi^{\pm}$};
      \node at (0.25,-2.5) {\scriptsize$N^D_{L \alpha}$};
      \node at (-0.1,-1.4) {\scriptsize$m_{l_j}$};
      \node at (0.25,-1.55) {\scriptsize$l_{j,R}$};
      \node[text=blue] at (0.6,-1.75) {\scriptsize$({\yeL})^{j\alpha}$};
      \node[text=blue] at (0.7,-3.15) {\scriptsize$({\yeL})^{i\alpha}$};

      \node at (1.75,-1.5) {\scriptsize$l_{j,L}$};
      \node at (1.75,-3.5) {\scriptsize$l_{i,R}$};
      \node at (4.15,-2.5) {\scriptsize$\gamma$};
      \node at (3.5,-2.1) {\scriptsize$\phi^{\pm}$};
      \node at (3.5,-2.9) {\scriptsize$\phi^{\pm}$};
      \node[scale=0.8] at (2.75,-2.25) {\scriptsize$N^S_{R \alpha}$};
      \node[scale=0.8] at (2.75,-2.75) {\scriptsize$N^D_{L \beta}$};
      \node[text=blue] at (3.1,-1.75) {\scriptsize$({\ylN})^{j\alpha}$};
      \node[text=blue] at (3.2,-3.15) {\scriptsize$({\yeL})^{i\beta}$};
\end{feynman}
    \end{tikzpicture}
  \end{minipage}
\caption{Feynman diagrams representing the process $l_{jL} \to l_{iR} ~\gamma$ where $i,j=1-3$ denote the three SM charged leptons. Exotic fermions in the loop are expressed in gauge bases with indices $\alpha, \beta=1,2$. 
This set of diagrams contributes to matrix $A^L_{ij}$. For $A^R_{ij}$, contribution comes from the process $l_{jR} \to l_{iL} ~\gamma$.}
\label{fig:lfvdiag}
\end{figure}
Contributions to $A^L$($A^R$) result from the six Feynman diagrams depicted in Fig.~\ref{fig:lfvdiag}. The exotic fermions in the loop therein are expressed in gauge bases. Making transformations to mass bases, $A^L$ can be written as\footnote{$m_{\phi}^2=m_{\phi_S}^2$ for charged fermion terms and $m_{\phi^{\pm}}^2$ for neutral fermion terms},
\beq
A^L~=~\frac{1}{32\pi^2 m_{\phi}^2}\left[C^{LL}+C^{RR}+C^{LR}+N^{LL}+N^{RR}+N^{LR}\right],
\eeq
where $C^{LL}$, $C^{RR}$ and $\CLR$ denote contributions from the four heavy $Z_2$-odd charged fermions namely, $\widetilde{\Psi}_{\tilde \kappa}^E \ni \{{\ED}_\kappa$, ${\ES}_\kappa$\} with $\kappa=1,2$ (the diagrams in the top panel of Fig.~\ref{fig:lfvdiag}) whereas $N^{LL}$, $N^{RR}$ and $\NLR$ denote contributions from the six heavy $Z_2$-odd neutral fermions namely, $\widetilde{\Psi}_{R\tilde \kappa}^N \ni \{{\ND}_{a\kappa}$, ${\ND}_{b\kappa}$, $\widetilde N^S_{\kappa}$\} (the diagrams in the bottom panel of Fig.~\ref{fig:lfvdiag}). The expressions for the six contributions to $A^L$ in terms of $Z_2$-odd fermion masses, mixings and Yukawa couplings are
\bea
C^{RR}_{ij}&=&-{m_l}_i \,\,\left[\ylE 
\left\{
U_R^{21}\,\,{\cal D}^f_{{\ED}}\left(U_R^{21}\right)^\dagger + 
U_R^{22}\,\,{\cal D}^f_{{\ES}}\left(U_R^{22}\right)^\dagger
\right\} \ylE^\dagger\right]^{ij}, \nonumber\\
C^{LL}_{ij}&=&- \left[\yeL 
\left\{
U_L^{11}\,\,{\cal D}^f_{{\ED}}\left(U_L^{11}\right)^\dagger + 
U_L^{12}\,\,{\cal D}^f_{{\ES}}\left(U_L^{12}\right)^\dagger
\right\} \yeL^\dagger\right]^{ij}
\,\,{m_l}_j, \nonumber\\
C^{LR}_{ij}&=&-\lfb \frac{-\lambda_3 v^2}{m_{\phi_s}^2} \rfb \,\,\left[\yeL 
\left\{
U_L^{11}\,\,{\cal M}_{{{\ED}}}{\cal D}^g_{{\ED}}\left(U_R^{21}\right)^\dagger + 
U_L^{12}\,\,{\cal M}_{{{\ES}}}{\cal D}^g_{{\ES}}\left(U_R^{22}\right)^\dagger
\right\} \ylE^\dagger\right]^{ij}, \nonumber\\
N^{RR}_{ij}&=&{m_l}_i\,\,\left[\ylN 
\left\{
U_N^{31}\,\,{\cal D}^{F_1}_{{\ND}_a}\left(U_N^{31}\right)^\dagger + 
U_N^{32}\,\,{\cal D}^{F_1}_{{\ND}_b}\left(U_N^{32}\right)^\dagger +
U_N^{33}\,\,{\cal D}^{F_1}_{\widetilde N^S}\left(U_N^{33}\right)^\dagger 
\right\} \ylN^\dagger\right]^{ij}, \nonumber\\
N^{LL}_{ij}&=&\left[\yeL 
\left\{
\lfb U_N^{11}\rfb^*\,\,{\cal D}^{F_1}_{{\ND}_a}\left(U_N^{11}\right)^T + 
\lfb U_N^{12} \rfb^*\,\,{\cal D}^{F_1}_{{\ND}_b}\left(U_N^{12}\right)^T +
\lfb U_N^{13}\rfb^*\,\,{\cal D}^{F_1}_{\widetilde N^S}\left(U_N^{13}\right)^T 
\right\} \yeL^\dagger\right]^{ij}\,{m_l}_j, \nonumber\\
N^{LR}_{ij}&=&-\left[\yeL 
\left\{
\lfb U_N^{11}\rfb^*\,{\cal M}_{{\ND}_a}{\cal D}^{F_2}_{{\ND}_a}\left(U_N^{31}\right)^\dagger + 
\lfb U_N^{12} \rfb^*\,{\cal M}_{{\ND}_b}{\cal D}^{F_2}_{{\ND}_b}\left(U_N^{32}\right)^\dagger \right. \right. \nonumber\\
&+& \left. \left. \lfb U_N^{13}\rfb^*\,{\cal M}_{\widetilde N^S}{\cal D}^{F_2}_{\widetilde N^S}\left(U_N^{33}\right)^\dagger 
\right\} \ylN^\dagger\right]^{ij},
\label{eqn:AL_num}
\eea
where $U_L^{lm}$, $U_R^{lm}$ and $U_N^{lm}$ represent the `$lm^{\text{th}}$' $2\times 2$ block of the $4\times 4$ mixing matrices $U_L^E$, $U_R^E$ in the exotic charged fermion sector and $6\times 6$ mixing matrix $U_N$ in the exotic neutral fermion sector, respectively (see Eq.~\ref{eqn:mass_basis}). The matrices ${\cal M}_Y={\rm Diag} \ltb m_{Y_1}, m_{Y_2} \rtb$ while ${\cal D}^{A}_{Y}={\rm Diag} \ltb A(\nu_{Y_1}), A(\nu_{Y_2})\rtb$ are defined in terms of the loop function $A(\nu)$ with $\nu_{Y_i}={m_{Y_i}^{2}}/{m_{\phi_S}^2}$:
\bea
&&f(x)~=~F_3(x){+} \frac{\lambda_3 v^2}{m_{\phi_s}^2} F_3(x){+}\frac{\lambda_3 v^2}{m_{\phi_s}^2} x F_3' (x), ~~g(x)~=~F_4(x) + x F_4'(x), \nonumber\\
&&F_1 \lfb x \rfb~=~\frac{\lfb 1-6x+3x^2+2x^3-6x^2 \ln x\rfb}{6(1-x)^4}, ~~F_2 \lfb x \rfb~=~\frac{\lfb 1-x^2+2x \ln x\rfb}{(1-x)^3},\nonumber\\
&&F_3 \lfb x \rfb~=~\frac{\lfb 2+3x-6x^2+x^3+6x \ln x \rfb}{6(1-x)^4}, ~~F_4 \lfb x \rfb~=~\frac{\lfb -3+4x-x^2-2 \ln x\rfb}{(1-x)^3}. \nonumber
\eea
At this point, one should note the following:
\begin{itemize}
\item The elements of $\NLR$ ($\CLR$) are enhanced by the TeV scale masses of the $Z_2$-odd neutral (charged) fermions namely, $m_{{\ND}_{a,\kappa}},~m_{{\ND}_{b,\kappa}}~{\rm and}~m_{\widetilde N^S_{\kappa}}$ ($m_{{\ED}_{\kappa}}~{\rm and}~m_{{\ES}_{\kappa}}$) where $\kappa~=~1,2$. Consequently, these contributions are orders of magnitude larger than those of $\NLL$ and $\NRR$ ($\CLL$ and $\CRR$), whose elements are proportional to the SM-charged lepton masses.
\item Eq.~\ref{eqn:AL_num} clearly shows that $\CLR$ is proportional to $\lambda_3$, while the $\lambda_3$ dependence of $\NLR$ results from the $\lambda_3$ dependence of $\ylN$ (see Eq.~\ref{eqn:ylN_nu}) which scales as $\lambda_3^{-1/2}$. Therefore, for large $\lambda_3$, dominant contributions to cLFV and $g-2$ result from $\CLR$ while $\NLR$ contributes dominantly for small $\lambda_3$.
\item $\CLR$ and $\NLR$ are matrices of rank $2$ as matrix product on the RHS of Eq.~\ref{eqn:AL_num} inlcudes a matrix of dimensions $2 \times 2$. Thus, in both the regions with either $\CLR$ or $\NLR$ as the dominant contributions, it is possible to address the issues of anomalous magnetic moments of electron as well as muon while keeping $Br(\mu \to e \gamma)$ highly suppressed.
\item To explain the discrepancy between the observed and the SM predicted values of electron and muon $g-2$, $C^{LR}\sim {\cal O}\lfb 0.1-1 \rfb$ is required (see Eq.~\ref{eqn:lfv_decay}). This can be easily obtained for $\lambda_3 \sim {\cal O}\lfb 1\rfb$ and $\yeL ,~\ylE \sim {\cal O}\lfb 1\rfb$, for sufficiently large mixings ({\em i.e.,} $U_{L(R)}^E\sim {\cal O}\lfb 0.1\rfb$) between the exotic fermions. 
\item In the small $\lambda_3$ region, obtaining $N^{LR}\sim {\cal O}\lfb 0.1-1 \rfb$ for $\yeL \sim {\cal O}\lfb 1 \rfb$ requires $\lambda_3 \sim {\cal O}\lfb 10^{-8} \rfb$. Then, $\ylN \sim {\cal O}\lfb 10^{-3}-10^{-1} \rfb$ and both $\Delta a_e$ and $\Delta a_\mu$ can be generated, keeping $Br(\mu \to e \gamma)$ suppressed.
\end{itemize}
It may be noted here that in appropriate limits, our analytical expressions for the anomalous magnetic moments and lepton flavor violating observables reproduce the known results in the literature. In particular, when the vector-like fermion sector is suitably simplified, our results reduce to those obtained for the scotogenic model \cite{Ma:2006km,Kubo:2006yx} as well as in related frameworks like \cite{Chen:2019nud}. 
The above discussion shows that anomalous magnetic moments of electron and muon, both, can be generated in the framework of the present model for $\lambda_3~\sim {\cal O}\lfb 1 \rfb$ or ${\cal O}\lfb 10^{-8}\rfb$. 
In the latter case, however, the structure of $\ylN$ is constrained from neutrino mass generation and therefore, in the $\NLR$ contribution, a simultaneous suppression of $Br(\tau \to e \gamma)$ and $Br(\tau \to \mu \gamma)$ is not guaranteed. Furthermore, the $N^{RR}$ (the texture of which is almost completely determined from the low energy observables in the neutrino sector) contribution to the cLFV processes is significantly enhanced for such a small $\lambda_3$. As a result, we only study large $\lambda_3$ region ({\em i.e.,} $\lambda_3 \sim {\cal O}\lfb 1 \rfb$) in the next part of our analysis.

While the Wilson coefficients $A^{L(R)}$ may be computed in a straightforward manner in terms of the general masses and Yukawa matrices, it is instructive to consider their approximate analytical form\footnote{Note that for such a simplified scenario, we have already derived the approximate compact expressions for the mixing matrices $U_N$, $U_{L(R)}^E$ and the parametrization of $\ylN$ in Eq.~\ref{eqn:ylN_nu}. Eq.~\ref{eqn:AL_simp} is obtained by substituting these approximate compact results in Eq.~\ref{eqn:AL_num}.} in the simplified scenario:
\bea
C^{RR}&\approx&-m_l~\ylE ~\left[ f({\rho}_L) \chi_E^2 \zRE^{\dagger} \zRE + f({\rho}_E) I_{n \times n} \right]~ {\ylE}^{\dagger},\nonumber\\
C^{LL}&\approx&-\yeL ~\left[ f({\rho}_L) I_{n \times n} +  f({\rho}_E) \chi_E^2 \zLE \zLE^{\dagger} \right]~ {\yeL}^{\dagger}m_l,\nonumber\\
C^{LR}&\approx& \lfb\frac{\lambda_3 v^2}{m_{\phi_s}^2}\rfb ~\chi_E~\yeL ~\Big\{
\big(\mE~g({\rho}_E)-\epsilon_E \mL~g({\rho}_L)\big) \zLE 
- \big(\mL~g({\rho}_L)-\epsilon_E\mE~g({\rho}_E)\big) \zRE 
\Big\}~ {\ylE}^{\dagger},\nonumber\\
N^{RR}&\approx&\frac{16 \pi^2 \mN {\lambda}_3^{-1}}{v^2 I \lfb {\omega}_N \rfb}~m_l~ U_{MNS}^*~D_{\sqrt{\nu}}~R~P~R^{\dagger}~D_{\sqrt{\nu}}~U_{MNS}^T,\nonumber\\
N^{LL}&\approx&\yeL ~\left[ F_1({\rho}_L) I_{n \times n} + 2  \chi_N^2 z_{LN} z_{LN}^{\dagger} \lfb F_1({\rho}_N)~-~F_1({\rho}_L)\rfb \right]~ {\yeL}^{\dagger} m_l,\nonumber\\
N^{LR}~&\approx&\frac{-4\pi}{v}\sqrt{\frac{2 \mN}{\lambda_3 I \lfb {\omega}_N \rfb}}~\yeL ~\lfb ~\mN F_2({\rho}_N) - \epsilon_N ~\mL F_2({\rho}_L) \rfb~{\chi}_N~z_{LN}~R^{\dagger}~D_{\sqrt{\nu}}~U_{MNS}^T, 
\label{eqn:AL_simp}
\eea
where $P=F_1({\rho}_N) I_{n \times n} + 2 \chi_N^2 z_{LN}^T z_{LN}^* \lfb F_1({\rho}_L)-F_1({\rho}_N)\rfb$, $\rho_{L(E)[N]}=\frac{\widetilde m_{L(E)[N]}^2}{m_{\phi}^2}$ and $m_l={\rm diag\lfb m_e, m_\mu,m_\tau\rfb}$ is a $3 \times 3$ diagonal matrix with the masses of the SM charged leptons as the diagonal entries. 
\vspace{0.25cm}
\subsection{The textures of the Yukawa matrices}
\label{sec:lfvparameters}
\vspace{0.25cm}
Our objective is to obtain general textures of the Yukawa matrices in the framework of the minimal model with only two generations of exotic $Z_2$-odd fermions that explain the anomalies of the SM charged lepton magnetic moments without contributing significantly to the cLFV observables. It has already been discussed in the previous section that for $\lambda_3~\sim {\cal O}\lfb 1 \rfb$, the contribution from $\CLR$ is a few orders of magnitude larger than the other contributions {\em i.e.,} 
\beq
\frac{1}{32 \pi^2 m_{\phi_S}^2}~C^{LR} \approx A^L.
\eeq
Putting in the expression for $\CLR$ as in Eq.~\ref{eqn:AL_simp},
\beq
\lfb \frac{\lambda_3 v^2 \chi_E}{32 \pi^2 m_{\phi_S}^4} \rfb ~\yeL ~\Big\{
\big(\mE~g({\rho}_E)-\epsilon_E \mL~g({\rho}_L)\big) \zLE 
- \big(\mL~g({\rho}_L)-\epsilon_E\mE~g({\rho}_E)\big) \zRE 
\Big\}~ {\ylE}^{\dagger} \approx A^L.
\label{eqn:CLR_AL}
\eeq
With two generations of exotic leptons, the structure of component matrices on the LHS is
 $(3 \times 2)(2 \times 2)(2 \times 3)$ which implies only two non-zero eigenvalues of $A^L$. 
 Thus, we have the freedom to 
 generate only two of the three anomalous magnetic 
 moments which we choose as $\Delta a_{e}$ and $\Delta a_{\mu}$. We further want to suppress 
 the elements in $A^L$ that lead to cLFV {\em viz.,} $A^L_{12}$ and $A^L_{21}$ for 
 Br($\mu \to e \gamma$), $A^L_{13}$ and $A^L_{31}$ for Br($\tau \to e \gamma$) and 
 $A^L_{23}$ and $A^L_{32}$ for Br($\tau \to \mu \gamma$)\footnote{For the corresponding 
 dominant contribution to $A^R$, the matrix elements giving rise to cLFV decays are 
 $A^R_{12}={A^L_{21}}^{*}$ for Br($\mu \to e \gamma$), $A^R_{13}={A^L_{31}}^{*}$ for Br($\tau \to e \gamma$) 
 and $A^R_{23}={A^L_{32}}^{*}$ for Br($\tau \to \mu \gamma$).} {\em i.e.,} we demand
\beq
\lfb \frac{\lambda_3 v^2 \chi_E}{32 \pi^2 m_{\phi_S}^4} \rfb ~\yeL ~{\cal Z} ~{\ylE}^{\dagger}\approx {\rm diag}\lfb -\frac{\Delta a_{e}}{2m_e},~-\frac{\Delta a_{\mu}}{2m_\mu},~-\frac{\Delta a_{\tau}}{2m_\tau}\rfb,
\label{eqn:CLR_3x3}
\eeq
where ${\cal Z}=~\Big\{
\big(\mE~g({\rho}_E)-\epsilon_E \mL~g({\rho}_L)\big) \zLE 
- \big(\mL~g({\rho}_L)-\epsilon_E\mE~g({\rho}_E)\big) \zRE 
\Big\}~$, 
$\Delta a_{l}~=~a_l^{\rm exp}-a_l^{\rm SM}$ and $a_l~=~(g-2)_l/2$.
At this point, we can obtain an expression for $\yeL$ {\em viz.,}
\beq
{\yeL}=\lfb\frac{32\pi^2 m_{\phi_s}^4}{\lambda_3 v^2 {\chi}_E }\rfb {\rm diag}\lfb -\frac{\Delta a_{e}}{2m_e},~-\frac{\Delta a_{\mu}}{2m_\mu},~-\frac{\Delta a_{\tau}}{2m_\tau} \rfb \times M^{RI},
\label{eqn:yeL_3x2}
\eeq
where $M^{RI}={\ylE}^* \lfb{\ylE}^{\dagger} {\ylE}^*\rfb^{-1} ~{\cal Z}^{-1}$ is a $3\times 2$ matrix which is the right inverse of the matrix $M={\cal Z}~\ylE^{\dagger}$
such that $M \times M^{RI}~=~{\rm diag}\lfb 1,1\rfb $. 
For arbitrary Yukawa matrices $\zLE,~\zRE$ and $\ylE$\footnote{Only those values of $\ylE$, $\zRE$ and $\zLE$ are accepted as solutions for which $M^{RI}$ exists {\em i.e.,} $\lfb{\ylE}^{\dagger} {\ylE}^*\rfb$ and $\lfb \mE~g({\rho}_E) {\zLE}-\mL~g({\rho}_L) \zRE \rfb$ are non-singular matrices.}, $\yeL$ resulting from Eq.~\ref{eqn:yeL_3x2} explains the anomalies related to SM charged leptons' magnetic moments as well as ensures null contributions to the cLFV processes from $\CLR$. 

However, it should be noted that the third rows in both $\yeL$ and $\ylE$ are redundant for determining $\Delta a_{e}$ and $\Delta a_{\mu}$. Therefore, in order to ensure vanishing of Br($\tau \to e \gamma$) and Br($\tau \to \mu \gamma$), we can assume them to be zero. 
Consequently, $\Delta a_{\tau}$ is zero too and instead of Eq.~\ref{eqn:CLR_3x3}, we only need to solve the $2\times 2$ equation,
\beq
\lfb \frac{\lambda_3 v^2 \chi_E}{32 \pi^2 m_{\phi_S}^4} \rfb ~\yeL^{2\times 2} ~{\cal Z} ~\lfb \ylE^{2\times 2}\rfb^{\dagger} \approx {\rm diag}\lfb -\frac{\Delta a_{e}}{2m_e},~-\frac{\Delta a_{\mu}}{2m_\mu}\rfb,
\label{eqn:CLR_2x2}
\eeq
where $\ylE^{2\times 2}$ and $\yeL^{2\times 2}$ are the $2\times 2$ submatrices of $\ylE$ and $\yeL$, respectively, comprising elements ${\ylE}^{i\alpha}$ and ${\yeL}^{i\alpha}$ where $i,\alpha=1,2$. Note that ${\cal Z}$ is a $2\times 2$ matrix which is a function of the matrices $\zRE$ and $\zLE$. Rearranging, we may write,
\beq
{\yeL^{2\times 2}}=\lfb\frac{32\pi^2 m_{\phi_s}^4}{\lambda_3 v^2 {\chi}_E}\rfb {\rm diag}\lfb -\frac{\Delta a_{e}}{2m_e},~-\frac{\Delta a_{\mu}}{2m_\mu} \rfb \times \lfb{\ylE^{2\times 2}}^{\dagger}\rfb^{-1} \times {\cal Z}^{-1}.
\label{eqn:yeL_2x2}
\eeq

We next specify the textures of the matrices $\ylE$, $\zRE$, $\zLE$ and $\zLN$. These are motivated to minimise the cLFV branching ratios that may arise from the off-diagonal elements of different contributions to $A^{L(R)}$, as in Eq.~\ref{eqn:AL_simp}.

\begin{itemize}
\item The off-diagonal elements of $C^{RR}$ can be suppressed if $\zRE$ is unitary and $\ylE \ylE^\dagger$ is diagonal. This leads to the following most general textures for $\zRE$ and $\ylE$ {\em viz.,}
  \beq
  \zRE = \xi_{\zRE}
  \lfb {\begin{array}{cc}
      {\rm exp}\ltb i\delta_{\zRE}^{11} \rtb{\rm cos}\theta_{\zRE} & {\rm exp}\ltb i\delta_{\zRE}^{12}\rtb{\rm sin} \theta_{\zRE}\\
      -{\rm exp}\ltb i\delta_{\zRE}^{21}\rtb{\rm sin}\theta_{\zRE} & {\rm exp}\ltb i\delta_{\zRE}^{22}\rtb{\rm cos}\theta_{\zRE} \\
  \end{array} } \rfb,
  \label{eqn:tex_zRE}
  \eeq
where $\xi_{\zRE}$ is in general complex and $\delta_{\zRE}^{ij}$ are real, subjected to $\delta_{\zRE}^{11}+\delta_{\zRE}^{22}-\delta_{\zRE}^{12}-\delta_{\zRE}^{21}=0$, and,
\bea
\ylE&=&
  \lfb {\begin{array}{cc}
      \xi_{lE}{\rm exp}\ltb i\delta_{lE}^{11} \rtb {\rm cos}\theta_{lE} & \xi_{lE}{\rm exp}\ltb i\delta_{lE}^{12} \rtb{\rm sin} \theta_{lE}\\
      -\eta_{lE}{\rm exp}\ltb i\delta_{lE}^{21} \rtb{\rm sin}\theta_{lE} & \eta_{lE}{\rm exp}\ltb i\delta_{lE}^{2} \rtb{\rm cos}\theta_{lE} \\
      0 & 0\\
  \end{array} } \rfb,
  \label{eqn:ylE}
  \eea
where $\xi_{lE}$, $\eta_{lE}$ are in general complex and $\delta_{lE}^{ij}$ are real, subjected to $\delta_{lE}^{11}+\delta_{lE}^{22}-\delta_{lE}^{12}-\delta_{lE}^{21}=0$. 
\item The off-diagonal elements of $C^{LL}$ can be suppressed if $\zLE$ is unitary and $\yeL \yeL^\dagger$ is diagonal. We assume $\zLE=\zRE$ so that $\zLE$ is unitary too. Although we do not have the freedom to choose $\yeL$ since it is already determined from Eq.~\ref{eqn:yeL_2x2} in terms of $\zLE,~\zRE$ and $\ylE$, the choice of $\ylE$ in Eq.~\ref{eqn:ylE} and unitary $\zRE,\zLE$ ensure that $\yeL \yeL^\dagger$ is diagonal (see Appendix~\ref{app:yeLyeLd}).

With the assumption of $\zLE=\zRE$, Eq.~\ref{eqn:CLR_2x2} is simplified further and can be expressed as,
\beq
{\cal Y} ~\yeL^{2\times 2} ~\zRE ~\lfb\ylE^{2\times 2}\rfb^{\dagger} \approx {\rm diag}\lfb -\frac{\Delta a_{e}}{2m_e},~-\frac{\Delta a_{\mu}}{2m_\mu}\rfb,
\label{eqn:CLR_2x2_simp}
\eeq
where ${\cal Y}=\lfb\frac{\lambda_3 v^2 \chi_E}{32 \pi^2 m_{\phi_S}^4}\rfb~\Big\{
\big(\mE~g({\rho}_E)-\epsilon_E \mL~g({\rho}_L)\big) 
- \big(\mL~g({\rho}_L)-\epsilon_E\mE~g({\rho}_E)\big)
\Big\}~$. 

These constitute four equations which can be expressed as,
\bea
\yeL^{11}\lfb\zRE^{11}\ylE^{11*}+\zRE^{12}\ylE^{12*}\rfb+\yeL^{12}\lfb\zRE^{21}\ylE^{11*}+\zRE^{22}\ylE^{12*}\rfb&=&-\frac{\Delta a_e}{2 m_e {\cal Y}}, \nn \\
\yeL^{21}\lfb\zRE^{11}\ylE^{21*}+\zRE^{12}\ylE^{22*}\rfb+\yeL^{22}\lfb\zRE^{21}\ylE^{21*}+\zRE^{22}\ylE^{22*}\rfb&=&-\frac{\Delta a_\mu}{2 m_\mu {\cal Y}}, \nn \\
\yeL^{11}\lfb\zRE^{11}\ylE^{21*}+\zRE^{12}\ylE^{22*}\rfb+\yeL^{12}\lfb\zRE^{21}\ylE^{21*}+\zRE^{22}\ylE^{22*}\rfb&=&0, \nn \\
\yeL^{21}\lfb\zRE^{11}\ylE^{11*}+\zRE^{12}\ylE^{12*}\rfb+\yeL^{22}\lfb\zRE^{21}\ylE^{11*}+\zRE^{22}\ylE^{12*}\rfb&=&0.
\label{eqn:yeL_equations}
\eea

For fixed values of $\zRE$ and $\ylE$, the equations have a unique solution and given as,
\bea
\yeL^{11}&=&\frac{D}{AD-BC} \lfb\frac{-\Delta a_e}{2 m_e {\cal Y}}\rfb; ~~~\yeL^{12}=\frac{-C}{D} ~\yeL^{11}, \nn \\
\yeL^{21}&=&\frac{-B}{AD-BC} \lfb\frac{-\Delta a_\mu}{2 m_\mu {\cal Y}}\rfb; ~~~\yeL^{22}=\frac{-A}{B} ~\yeL^{21},
\label{eqn:yeL_solutions}
\eea
where,
\bea
A=\zRE^{11}\ylE^{11*}+\zRE^{12}\ylE^{12*}; ~~~~~~B=\zRE^{21}\ylE^{11*}+\zRE^{22}\ylE^{12*}, \nn \\ 
C=\zRE^{11}\ylE^{21*}+\zRE^{12}\ylE^{22*}; ~~~~~~D=\zRE^{21}\ylE^{21*}+\zRE^{22}\ylE^{22*}.
\label{eqn:yeL_equations_coefficients}
\eea

It is interesting to note here that corresponding to deviations $\Delta a_e \pm 2\sigma_e$ and $\Delta a_\mu \pm 2\sigma_\mu$, the deviations in solutions for $\yeL$ in Eq.~\ref{eqn:yeL_solutions} 
are given as,
\bea
\delta\yeL^{11}&=&\frac{D}{AD-BC} \lfb\frac{\pm \sigma_e}{m_e {\cal Y}}\rfb; ~~~\delta\yeL^{12}=\frac{-C}{D} ~\delta\yeL^{11}, \nn \\
\delta\yeL^{21}&=&\frac{-B}{AD-BC} \lfb\frac{\pm \sigma_\mu}{m_\mu {\cal Y}}\rfb; ~~~\delta\yeL^{22}=\frac{-A}{B} ~\delta\yeL^{21},
\label{eqn:yeL_range}
\eea
where $\sigma_e=0.36 \times 10^{-12}$ and $\sigma_\mu=0.48 \times 10^{-9}$. For TeV scale masses and ${\cal O}\lfb 1\rfb$ $\ylE$ and $\zRE$, these deviations are of ${\cal O}\lfb 0.001-0.1\rfb$. The expressions for $\delta\yeL^{12}$ and $\delta\yeL^{22}$ ensure that $\CLR$ contribution to Br($\mu \to e \gamma)$ is still zero. However, if we relax these expressions such that 
Br$(\mu \to e \gamma)$ is less than its upper bound, 
we see that
\bea
|A~\delta\yeL^{11}+B~\delta\yeL^{12}|^2 + |C~\delta\yeL^{21}+D~\delta\yeL^{22}|^2 &\lesssim& {\cal O}\lfb 10^{-10}\rfb, \nn \\ \nn \\
\implies A~\delta\yeL^{11}+B~\delta\yeL^{12}, ~~~~ C~\delta\yeL^{21}+D~\delta\yeL^{22} &\lesssim& {\cal O}\lfb 10^{-5}\rfb.
\label{eqn:yeL_range_Br}
\eea
\item To suppress the off-diagonal contribution to $A^{L(R)}$ from $\NLL,~\NLR$ and $\NRR$, we note that $\NLL$ is already suppressed to some extent as $\yeL \yeL^\dagger$ is ensured to be diagonal. To constrain $\NRR$ and $\NLR$, it is important to remind that the texture of $\ylN$ is determined by Eq.~\ref{eqn:ylN_nu} in terms of the neutrino oscillation data and matrix $R$ which is a $3\times 2$ matrix (for the minimal model with two generations of $Z_2$-odd fermions) satisfying $RR^T~=~{\rm diag}\lfb 0,1,1\rfb$ for NH and ${\rm diag}\lfb 1,1,0\rfb$ for IH. The most general parametrization of $R$ in terms of a complex angle $\theta_R$ \cite{Ibarra:2003up} is as follows,
\begin{equation}
R=\begin{cases} \left( \begin{array}{ccc}
  0 & 0 \\
\cos \theta_R & -\sin \theta_R\\
\zeta \sin \theta_R & \zeta \cos \theta_R
\end{array} \right) \mathrm{for~NH},~~~~
\left( \begin{array}{ccc}
  \cos \theta_R & -\sin \theta_R \\
  \zeta \sin \theta_R & \zeta \cos \theta_R\\
  0 & 0
\end{array} \right) \mathrm{for~IH}~, \end{cases}
\end{equation}
where $\zeta = \pm$1. Throughout this analysis, we take $\zeta=+1$ as $\zeta=-1$ will not yield any physically different texture of $\ylN$.

Eq.~\ref{eqn:AL_simp} shows that the texture of $N^{RR}$ is dominantly determined by the neutrino oscillation data and hence, the cLFV contributions can not be suppressed by choosing the Yukawas. However, $N^{RR}$ being inversely proportional to $\lambda_3$, its  contributions to cLFV decays become significant only in the small $\lambda_3$ region\footnote{It has already been estimated earlier in this section that in the small $\lambda_3$ region, one can explain the $g-2$ anomalies for $\lambda_3\sim {\cal O}\lfb 10^{-8}\rfb$. The cLFV decay branching ratios of the SM charged leptons resulting from the $N^{RR}$ for such a small $\lambda_3$ can be easily estimated from Eq.~\ref{eqn:AL_simp} to be of the order of $10^{-12}$ which is larger than the experimental bound on ${\rm Br}\lfb \mu \to e \gamma\rfb$. We find it challenging to satisfy bounds from cLFV observables as well as explain SM charged lepton $g-2$ anomalies simultaneously in the low $\lambda_3$ region in the minimal model {\em i.e.,} with only two generations of exotic $Z_2$-odd fermions.}. For $\lambda_3\sim {\cal O}\lfb 1\rfb$, the elements of $N^{RR}$ are already suppressed (by the small neutrino masses) to contribute significantly to the cLFV decays. One can naively estimate the branching ratio of process $\mu \to e \gamma$ resulting from the non-zero off-diagonal elements of $N^{RR}$ to be of ${\cal O}\lfb 10^{-28} \rfb$ or less for $\lambda_3\sim {\cal O}\lfb 1\rfb$.    

The texture of $\NLR$ (see Eq.~\ref{eqn:AL_simp}) is partially determined by the neutrino oscillation data. However, we have the freedom to choose the texture of $\zLN$ \footnote{Unlike the case of neutrino mass generation in Sec.~\ref{sec:numass} where only the gauge state $\NSR$ contributes and dependence on $\zLN$ upto ${\cal O}\lfb {\zLN} \rfb$ is absent in the {\em simplified scenario}, here, the dependence on $\zLN$ is more pronounced.}. It is not possible to suppress all the 6 complex off-diagonal elements of $\NLR$ by properly choosing the texture of $\zLN$. Since the experimental bounds on the cLFV transition rates of $\mu$ to $e$ are the strongest, we try to suppress the 12 and 21 elements of $\NLR$ by considering,
  \beq
  \zLN=\lfb {\begin{array}{cc}
      \yeL^{11} & \yeL^{12}\\
      \yeL^{21} & \yeL^{22}\\
  \end{array} } \rfb^{-1}
  \lfb {\begin{array}{cc}
      d_1^\prime & 0\\
      0 & d_2^\prime\\
  \end{array} } \rfb
  \left [\lfb {\begin{array}{cc}
      \ylN^{11} & \ylN^{12}\\
      \ylN^{21} & \ylN^{22}\\
    \end{array} } \rfb^{\dagger} \right ]^{-1},
  \label{eqn:zLN}
  \eeq
where $d_1^\prime,~d_2^\prime$ are complex numbers and $\yeL^{ij}$ and $\ylN^{ij}$ are the $ij^{\rm th}$ elements of $\yeL$ and $\ylN$, respectively. To normalize the elements of $\zLN$ defined in  Eq.~\ref{eqn:zLN} to unity, we define $d^\prime_1(d^\prime_2)$ as
\beq
d^\prime_1(d^\prime_2)=\frac{d_1(d_2)}{{\rm Max}\lfb |\zLN^{11}|,|\zLN^{12}|,|\zLN^{21}|,|\zLN^{22}|\rfb},
\label{eqn:d1d2}
\eeq
where $d_1,~d_2$ are free parameters. It is important to note that the structure of $\zLN$ in Eq.~\ref{eqn:zLN} is not unitary and hence, the  $\mu$ to $e$  transition rates resulting from the sub-leading term (the term proportional to $\chi^2_N\zLN \zLN^\dagger$) in $N^{LL}$ are not suppressed and are proportional to $d_1 d_2$. These contributions can be suppressed by choosing one of $d_1$ or $d_2$ to be very small.    
\end{itemize}

\subsection{Numerical analysis}
\label{sec:numerical}
\vspace{0.25cm}
The phenomenology of the {\em simplified scenario} (with only two generations of exotic $Z_2$-odd fermions), which is consistent with the neutrino oscillation data, $g-2$ anomalies and the experimental bounds on the cLFV observables, is determined in terms of the following free parameters:
\bea
   && {\rm Scalar~sector~Parameters:}~ ~\mu,\lambda_\mu,~\lambda_3, \nonumber\\
   && {\rm Mass~Parameters:}~ \mN,~\mL,~\mE,~\muphi, \nonumber\\
&& {\rm Yukawa~Parameters:}~ \xi_{\zRE},~\xi_{lE},~\eta_{lE},~d_2,~\theta_R,~\theta_{\zRE},~\theta_{lE},~\delta_{\zRE}^{ij},~\delta_{lE}^{ij},
\label{eqn:freeparam}
\eea
where $i,j=1,2$ and $\sum_{i,j} (-1)^{i+j}\delta_{\zRE(lE)}^{ij}=1$. 
Upto now, all the results are valid for complex Yukawa matrices, in general. As a further simplification, we assume all the input Yukawa matrices to be real ({\em i.e,} $\delta_{\zRE}^{ij},~\delta_{lE}^{ij}=0$). Note that we have kept the parameter $d_1$ in the defining Eq.~\ref{eqn:zLN} as zero. The neutrino oscillation (NO) data and the experimental bounds on LFV processes that we use to perform the numerical analysis are summarised below. \\

{\noindent \bf \em Neutrino Oscillation Data:} $U_{MNS}$ is parameterized as:
\begin{equation*}
U_{\rm PMNS} = \left( \begin{array}{ccc}
c_{12} c_{13} & s_{12} c_{13} & s_{13} e^{-i\delta} \\
-s_{12} c_{23} -c_{12} s_{13} s_{23} e^{i\delta} & c_{12} c_{23} -s_{12} s_{13} s_{23} e^{i\delta} & c_{13} s_{23} \\
s_{12} s_{23} -c_{12} s_{13} c_{23} e^{i\delta} & -c_{12} s_{23} -s_{12} s_{13} c_{23} e^{i\delta} & c_{13} c_{23}
\end{array} \right) \times \mathrm{diag}(e^{-i \phi/2},e^{-i \phi^\prime/2},1),
\end{equation*}
where $c_{ij}=\cos\theta_{ij}$ and $s_{ij}=\sin\theta_{ij}$; $\theta_{12},\theta_{13}$ and $\theta_{23}$ are the light neutrino mixing angles, $\delta$ is the Dirac CP phase, and $\phi$ and $\phi^\prime$ are the Majorana phases. For simplicity, we take the Majorana phases to be zero throughout this work. The following best fit values and $3\sigma$ range for the neutrino oscillation parameters \cite{Esteban:2018azc} are used:
\begin{itemize}
\item For NH:
$\theta_{12}=33.82^\circ~[31.61^\circ \to 36.27^\circ]$,~
$\theta_{13}=8.60^\circ~[8.22^\circ \to 8.98^\circ]$,~
$\theta_{23}=48.6^\circ~[41.1^\circ \to 51.3^\circ]$,~
$\Delta m_{21}^2 \times 10^5 ~\rm eV^{-2}=7.39~[6.79 \to 8.01]$,~
$\Delta m_{31}^2 \times 10^3 ~\rm eV^{-2}=2.528~[2.436 \to 2.618]$~and~
$\delta=221^\circ~[144^\circ \to 357^\circ]$~,
\item For IH:
$\theta_{12}=33.82^\circ~[31.61^\circ \to 36.27^\circ]$,~
$\theta_{13}=8.64^\circ~[8.26\circ \to 9.02^\circ]$,~
$\theta_{23}=48.8^\circ~[41.4^\circ \to 51.3^\circ]$,~
$\Delta m_{21}^2 \times 10^5 ~\rm eV^{-2}=7.39~[6.79 \to 8.01]$,~
$\Delta m_{32}^2 \times 10^3 ~\rm eV^{-2}=-2.510~[-2.601 \to -2.419]$~and~
$\delta=282^\circ~[205^\circ \to 348^\circ]$~,
\end{itemize}
where $\Delta m_{ij}^2=m_i^2-m_j^2$. Following the description in Sec.~\ref{sec:numass}, the lightest neutrino mass is fixed to zero in this model. Consequently, we obtain $\Sigma m_\nu \sim 0.06 (0.1)$ eV for NH (IH), which is consistent with current cosmological bounds \cite{Planck:2018vyg}. The effective
mass relevant for neutrinoless double beta decay is found to be $m_{\beta\beta} \sim 0.0036 (0.037)$ eV for NH (IH), under the assumption of vanishing Majorana phases. It should be noted, however, that $m_{\beta\beta}$ depends on the unknown Majorana phases. Varying the Majorana phases would lead to $m_{\beta\beta} \sim (1.5 - 3.7)\times 10^{-3}$ eV (NH) and $(0.018 - 0.05)$ eV (IH). These values lie below the current exclusion limits \cite{KamLAND-Zen:2024eml}. The projected sensitivities of next-generation experiments are expected to reach $(0.009-0.021)$ eV \cite{LEGEND:2021bnm} and $(0.0047-0.0203)$ eV \cite{nEXO:2021ujk}. Thus, the inverted hierarchy
scenario lies within the reach of upcoming experiments, while the normal hierarchy case remains more challenging to probe.

\vspace{0.5cm}
{\noindent \bf \em {\noindent \bf \em Experimental bounds on the cLFV observables:} :}
\label{sec:Expt_LFV}
Lepton flavour violation in the charged fermion sector is yet to be observed. In the absence of any observation, there are various  experimental limits on different lepton flavour violating transitions. 
Among the LFV radiative decays $\ell_\alpha \to \ell_\beta \gamma$, the most stringent bound comes from the MEG II experiment,
which reports $\text{Br}(\mu \to e\gamma) < 1.5 \times 10^{-13}$ at $90\%$ C.L.~\cite{MEGII:2025gzr}. 

The most stringent bound on $\ell_\alpha \to \ell_\beta \ell_\gamma \ell_\delta$ is $Br(\mu^+ \to e^+e^+e^-) \le 1.0\times 10^{-12}$ from SINDRUM collaboration \cite{Bellgardt:1987du}. In case of $\mu \to e$ conversion in nuclei, the most stringent bound is on the conversion rate in Gold which is $\le 7\times 10^{-13}$ from SINDRUM-II collaboration \cite{Bertl:2006up}. The COMET \cite{Kuno:2013mha} collaboration at J-PARC and the Mu2e collaboration at FNAL, both have projected a future sensitivity of $1.0\times 10^{-16}$ on $\mu \to e$ conversion rate in Al. All these experimental bounds are summarised in Table~\ref{table:low_energy}.
\begin{figure}[htb!]
\centering 
\includegraphics[width=.8\textwidth]{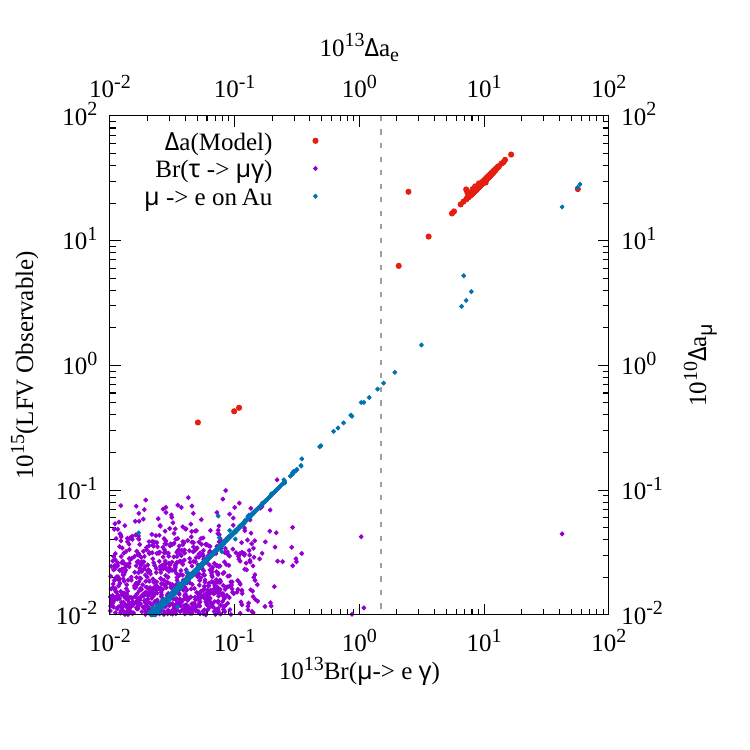}
\caption[Low energy observables $\Delta a_{l}$, Br$\lfb l_j \to l_i \gamma\rfb$ and $\mu \to e$ conversion rate on Au are presented for $10^4$ randomly generated parameter points ($10^4$ points for each observable). The points in red represent ($\Delta a_{e}, \Delta a_{\mu}$), the ones in violet represent (Br$\lfb \mu \to e \gamma\rfb$, Br$\lfb \tau \to \mu \gamma\rfb$) and the ones in blue show $\mu \to e$ conversion rate on Au. The experimental limit on Br$\lfb \mu \to e \gamma\rfb$ is shown by the dashed line. The experimental limits Br$\lfb \tau \to \mu \gamma\rfb$ and $\mu \to e$ conversion on Au are much weaker and hence, not shown.]
{Low energy observables $\Delta a_{l}$, Br$\lfb l_j \to l_i \gamma\rfb$ and $\mu \to e$ conversion rate on Au are presented for $10^4$ randomly generated parameter points ($10^4$ points for each observable). The points in red represent ($\Delta a_{e}, \Delta a_{\mu}$), the ones in violet represent (Br$\lfb \mu \to e \gamma\rfb$, Br$\lfb \tau \to \mu \gamma\rfb$) and the ones in blue show $\mu \to e$ conversion rate on Au. The experimental limit on Br$\lfb \mu \to e \gamma\rfb$ is shown by the dashed line. The experimental limits Br$\lfb \tau \to \mu \gamma\rfb$ and $\mu \to e$ conversion on Au are much weaker and hence, not shown.}
\label{fig:validation}
\end{figure}
\vspace{0.5cm}
\subsection{Results}
\vspace{0.25cm}
\begin{table}[h]
\scalebox{0.8}{
\begin{tabular}{||c||c||}
\hline\hline
\multicolumn{2}{||c||}{\textbf{Mass hierarchy 1 (MH1)}}\\  
\hline\hline 
\textbf{Mass parameters} & \textbf{Yukawa parameters} \\
(GeV) & $\xi_{\zRE}=1$, $\theta_{\zRE}=35^0$, $\xi_{lE}=0.5$, $\eta_{lE}=1$, $\theta_{lE}=30^0$, $d_2=0.02$, $\theta_{R}=-55^0$ \\ \hline\hline
\begin{minipage}{0.3\textwidth}
\vspace{-4cm}
\begin{center}
  $\mN=500$\\
  $\mE=600$\\
  $\mu_{\Phi}=650$\\
  $\mL=850$\\
\end{center}
\end{minipage}
&
\begin{minipage}{0.9\textwidth}
\begin{center}
\vspace{0.5cm} 
$\zRE:{\begin{pmatrix} 0.8192 & 0.5736 \\ -0.5736 & 0.8192 \end{pmatrix}}$,~
$\ylE:{\begin{pmatrix} 0.4330 & 0.25 \\-0.5 & 0.8660 \\ 0 & 0  \end{pmatrix}}$\\
$\yeL:{\begin{pmatrix} -0.06376 & 0.005577 \\ 0.03757 & 0.4296 \\ 0 & 0 \end{pmatrix}}$\\
$\ylN:{\begin{pmatrix} 1.744\cdot 10^{-7} & 4.072\cdot 10^{-6}\\ 
                      -7.753\cdot 10^{-6} & 8.995\cdot 10^{-6} \\  
                      -1.024\cdot 10^{-5} & 1.86\cdot 10^{-6} \end{pmatrix}} \text{(NH)}$;~~~
      ${\begin{pmatrix} 3.002\cdot 10^{-7} & 1.489\cdot 10^{-5}\\ 
                       -9.674\cdot 10^{-6} & -1.426\cdot 10^{-6} \\  
                        1.158\cdot 10^{-5} & -1.714\cdot 10^{-6} \end{pmatrix}} \text{(IH)}$ \\
$\zLN:{\begin{pmatrix} 1.749\cdot 10^{-3} & -7.492\cdot 10^{-5} \\ 
                           2\cdot 10^{-2} & -8.566\cdot 10^{-4} \end{pmatrix}} \text{(NH)}$;~~~
      ${\begin{pmatrix} 1.749\cdot 10^{-3} & -3.527\cdot 10^{-5} \\ 
                           2\cdot 10^{-2} &  -4.0320\cdot 10^{-4} \end{pmatrix}} \text{(IH)}$ \\
\vspace{0.5cm}         
\textbf{Physical masses (GeV):}\\
\{$\phi_P,~\phi^\pm,~\phi_S$\} $\sim$ \{$650,~673,~737$\}\\
\{${\widetilde N^S},~{\widetilde N^D_{a}},~{\widetilde N^D_{b}}$\} $\sim$ \{$500,~850,~850$\}\\
\{${\widetilde E^S},~{\widetilde E^D}$\} $\sim$ \{$511,~939$\}
\vspace{0.5cm} 
\end{center}
\end{minipage}
\\
\hline\hline
\end{tabular}
}
\caption{Input and output parameters for mass hierarchy, MH1, alongwith the physical mass spectrum that
generates tiny neutrino masses and $\Delta a_\ell$ for electron and muon while also satisfying the experimental constraints on the lepton flavor violating (LFV) observables. Yukawa couplings determined from the neutrino sector have been specified for normal hierarchy (NH) and inverted hierarchy (IH) of the neutrino mass spectrum. 
All the Yukawa parameters have been determined upto three decimal places to ensure $\zRE \sim$ unitary, $\ylE \ylE^{\dagger} \sim$ diagonal and Br$\lfb\mu \to e \gamma \rfb$ is within the constrained value.}
\label{table:input_para_1}
\end{table}
\begin{table}[h]
\scalebox{0.8}{
\begin{tabular}{||c||c||}
\hline\hline 
\multicolumn{2}{||c||}{\textbf{Mass hierarchy 2 (MH2)}}\\  
\hline\hline 
\textbf{Mass parameters} & \textbf{Yukawa parameters} \\
(GeV) & $\xi_{\zRE}=0.3$, $\theta_{\zRE}=40^0$, $\xi_{lE}=0.9$, $\eta_{lE}=2$, $\theta_{lE}=75^0$, $d_2=0.017$, $\theta_{R}=-30^0$ \\\hline\hline
\begin{minipage}{0.3\textwidth}
\vspace{-4cm}
\begin{center}
  $\mN=470$\\
  $\mL=500$\\
  $\mE=1100$\\
  $\mu_{\Phi}=1200$
\end{center}
\end{minipage}
&
\begin{minipage}{0.9\textwidth}
\begin{center}
\vspace{0.5cm} 
$\zRE:{\begin{pmatrix} 0.2298 & 0.1928 \\ -0.1928 & 0.2298 \end{pmatrix}}$,~
$\ylE:{\begin{pmatrix} 0.2329 & 0.8693 \\-1.932 & 0.5176 \\ 0 & 0  \end{pmatrix}}$\\
$\yeL:{\begin{pmatrix} -0.4532 & -0.3174 \\ -1.940 & 2.770 \\ 0 & 0 \end{pmatrix}}$\\
$\ylN:{\begin{pmatrix} 2.927\cdot 10^{-6} & 5.633\cdot 10^{-6}\\ 
                      -5.024\cdot 10^{-6} & 1.780\cdot 10^{-5} \\ 
                      -1.324\cdot 10^{-5} & 9.368\cdot 10^{-6} \end{pmatrix}} \text{(NH)}$;~
     ${\begin{pmatrix} 1.023\cdot 10^{-5} & 2.082\cdot 10^{-5}\\ 
                      -1.459\cdot 10^{-6} & 4.356\cdot 10^{-5} \\ 
                       1.522\cdot 10^{-5} & -1.004\cdot 10^{-5} \end{pmatrix}} \text{(IH)}$\\
$\zLN:{\begin{pmatrix} -1.191\cdot 10^{-2} & 6.187\cdot 10^{-3} \\ 
                        1.700\cdot 10^{-2} & -8.833\cdot 10^{-3} \end{pmatrix}} \text{(NH)}$;~
      ${\begin{pmatrix} -1.191\cdot 10^{-2} & 5.850\cdot 10^{-3} \\ 
                        1.700\cdot 10^{-2} & -8.353\cdot 10^{-3} \end{pmatrix}} \text{(IH)}$ \\
\vspace{0.5cm}         
\textbf{Physical masses (GeV):}\\
\{$\phi_P,~\phi^\pm,~\phi_S$\} $\sim$ \{$1200,~1213,~1250$\}\\
\{${\widetilde N^S},~{\widetilde N^D_{a}},~{\widetilde N^D_{b}}$\} $\sim$ \{$470,~500,~500$\}\\
\{${\widetilde E^D},~{\widetilde E^S}$\} $\sim$ \{$496,~1105$\}
\vspace{0.5cm} 
\end{center}
\end{minipage}
\\
\hline\hline
\end{tabular}
}
\caption{Same as Table~\ref{table:input_para_1}, for mass hierarchy, MH2.}
\label{table:input_para_2}
\end{table}

For numerical evaluation of the $(g-2)$ and cLFV observables, we have implemented\footnote{Note that we have already derived the analytical results for the $\Delta a_l$ and LFV radiative decays, $\ell_\alpha \to \ell_\beta \gamma$, in this section. We used these analytical results to validate our model implementation.} the model in {\texttt{SARAH}} \cite{Staub:2013tta,Staub:2015kfa}, generated modules for {\texttt{SPheno}}, and used {\texttt{SPheno}} \cite{Porod:2003um,Porod:2011nf} for the numerical evaluation. To validate our approach of constraining the parameter-space of the model, we randomly generate $10^4$ parameter points by varying the free parameters listed in Eq.~\ref{eqn:freeparam}. The mass parameters, namely, $\mN,~\mL,~\mE$ and $\muphi$, are varied in the range [400--1200] GeV. To avoid a charged particle as the lightest $Z_2$-odd particle (L$Z_2$OP), we ensure $\mN$ is always smaller than the other mass parameters. The Yukawa parameters in Eq.~\ref{eqn:freeparam} are also randomly generated such that
\beq
\zRE^{ij}~\in~[0.2-1.0],~~~\ylE^{ij} ~\in~[0.2-1.0],~~~\zLN^{ij}~\in~[0.0-0.1].
\label{eqn:range}
\eeq
Note that $\yeL$ is inversely dependent on $\zRE$ and $\ylE$ in Eq.~\ref{eqn:yeL_3x2}. The range of $\zRE^{ij}$ and $\ylE^{ij}$ in Eq.~\ref{eqn:range} is motivated to obtain $\yeL^{ij}\sim{\cal O}\lfb 1 \rfb$ or less. For $\zLN$ (see Eq.~\ref{eqn:zLN} and the discussion after that), we choose $d_1=0$ and varied $d_2$ such that $\zLN^{ij}$ falls in the range defined in Eq.~\ref{eqn:range}. The range of $\zLN^{ij}$ is motivated to avoid the constraints from dark-matter direct detection experiments which will be discussed in the next section in detail.
For the $10^4$ parameter points so generated, the numerical values of low energy observables $\Delta a_{l}$, Br$\lfb l_j \to l_i \gamma\rfb$ and $\mu \to e$ conversion rate on Au are presented for $10^4$ randomly generated parameter points in Fig.~\ref{fig:validation}\footnote{The scatter plot shown here was motivated by the conventional discrepancy in $a_\mu$ \cite{Aoyama:2020ynm}. It must be noted, however, that the model can generically accomodate values more compatible with the new results for $\Delta a_\mu$.}. The $10^4$ points are plotted for three pairs of observables each {\em viz.,} in red for ($\Delta a_{e}, \Delta a_{\mu}$), in violet for (Br$\lfb \mu \to e \gamma\rfb$, Br$\lfb \tau \to \mu \gamma\rfb$) and in blue for (Br$\lfb \mu \to e \gamma\rfb$, $\mu \to e$ conversion on Au). 
The values of Br$\lfb \mu \to e \gamma\rfb$ lie well below the upper-bound of $1.5 \times 10^{-13}$, shown by the dashed line. The experimental limits Br$\lfb \tau \to \mu \gamma\rfb$ and $\mu \to e$ conversion on Au are much weaker and hence, not shown. 
We see the consistency of results with the electron and muon $(g-2)$ anomalies as well as experimental bounds on the cLFV observables. In Tables~\ref{table:input_para_1} and \ref{table:input_para_2}, we list down two sets of parameter points derived from this random scan.
The two parameter space points represent two different mass hierarchies for the exotic particles, henceforth denoted as MH1 and MH2. The numerical values for low energy observables ($\Delta a_{e}, \Delta a_{\mu}$ and the cLFV branching ratios) pertaining to these two points are listed in Table~\ref{table:low_energy}.

\begin{table}[h]
\hspace{-1cm}
\scalebox{0.9}{
\begin{tabular}{|c|c|c|c|}
\hline
{\bf Low Energy} & {\bf Expt. limit} & {\bf Model prediction for MH1} & {\bf Model prediction for MH2} \\
\bf Observable && {\bf NH (IH)} & {\bf NH (IH)}\\
\hline
$T$ & $-$ & $-4.19 \times 10^{-2} ~(-4.19  \times 10^{-2})$ & $0.787 \times 10^{-2} ~(0.787 \times 10^{-2})$
\\
$S$ & $-$ & $-2.61 \times 10^{-2} ~(-2.61 \times 10^{-2})$ & $-2.81 \times 10^{-3} ~(-2.81 \times 10^{-3})$
\\
$U$ & $-$ & $-5.56 \times 10^{-4} ~(-5.56 \times 10^{-4})$ & $-5.97 \times 10^{-4} ~(-5.97 \times 10^{-4})$
\\
$\rho$ & $-$ & $-4.18 \times 10^{-5} ~(-4.18 \times 10^{-5})$ & $-4.05 \times 10^{-4} ~(-4.05 \times 10^{-4})$
\\\hline
$10^{12}\Delta a_{e}$ & $-0.88 \pm 0.36$ \cite{Parker:2018vye} & $-1.07 ~(-1.07)$ & $-0.96 ~(-0.96)$
\\
$10^{9}\Delta a_{\mu}$ & $0.38\pm 0.64$ \cite{Aliberti:2025beg,Muong-2:2025xyk} & $3.03 ~(3.03)$ & $2.74 ~(2.74)$
\\\hline
Br$\lfb \mu^\pm \to e^\pm \gamma\rfb$ & $1.5 \times 10^{-13}$ \cite{MEGII:2025gzr} & $7.09 \times 10^{-14} ~(7.09 \times 10^{-14})$ & $1.82 \times 10^{-14} ~(1.82 \times 10^{-14})$
\\
Br$\lfb\tau^\pm \to e^\pm \gamma\rfb$ & $3.3\times 10^{-8}$ \cite{BaBar:2009hkt} & $1.12 \times 10^{-29} ~(1.74 \times 10^{-29})$ & $1.15  \times 10^{-25} ~(1.15  \times 10^{-25})$
\\
Br$\lfb\tau^\pm \to \mu^\pm \gamma\rfb$ & $4.4\times 10^{-8}$ \cite{Belle:2021ysv} & $1.46 \times 10^{-18} ~(1.86 \times 10^{-18})$ & $2.48  \times 10^{-16} ~(3.08  \times 10^{-16})$
\\\hline
Br$\lfb\mu^+ \to e^+e^+e^-\rfb$ & $1.0\times 10^{-12}$ \cite{Bellgardt:1987du} & $4.98 \times 10^{-16} ~(4.98 \times 10^{-16})$ & $1.29 \times 10^{-16} ~(1.29 \times 10^{-16})$
\\
Br$\lfb\tau^- \to e^-e^+e^-\rfb$ & $2.7\times 10^{-8}$ \cite{Hayasaka:2010np} & $1.35 \times 10^{-31} ~(2.15 \times 10^{-31})$ & $1.38 \times 10^{-27} ~(1.39 \times 10^{-27})$
\\
Br$\lfb\tau^- \to \mu^-\mu^+\mu^-\rfb$ & $2.1\times 10^{-8}$ \cite{Hayasaka:2010np} & $3.77 \times 10^{-21} ~(4.77 \times 10^{-21})$ & $6.39 \times 10^{-19} ~(7.92 \times 10^{-19})$
\\
Br$\lfb\tau^- \to e^-\mu^+\mu^-\rfb$ & $2.7\times 10^{-8}$ \cite{Hayasaka:2010np} & $2.63 \times 10^{-32} ~(4.12 \times 10^{-32})$ & $2.72 \times 10^{-28} ~(3.00 \times 10^{-28})$
\\
Br$\lfb\tau^- \to \mu^-e^+e^-\rfb$ & $1.8\times 10^{-8}$ \cite{Hayasaka:2010np} & $1.73 \times 10^{-20} ~(2.20 \times 10^{-20})$ & $2.94 \times 10^{-18} ~(3.64 \times 10^{-18})$
\\
Br$\lfb\tau^- \to e^+\mu^-\mu^-\rfb$ & $1.7\times 10^{-8}$ \cite{Hayasaka:2010np} & $3.49 \times 10^{-43} ~(4.53 \times 10^{-43})$ & $4.43 \times 10^{-40} ~(5.75 \times 10^{-40})$
\\
Br$\lfb\tau^- \to \mu^+e^-e^-\rfb$ & $1.5\times 10^{-8}$ \cite{Hayasaka:2010np} & $1.26 \times 10^{-45} ~(1.83 \times 10^{-44})$ & $1.60 \times 10^{-42} ~(2.31 \times 10^{-41})$
\\\hline
$\mu \to e$ on Pb & $4.6\times 10^{-11}$ \cite{SINDRUMII:1996fti} & $3.08 \times 10^{-16} ~(3.08 \times 10^{-16})$ & $7.96 \times 10^{-17} ~(7.95 \times 10^{-17})$
\\
$\mu \to e$ on Ti & $4.3\times 10^{-12}$ \cite{SINDRUMII:1993gxf} & $3.99 \times 10^{-16} ~(3.99 \times 10^{-16})$ & $1.04 \times 10^{-16} ~(1.04 \times 10^{-16})$
\\
$\mu \to e$ on Au & $7.0\times 10^{-13}$ \cite{Bertl:2006up} & $3.28 \times 10^{-16} ~(3.28 \times 10^{-16})$ & $8.46 \times 10^{-17} ~(8.46 \times 10^{-17})$
\\
$\mu \to e$ on Al\footnotemark & $10^{-15} - 10^{-18}$ \cite{Kuno:2013mha} & $2.22 \times 10^{-16} ~(2.22 \times 10^{-16})$ & $5.79 \times 10^{-17} ~(5.79 \times 10^{-17})$
\\
\hline
\end{tabular}}
\caption{Values of various low energy observables at the parameter sets of Tables~\ref{table:input_para_1} and \ref{table:input_para_2} for normal hierarchy (inverted hierarchy) of the neutrino mass spectrum.}
\label{table:low_energy}
\end{table}
\footnotetext[\thefootnote]{future sensitivity}
\section{Dark matter and feasible parameter space}
\label{sec:dm}
After 
the discussion of tiny neutrino masses and $\Delta a_\ell$ for electron and muon, as well as the experimental constraints on the lepton flavor violating (LFV) observables in sections \ref{sec:numass} and \ref{sec:lfv}, we next search for a candidate for cosmologically viable dark matter within the framework of this model. The lightest $Z_2$-odd particle (L$Z_2$OP) in the model, being stable and weakly interacting, can be a potential candidate for dark matter if it is neutral and satisfies the measured relic density and the constraints from dark matter direct and indirect detection experiments. The model gives rise to the possibility of both fermionic dark matter when a neutral exotic fermion is the L$Z_2$OP, and scalar dark matter when a neutral exotic scalar is the L$Z_2$OP.

We first take up the case of the scalar DM which is the CP-odd exotic scalar ($\phi_P$). In the course of this work, we have assumed a small mass splitting, $|m_{\phi_S}-m_{\phi_P}|$, which has been crucial in addressing the issues of neutrino masses as well as the bounds on various LFV processes. Since $\lambda_3 \sim {\cal O}\lfb 1\rfb$, the smallness of $|m_{\phi_S}-m_{\phi_P}|$ can only be assured by considering $\muphi >> v$. Therefore, we only consider scalar DM of mass $m_{\phi_P} \gtrsim 246$~GeV. For such a DM, the dominant annihilation channels typically include final states with electroweak gauge bosons, which can lead to an efficient depletion of the relic abundance for moderate dark matter masses. For larger masses, however, there can be a possibility of achieving relic density value consistent with observations.
Concerning direct detection, the small value of $|m_{\phi_S}-m_{\phi_P}|$ implies two-component DM. Although both have a considerable coupling with $Z$, direct detection can proceed via inelastic scattering which can significantly weaken current bounds. A detailed analysis of this possibility is beyond the scope of the present work.
  
We now discuss the scenario of a fermionic DM. The mass and mixings of DM are determined from the mass matrix in Eq.~\ref{eqn:mass_matrices} which imply that the DM can be dominantly an SU(2) singlet, an SU(2) doublet or an admixture of the two. This nature affects the DM phenomenology crucially. The parameters relevant for studying this DM are $\mL$, $\mN$ and the parameter $d_2$ that defines their mixing $\zLN$\footnote{$\zLN$ corresponds to lagrangian term $\overline{L_L}\tilde H N^S_R$, responsible for the exotic neutral fermions' doublet-singlet mixing and their interaction with the Higgs.}. Guided by the two parameter points in Tables~\ref{table:input_para_1} and \ref{table:input_para_2}, we explore the region around them for a viable DM candidate, by varying the parameters $\mN$ and $d_2$ while the other free parameters remain the same as per the table. For the points in the extended region, the Yukawa matrices $\ylN$ and $\zLN$ are modified slightly and may affect some cLFV branching ratios, their effect being negligible though. The Yukawa couplings more significant for the cLFV decays {\em viz.,} $\ylE$ and $\yeL$ remain unchanged. Thus, we investigate two mass hierarchies for studying dark matter {\em viz.,} MH1 and MH2. 
\vspace{0.25cm}
\subsection{Relic density}
\begin{figure}[h!]
\makebox[\textwidth][c]{
\scalebox{1.4}{
\includegraphics[width=7cm,height=6cm]{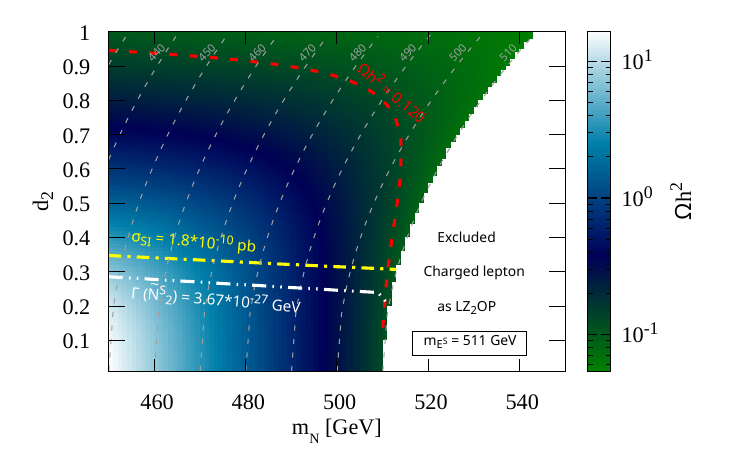} 
\includegraphics[width=7cm,height=6cm]{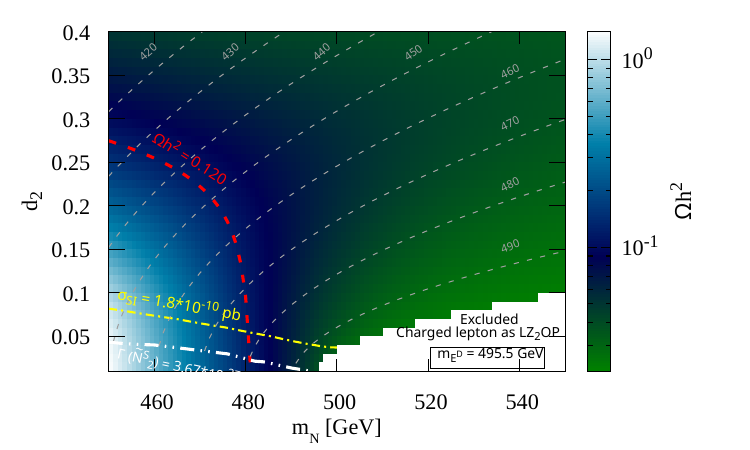}}} 
\caption[Relic density of the lightest exotic neutral fermion as DM, as a function of the mass parameter $\mN$ and scaling factor $d_2$ of coupling $\zLN$. The scale on the right side of the plot measures the abundance throughout the plane. The figures correspond to MH1 (left) and MH2 (right), as per Tables~\ref{table:input_para_1} and \ref{table:input_para_2}. The dashed curves correspond to observational bounds {\em viz.,} $\Omega h^2=0.120$~\cite{Planck:2018vyg} (red), $\sigma_{SI}=1.8 \times 10^{-10}$~pb~\cite{Amole:2019fdf, LZ:2022lsv} (yellow) and $\Gamma ~(\NS_2) = 3.67 \times 10 ^ {-27} $~GeV (white). The allowed regions are explained in the text. The region in white is excluded as charged lepton becomes the L$Z_2$OP. In the feasible region, $\Delta m ({\NS}_2, {\NS}_1) \sim 2.85$ GeV (left) and $1.25$ GeV (right).]{Relic density of the lightest exotic neutral fermion as DM, as a function of mass paramter $\mN$ and scaling factor $d_2$ of coupling $\zLN$. The scale on the right side of the plot measures the abundance throughout the plane. The figures correspond to MH1 (left) and MH2 (right), as per Tables~\ref{table:input_para_1} and \ref{table:input_para_2}. The dashed curves correspond to observational bounds {\em viz.,} $\Omega h^2=0.120 \pm 0.001$~\cite{Planck:2018vyg} (red), $\sigma_{SI}=1.8 \times 10^{-10}$~pb~\cite{Amole:2019fdf, LZ:2022lsv} (yellow)\footnotemark ~and $\Gamma ~(\NS_2) = 3.67 \times 10 ^ {-27} $~GeV (white)\footnotemark. The allowed regions are explained in the text. The region in white is excluded as charged lepton becomes the L$Z_2$OP.}
\label{fig:dm_main}
\end{figure}

\vspace{0.25cm}
The thermal relic abundance of DM is governed by the Boltzmann equation which describes the evolution of its number density as a function of its various interactions as well as universe's expansion. Considering the large spectrum of $Z_2$-odd particles, the DM relic abundance gets contribution not only from its annihilations but also from the annihilation of heavier particles which subsequently decay to DM. This is called coannihilation and it is relevant provided the mass-difference between DM and the heavier species is small. The corresponding Boltzmann equation is \cite{Edsjo:1997bg},
\bea
\frac{dn_i}{dt} &=& -3 H n_i - \sum_{i,j=1}^N \langle \sigma_{ij}v_{ij}\rangle \left(n_in_j-n_i^{eq}n_j^{eq}\right) \nonumber \\
&& -\sum_{j\neq i} \left[\langle {\sigma'}_{X_{ij}}v_{ij}\rangle \left(n_in_X-n_i^{eq}n_X^{eq}\right)
                   - \langle {\sigma'}_{X_{ji}}v_{ij}\rangle \left(n_jn_X-n_j^{eq}n_X^{eq}\right)\right] \nonumber \\
&& -\sum_{j\neq i} \left[\Gamma_{ij} \left(n_i-n_i^{eq}\right) - \Gamma_{ji} \left(n_j-n_j^{eq}\right)\right],
\eea
where $n_i$ denotes the number density of DM or one of the coannihilating species. There will be one such equation for each of the coannihilating species including DM. Denoting $Z_2$-odd particles by $\chi_{i,j}$ and SM particles by $X,Y$, the terms other than the Hubble expansion term correspond to annihilations $\chi_i\chi_j \to X X$, $\chi_i$ to $\chi_j$ conversion via scatterings $\chi_iX \to \chi_j Y$ and decays $\chi_i \to \chi_j X$, respectively. Other symbols are as follows: $\langle \sigma v\rangle$ denotes the thermally averaged cross-section, subscript `$eq$' denotes equilibrium and $H$ is the Hubble parameter. The processes leading to DM annihilation as well as coannihilation in the present model can be classified as
\begin{itemize}
	\item{Annihilations to SM fermions, gauge bosons or Higgs via s-channel mediation by SM gauge bosons or Higgs}
	\item{Annihilations to SM fermions via t-channel exchange of inert scalars}
	\item{Annihilations to SM bosons via t-channel mediation by $Z_2$-odd leptons.}
	\end{itemize}
The gauge coupling with $Z$-boson turns out to be the most critical in the annihilation process, affected through the doublet component in DM. This is why the singlet-doublet mixing of DM is crucial. On the other hand, the Yukawa couplings $\ylN$, $\ylE$ or $\yeL$ affect only a few t-channel mediated processes and thus, are less significant. The relic abundance of DM is evaluated numerically using {\texttt{micrOMEGAs}} \cite{Belanger:2013oya}. It was found that the normal and inverted hierarchies with respect to the neutrino masses do not lead to qualitatively different results for dark matter. Henceforth, we discuss our results with respect to the normal hierarchy only. 
The effect of varying $\mN$ and $d_2$ ($\zLN$) around the parameter points, MH1 and MH2, on dark matter relic density and other related observables is depicted in Fig.~\ref{fig:dm_main}, left and right, respectively. The other important parameter for dark matter study {\em viz.,} $\mL$ is fixed in these plots ($850$ GeV for MH1 and $500$ GeV for MH2). The value of $\mN$ relative to the fixed parameter $\mL$ has an impact on the nature of DM.
To understand this,
we consider the mixings of the $Z_2$-odd neutral fermions given by equations \ref{eqn:UN0_2} and \ref{eqn:UN1}.
Following that, we see
\begin{itemize}
	\item{In the $\mN < \mL$ region: The DM is $\NS$ {\em i.e.,} dominantly a singlet. With an increase in $\mN$, the splitting $|\mL-\mN|$ decreases and the doublet component in DM increases. At the same time, with an increase in $d_2$ ($\zLN$), the doublet component in the singlet-DM increases.}
\item{In the $\mN > \mL$ region: The DM is $\ND$ {\em i.e.,} dominantly a doublet. With an increase in $\mN$, the splitting $|\mL-\mN|$ increases and the doublet component in DM increases further. At the same time, with an increase in $d_2$ ($\zLN$), the singlet component in the doublet-DM increases.}
\end{itemize} 
\addtocounter{footnote}{-1}
\footnotetext[\thefootnote]{within 1$\sigma$ uncertainty of the constraint.}
\addtocounter{footnote}{1}
\footnotetext[\thefootnote]{from the perspective of constraints from BBN and CMB, as explained later in the text, we consider individual components of DM ($\NS$).}
This is indeed what we observe for both the plots in Fig.~\ref{fig:dm_main}. Since the doublet component causes the DM to annihilate more, 
the relic density is maximum for the bottom left corner in region $\mN < \mL$ and minimum for the bottom right corner in region $\mN > \mL$. The region in the white is where the lightest $Z_2$-odd particle is charged and therefore, excluded from the feasible region.

The set of points satisfying the relic density value of $\Omega h^2=0.120$ are shown by the dashed curves in red.
For the two different scenarios depicted here, the shapes of the relic density curves have a common interpretation.
In the regions of small $d_2$ ($\zLN$) and small $\mN$ ($\mN < \mL$), the DM is strictly a singlet and therefore, over-abundant in the absence of any considerable interactions. An increase in $d_2$ ($\zLN$) as well as $\mN$ increases the doublet component in DM and therefore, enhances its annihilations through a stronger coupling with Z boson. 
However, an increase in $\mN$ also leads to coannihilations with other $Z_2$-odd particles ($\ES$, $\ED$, $\ND$ or $\Phi$'s) which become increasingly efficient with the decrease in splitting $|\mN-\mL|$ whereas the decay of heavier species to DM becomes less efficient. 
These together lead to an over-decrease in DM density which is avoided by a sharply decreasing singlet-doublet coupling $d_2$ ($\zLN$), for large $\mN$.
This explains why the relic density is satisfied for a higher value of $d_2$ ($\zLN$) at low values of $\mN$
and an almost vertically falling relic density curve as higher values of $\mN$ are approached.
In brief, the DM relic abundance can be determined either via sizeable self-annihilation cross-sections - requiring a large $d_2$ ($\zLN$) - or through coannihilation processes involving heavier states. 

The qualitative differences between the two figures can be understood as follows.\\

{\noindent \bf \em Mass hierarchy-1 (MH1)}: This scenario is depicted in Fig.~\ref{fig:dm_main} (left). The relevant mass parameters other than $\mN$ are $\mE=600$~GeV, $\muphi=650$~GeV and $\mL=850$~GeV. The lightest $Z_2$-odd charged particle (L$Z_2$OCP) is $\ES$ with $m_{\ES} \sim 511$~GeV. The part of the plane is excluded where $\mN$ value is such that $\ES$ becomes the L$Z_2$OP (shown as white region in the plot). Thus, $\mN$ is varied in the range $450-550$~GeV (for a given $\mN$, the mass of DM decreases with increase in $d_2$ ($\zLN$)). Since $\mL \sim 850$~GeV, the DM is dominantly singlet ($\NS$) in the whole of the depicted plane. 
At $\mN \sim \mE$, the DM co-annihilates with the next heavier state, $\ES$, and gets contribution from its decays as well. The DM self-annihilations are quite suppressed and the annihilations of $\ES$ determine the relic density. From Table \ref{table:input_para_1}, we observe that $\ES$ has a large doublet component, due to large $\zRE$. Consequently, the dominant channel for relic density is, $\bar{\ES} \ES \to W^+ W^-$. In addition to the s-channel process, the process via t-channel vector-like leptons contributes significantly to this channel due to large multiplicity of these mediators.

{\noindent \bf \em Mass hierarchy-2 (MH2)}: This scenario is depicted in Fig.~\ref{fig:dm_main} (right). The values of the other relevant mass parameters are $\mL=500$~GeV, $\mE=1100$~GeV and $\muphi=1200$~GeV. Here, the L$Z_2$OCP is a doublet, $\ED$ with $m_{\ED} \sim 496$~GeV. Again, $\mN$ is varied in the range $450-550$~GeV. The region where $m_{\ND}$ exceeds $m_{\ED}$ is excluded. Unlike MH1, the plane here is divided into two regions {\em viz.,} $\mN < \mL$ and $\mN > \mL$, although the curve explaining the observed value of relic density falls within the region $\mN < \mL$. As a result, the DM is again, singlet-dominated. However, for a given $\mN$ and $d_2$ ($\zLN$), the DM in this scenario has a smaller mass-splitting $|\mN-\mL|$ and therefore, a larger doublet component compared to DM of Fig.~\ref{fig:dm_main} (left). 

The coannihilating partners of this DM are $\ND$ and $\ED$. From Tables \ref{table:input_para_1} and \ref{table:input_para_2}, we observe that the scenario for MH2 admits relatively large values for the Yukawa couplings involving two vector-like leptons viz., $\ylE$ and $\yeL$. 
Consequently, the dominant co-annihilating processes for DM occur via inert scalars in t-channel. Following the specific structure of these Yukawa couplings, the dominant processes are, $\bar{\ED} \ED \to \mu^+ \mu^-$ and $\bar{\ND} \ND \to \mu^+ \mu^-$. The coannihilations in this scenario are much more efficient than those in MH1, 
as depicted by the relic density curve falling much before the excluded region in white. Further, the values of $d_2$ ($\zLN$) satisfying the relic density are also significantly lower than those in the MH1. It may be noted that the region to the right of the relic density curve in red is under-abundant and therefore, in principle, allowed unless excluded by charged lepton as the L$Z_2$OP.

Thus, from Fig.~\ref{fig:dm_main}, we see how the relic density of dark matter is satisfied for two very different mass hierarchies in this model.\\

{\noindent \bf \em Constraints:} It must be noted that for the two scenarios presented here, the output Yukawa couplings $\ylN$, $\yeL$ and $\zLN$ were determined as per equations \ref{eqn:ylN_nu}, \ref{eqn:CLR_2x2_simp} and \ref{eqn:zLN} such that the various low energy observables (sec.~\ref{sec:numass} and \ref{sec:lfv}) are explained by default. Nevertheless, the values of these observables were computed numerically and it was found that in the whole plane of $(\mN,d_2)$, as shown in Fig.~\ref{fig:dm_main} (left and right), the anomalies in electron and muon $g-2$ were explained while the various cLFV processes were within the experimental bounds. In addition, there are cosmological constraints as well as constraints from direct detection of dark matter which are discussed in the following subsections.
\vspace{0.25cm}
\subsection{Direct and Indirect detection}
\label{subsec:dd}
\vspace{0.25cm}
There are several experiments \cite{XENON:2018voc,LUX:2018akb,PICO:2019vsc,XENON:2019rxp} looking for direct detection (DD) of dark matter via its scattering off the nuclei of target materials. Provided with an effective lagrangian that describes the DM interaction with the quarks, the DM-nucleus cross-section is estimated by taking into account the hadronic matrix elements. The differential DM-nucleus cross section can be expressed as \cite{Cerdeno:2010jj}
\begin{equation}
  \frac{d\sigma}{dE_R}=\frac{m_{nuc}}{2\mu_N^2v^2}\left(\sigma_0^{SI}
  F^2_{SI}(E_R) + \sigma_0^{SD} F^2_{SD}(E_R)\right)\ ,
\end{equation}
where $E_R$ is the recoil energy, $\sigma_0^{SI,\, SD}$ are the spin-independent (SI) and spin-dependent (SD) cross
sections at zero momentum transfer, $F(E_R)$ are the form factors that include the dependence on the momentum-transfer, $m_N$ is the mass of the nucleus, $\mu_N$ is the reduced mass of the DM and the nucleus and $v$ is the velocity of the incoming particle. The source of the different contributions lies in the nature of coupling to quarks. A scalar- or a vector-type current {\em i.e.,} $\bar{q} q$ or $\bar{q} {\gamma}_{\mu} q$ keeps the DM-nucleon interaction spin-independent. On the other hand, currents of type $\bar{q} \gamma_5 q$, $\bar{q} {\gamma}_{\mu} \gamma_5 q$ or $\bar{q} {\sigma}_{\mu \nu} q$ introduce spin-dependence.

In case of a Majorana fermion DM, which is the case here, the current coupling to DM can be $\bar{\chi} \chi$, $\bar{\chi} {\gamma}_5 \chi$, or $\bar{\chi} {\gamma}_{\mu}{\gamma}_5\chi$. Thus, effective operators for DM-nucleus interaction are $\bar{\chi} \chi \bar{q} q$, $\bar{\chi} \chi \bar{q} {\gamma}_5 q$, $\bar{\chi} {\gamma}_5 \chi \bar{q} q$, $\bar{\chi} {\gamma}_5 \chi \bar{q} {\gamma}_5 q$, $\bar{\chi} {\gamma}^{\mu}{\gamma}_5\chi \bar{q} {\gamma}_{\mu} q$ and $\bar{\chi} {\gamma}^{\mu}{\gamma}_5\chi \bar{q} {\gamma}_{\mu} {\gamma}_5 q$. The operator $\bar{\psi} {\gamma}_5 \psi$ vanishes in the zero-momentum transfer limits while only spatial and temporal components of operators $\bar{\psi} {\gamma}_{\mu} {\gamma}_5 \psi$ and $\bar{\psi} {\gamma}_{\mu} \psi$ remain \cite{Belanger:2008sj}. Thus, the only dominant contributions to DM-nucleus interaction are $\bar{\chi} \chi \bar{q} q$ and $\bar{\chi} {\gamma}^{\mu}{\gamma}_5\chi \bar{q} {\gamma}_{\mu} {\gamma}_5 q$ that correspond to SI and SD interactions, respectively. 

Upper bounds exist on the cross-section of DM interaction with the nucleus as provided in Ref.~\cite{LZ:2022lsv}. In the model here, the contribution to SI cross-section (${\sigma}_{SI}$) arises from DM coupling to Higgs whereas the contribution to SD cross-section (${\sigma}_{SD}$) arises from axial-vector coupling to Z. Due to Majorana nature of DM, vector coupling to Z and thus, the operator $\bar{\chi} {\gamma}^{\mu} \chi \bar{q} {\gamma}_{\mu} q$ is absent. In this way, a significant contribution to ${\sigma}_{SI}$, the bound on which is the more stringent one, is avoided. Only the stricter bound {\em i.e.} the one on $\sigma_{SI}$ is presented in Fig.~\ref{fig:dm_main}, shown by the dashed vurve in yellow. The SI cross-section {\em i.e.,} $\sigma_{SI}$ that depends upon the coupling of DM to Higgs arises from the interaction term, $\zLN^{\alpha\beta} \bar{\NDL}_{\alpha} \tilde{H} {\NSR}_{\beta}$, as mentioned earlier. The dependence on $d_2$ ($\zLN$) is, therefore, obvious. The coupling increases with an increase in $d_2$ ($\zLN$). The coupling, however, also depends upon the ratio of singlet-doublet components in DM and is maximum for $\mN \sim \mL$ {\em i.e.,} for an equal singlet-doublet admixture. Thus, with decreasing $|\mN - \mL|$, the coupling increases and the constraint is satisfied for a smaller value of $d_2$ ($\zLN$).

Although the shape of the curve is same for both Fig.~\ref{fig:dm_main} (left) (MH1) and Fig.~\ref{fig:dm_main} (right) (MH2), we see that for a given $\mN$ and $d_2$ ($\zLN$), the DM of MH1 has a smaller doublet component than the DM of MH2 (since $\mL=850$ GeV for MH1 and $500$ GeV for MH2, therefore, splitting $|\mN-\mL|$ is larger for MH1 for the same point in $\mN$-$d_2$ ($\zLN$) plane). As a result, the bound on $\sigma_{SI}$ is allowed for a larger $d_2$ ($\zLN$) for Fig.~\ref{fig:dm_main} (left) (MH1) in contrast to Fig.~\ref{fig:dm_main} (right) (MH2).

Before concluding this section, we briefly discuss prospects with respect to indirect detection. We have seen that DM self-annihilations are quite suppressed and relic density is controlled by the annihilations of the next heavier states (coannihilating partners $\ES$, $\ED$ and $\ND$) to SM against annihilations as well as decays of these coannihilating partners to DM. For the benchmark scenarios viz., MH1 and MH2, these heavier states annihilate dominantly to $W^+W^-$ (MH1) and $\mu^+\mu^-$ (MH2) while their decays include $\ES, \ED \to W^* \NS$ and $\ND \to W^* \ED, Z^* \NS$, $W^*$ and $Z^*$ being the off-shell bosons. These can contribute to gamma-ray and cosmic ray signals which can be constrained by Fermi-LAT \cite{Fermi-LAT:2015att} and CTA \cite{CTA:2020qlo}. However, the corresponding co-annihilators are no longer present, and the DM self-annihilations are highly suppressed due to velocity effects and are expected to lie below current bounds from the indirect detection experiments.
\vspace{0.25cm}
\subsection{Constraints from BBN and CMB}
\label{subsec:bbn}
\vspace{0.25cm}
Upto now we have referred to the exotic neutral fermions as $\NS$ and $\ND$, considering the mass-degeneracy amongst the singlet states as well as amongst the doublet states. However, in the exact diagonalisation, the small mass-splitting between the two generations of $\NS$ {\em viz.,} $\NS_1$ and $\NS_2$ may become crucial for dark matter relic density as well as BBN and CMB observations. It is so because $\NS_2$ can decay into SM leptons alongwith the DM. In order to avoid any conflict with the BBN and CMB observations, we demand that $\NS_2$ have a lifetime larger than the age of the universe or much smaller than the time of BBN {\em i.e.,}
\beq
\Gamma ~(\NS_2) ~\lesssim~ 10 ^ {-42} ~\textnormal{GeV} ~\cup ~\Gamma ~(\NS_2) ~\gg 3.67 \cdot 10 ^ {-27} ~\textnormal{GeV}.
\eeq
Decay width $\Gamma$($\NS_2$) is a function of the mass-splitting between $\NS_1$ and $\NS_2$ ($\Delta m_{phy}$) which, again, depends upon the mixing parameter $d_2$ ($\zLN$) and $\mN$. The possibility of ${\NS}_2$ being equally stable as DM is less as it requires a negligibly small $d_2$ ($\zLN$). In Fig.~\ref{fig:dm_main}, the white dashed line corresponds to $\Gamma ~(\NS_2) = 3.67 \cdot 10^{-27}$~GeV. The allowed region lies above the white dashed line. Thus, in this model, $\NS_2$ decays before BBN into DM via 3-body decay through off-shell $Z$ boson. As the width is directly proportional to the doublet component in $\NS$, the value of $d_2$ ($\zLN$) satisfying the bound of $3.67 \cdot 10 ^ {-27}$~GeV falls with an increasing $\mN$.
\vspace{0.25cm}
\subsection{Feasible parameter space}
\vspace{0.25cm}
The free parameters of the model were listed in Eq.~\ref{eqn:freeparam}, after defining the remaining parameters to address the issues of tiny neutrino masses and $e$ and $\mu ~(g-2)$ within the purview of the constraints from LFV decays. This set of free parameters spans a large space. However, we see that having a suitable DM candidate (charge neutral and stable on cosmological time scales) with correct relic density while also satisfying cosmological constraints and constraints on its interaction with the nucleons, sets the parameters competing against each other. After studying the different scenarios as shown in Fig.~\ref{fig:dm_main}, we summarise the following results for the feasible parameter space:
\begin{itemize}
\item{Pair ($\mN,d_2$) satisfying $\Omega~h^2=0.120$ is given by the red-dashed curve. Region below the curve is over-abundant while the region above the curve is under-abundant.}
\item{Relic density observations are satisfied for both the mass-hierarchies. In both the scenarios, the DM is a singlet-doublet admixture whose  relic abundance is determined either via sizeable self-annihilation cross-sections - requiring a large $d_2$ ($\zLN$) or, through coannihilation processes involving heavier states.}
\item{The bound on direct detection cross-section, $\sigma_{SI}$, is shown by the dashed curve in yellow and the region above the curve is excluded. For the same point in $\mN$-$d_2$ ($\zLN$) plane, the DM of MH1 has a smaller doublet component than the DM of MH2. As a result, a larger $d_2$ ($\zLN$) is allowed for MH1.} 
\item{The direct-detection constraint severly limits the strength of DM coupling to the SM, thereby, suppressing the DM self-annihilation cross-section. Consequently, determination of relic density must be governed by coannihilation processes alone, with heavier, nearly degenerate states.}
\item{The nature and efficiency of coannihilations differ, however, for the two mass-hierarchies. 
For MH1 {\em viz.,} $\mN < \mE < \muphi < \mL$, DM coannihilates with the singlet-dominated heavy charged leptons $\ES$ whereas for MH2 {\em viz.,} $\mN < \mL < \mE < \muphi$, it coannihilates with the doublet-dominated $\ND$ and $\ED$. 
In the latter case, the coannihilations are much more efficient and DM density and various constraints are explained for a comparatively smaller value of the parameter $d_2$ ($\zLN$).} 
\item{The bound from BBN and CMB {\em viz.,} $\Gamma({\NS}_2) ~\gg 3.6 \cdot 10^{-27}$ GeV is given by the dashed curve in white. Region above the curve is allowed. Although not depicted in the plot, $d_2$ ($\zLN$) below a certain value is also allowed, corresponding to $\Gamma ~(\NS_2) ~\lesssim~ 10 ^ {-42}$ GeV. For MH1, this constraint implies $d_2 \geq 0.24$ or $d_2 \leq 0.03$, for $\mN=510$ GeV. For MH2, it implies $d_2 \geq 0.026$ or $d_2 \leq 0.001$, for $\mN=481$ GeV.}
\item{The feasible parameter space corresponds to part of the red-dashed curve lying between the yellow and white curves and the parameter space to its right.}
\end{itemize}
Thus, we obtain a feasible region in the parameter space of the model explaining the observations of neutrino masses, anomalous magnetic moments of charged leptons and dark matter relic density while surviving the cLFV bounds as well as the cosmological and direct detection bounds on DM.
\section{Collider phenomenology}
\label{sec:collider}
\vspace{0.25cm}
At the LHC, it would be expected that the QCD-driven pair production
of the quarks would constitute the dominant processes as far as the
exotics are concerned. However, note that none of the low-energy
observables that motivates this study, {\em viz.,}  neutrino mass
generation (see Sec.~\ref{sec:numass}), anomalous magnetic moments and
lepton flavour violation (see Sec.~\ref{sec:lfv}) and the dark matter
relic density (see Sec.~\ref{sec:dm}) are significantly affected by
the presence of the exotic quarks. Indeed, the only low-energy theatre
where these could have been expected to play a dominant role is that of
flavour anomalies in the $B$-sector, an aspect that we are not
addressing here. On the other hand, note that even a semblance of
gauge coupling unification seemingly calls
  for such quarks to be much heavier\footnote{There is a
    caveat, though. If we were to admit large mass-splittings between
    quark fields with identical quantum numbers, it is possible to
    consider a spectrum that enables one to address, say the
    long-standing $2.9\sigma$ discrepancy in the forward-backward
    asymmetry in bottom-quark production at the $Z$-peak while
    maintaining the possibility of
    unification~\cite{Choudhury:2001hs}.}  (see
Sec.~\ref{sec:unification}), and beyond the reach of the LHC. Given
this, we do not investigate the exotic-quark sector and restrict
ourselves to the more difficult case of the leptons and the
(pseudo-)scalars.

The $Z_2$ symmetry stipulates that the exotics
can only decay into a lighter
exotic accompanied by
one or more SM particles (fermions or
bosons). Thus, the pair production and subsequent decay of the
exotics at collider experiments result in jets and leptons associated
with an imbalance in momentum in the transverse direction, largely stemming from the lightest (and, hence, stable and invisible)
$Z_2$-odd particle in the final state.
\begin{figure}[h!]
\begin{center}
\includegraphics[width=0.9\linewidth]{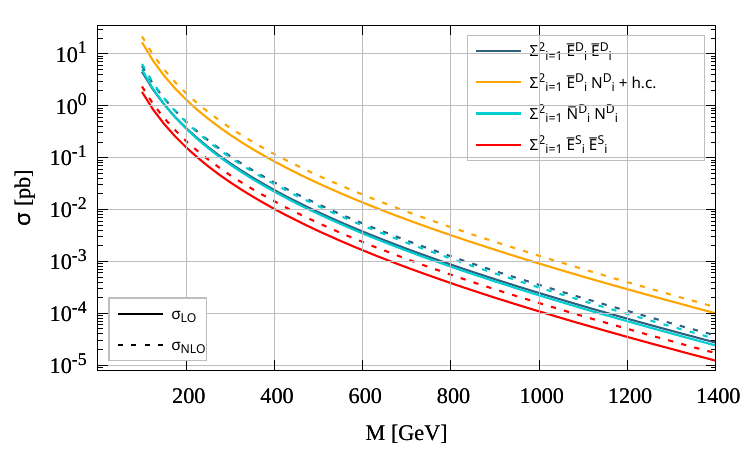}\\
\caption{Cross-section (in pb) for the production of $Z_2$-odd leptons
  at 13 TeV LHC. In the limit of zero singlet-doublet mixing, the
  particles produced in each pair are mass-degenerate and thus, mass
  on the $x$-axis corresponds to physical mass of the particles
  produced in each pair.}
\label{fig:production_plots}
\end{center}
\end{figure}
\vspace{0.25cm}
\subsection{Production at the LHC}
\vspace{0.25cm}
The gauge interactions of the exotic leptons and scalars facilitate 
the pair- and associated production of the $Z_2$-odd exotics at the LHC 
through Drell-Yan processes ({\em i.e.,} s-channel $W^\pm/Z/\gamma$ mediation)\footnote{Note that the Higgs
  mediated $s$-channel diagrams also contributes to the production of
  $Z_2$-odd exotics. Nevertheless, these contributions are subdued by
  the Yukawa couplings of light quarks and are thus not taken into
  consideration in this analysis.}. The production cross-sections are 
  computed at the leading order (LO) and next-to-leading order (NLO) in {\texttt{MG5{\_}aMC{\_}v2.9.9}} using 
  {\texttt{NN23LO1}} as the parton distribution function with the renormalization 
  and factorisation scales set at $m_Z$. The NLO computation of the cross-section 
  has been performed using the Universal {\sc FeynRules} Output (UFO) model \cite{AH:2023hft}. 
  We find that the LO and NLO K-factors vary between $1.28-1.46$ as a function of the exotic lepton mass. 
Fig.~\ref{fig:production_plots} demonstrates the LO and NLO production cross-sections\footnote{The calculation 
of exotic lepton pair production cross sections depicted in  Fig.~\ref{fig:production_plots} 
does not take into account the small doublet-singlet mixing necessary to explain anomalies in charged
lepton magnetic moments and dark matter relic density.} of the exotic leptons 
plotted against their masses. In the limit 
of zero singlet-doublet mixing, the particle spectrum simply corresponds to $E^D_i=
E^D_{L,i}+E^D_{R,i}$, $E^S_i=E^S_{L,i}+E^S_{R,i}$ and $N^D_i = N^D_{L,i}+N^D_{R,i}$ 
where $i=1,2$. Further, as gauge bosons couple only to gauge states of same generation, 
production of a pair of doublets (singlets) of two different generations will not be there. 
The pair-production cross-sections for these exotic leptons vary from a few picobarns
to a fraction of a femtobarn as we vary their masses from 100 GeV to a TeV. 
Hence, a substantial quantity of the exotic leptons is anticipated to be 
produced at the LHC, offering a potential avenue for testing the model in collider 
experiments. The nature of final state signatures, though, depends 
crucially on how these exotics decay, which we discuss next.
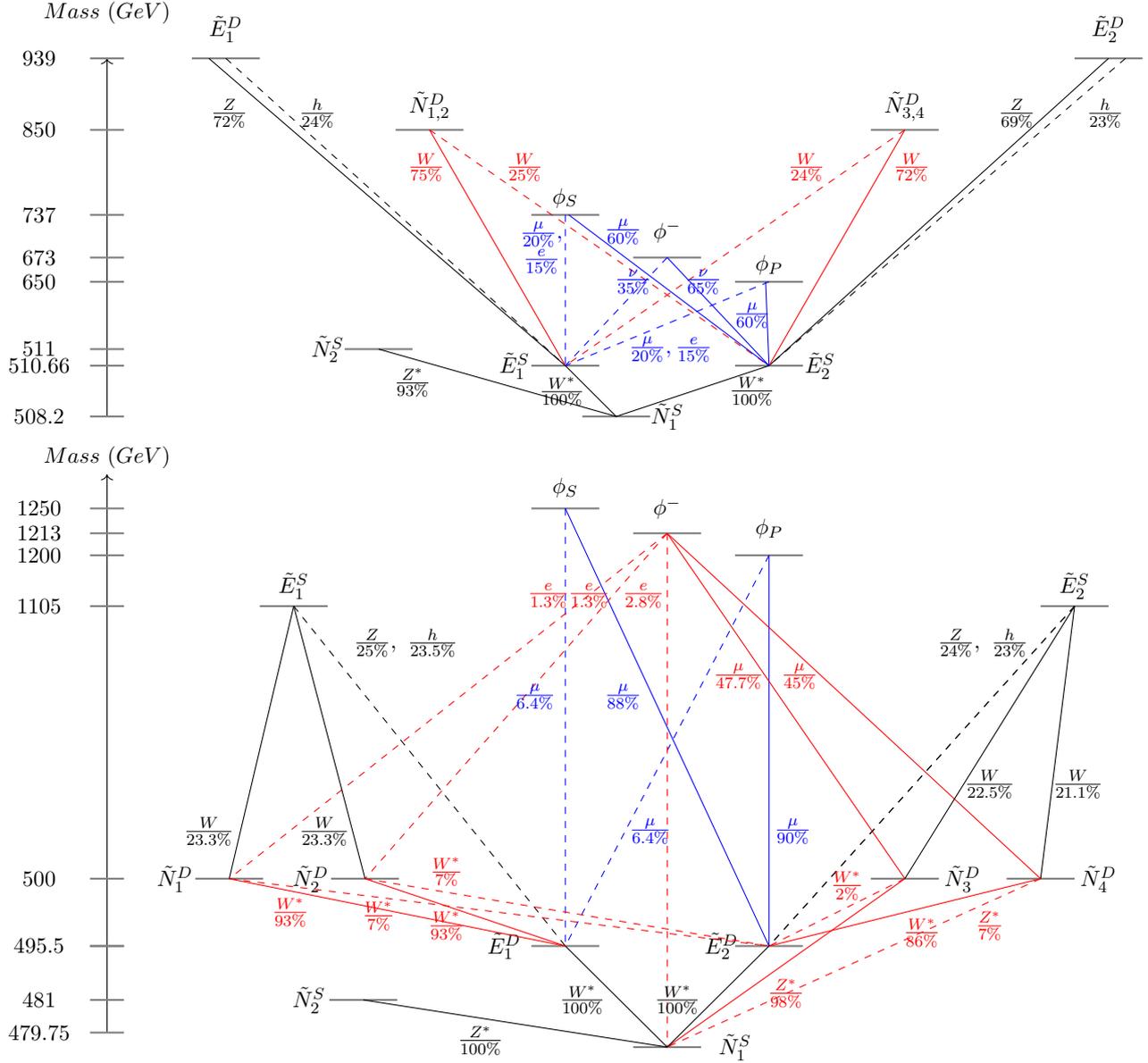
\begin{figure}[ht!]
\centering
    \begin{subfigure}[c]{\textwidth}
\hspace{-0.5cm}
\begin{tikzpicture}[line width=1.4 pt, scale=10,every node/.style={scale=0.85}]
\draw[linearrow,black,thin] (-0.75,0.45) --(-0.75,0.98154);
\node at (-0.75,1.05) {$Mass ~(GeV)$};

\draw[gray, thick] (-0.775,0.98154) -- (-0.725,0.98154);
\draw[gray, thick] (-0.775,0.8754) -- (-0.725,0.8754);
\draw[gray, thick] (-0.775,0.74932) -- (-0.725,0.74932);
\draw[gray, thick] (-0.775,0.68597) -- (-0.725,0.68597);
\draw[gray, thick] (-0.775,0.65) -- (-0.725,0.65);
\draw[gray, thick] (-0.775,0.55) -- (-0.725,0.55);
\draw[gray, thick] (-0.775,0.52542) -- (-0.725,0.52542);
\draw[gray, thick] (-0.775,0.45) -- (-0.725,0.45);

\node at (-0.85,0.98154) {$939$};
\node at (-0.85,0.8754) {$850$};
\node at (-0.85,0.74932) {$737$};
\node at (-0.85,0.68597) {$673$};
\node at (-0.85,0.65) {$650$};
\node at (-0.85,0.55) {$511$};
\node at (-0.85,0.52542) {$510.66$};
\node at (-0.85,0.45) {$508.2$};

\draw[gray, thick] (-0.625,0.98154) -- (-0.525,0.98154); 
\node at (-0.575,1.025) {$\tilde{E}^D_1$};

\draw[gray, thick] (0.675,0.98154) -- (0.775,0.98154); 
\node at (0.725,1.025) {$\tilde{E}^D_2$};


\draw[gray, thick] (-0.325,0.8754) -- (-0.225,0.8754); 
\node at (-0.275,0.91) {$\tilde{N}^D_{1,2}$};

\draw[gray, thick] (0.375,0.8754) -- (0.475,0.8754); 
\node at (0.425,0.91) {$\tilde{N}^D_{3,4}$};


\draw[gray, thick] (-0.125,0.74932) -- (-0.025,0.74932); 
\node at (-0.075,0.775) {$\phi_{S}$};

\draw[gray, thick] (0.025,0.68597) -- (0.125,0.68597); 
\node at (0.075,0.725) {$\phi^-$};

\draw[gray, thick] (0.175,0.65) -- (0.275,0.65); 
\node at (0.225,0.675) {$\phi_{P}$};

\draw[gray, thick] (-0.125,0.52542) -- (-0.025,0.52542); 
\node at (-0.15,0.52542) {$\tilde{E}^S_1$};

\draw[gray, thick] (0.175,0.52542) -- (0.275,0.52542); 
\node at (0.3,0.52542) {$\tilde{E}^S_2$};

\draw[gray, thick] (-0.4,0.55) -- (-0.3,0.55); 
\node at (-0.425,0.55) {$\tilde{N}^S_2$};

\draw[gray, thick] (-0.05,0.45) -- (0.05,0.45); 
\node at (0.075,0.45) {$\tilde{N}^S_1$};

\draw[majorana,black,thin] (-0.6,0.98154) --(-0.075,0.52542); 
\draw[scalar2,black,thin] (-0.575,0.98154) --(-0.075,0.52542); 
\draw[majorana,black,thin] (0.725,0.98154) --(0.225,0.52542); 
\draw[scalar2,black,thin] (0.75,0.98154) --(0.225,0.52542); 
\draw[majorana,red,thin] (0.425,0.8754) --(0.225,0.52542); 
\draw[scalar2,red,thin] (0.425,0.8754) --(-0.075,0.52542); 
\draw[majorana,red,thin] (-0.275,0.8754) --(-0.075,0.52542); 
\draw[scalar2,red,thin] (-0.275,0.8754) --(0.225,0.52542); 
\draw[majorana,blue,thin] (-0.07,0.74932) --(0.225,0.52542); 
\draw[scalar2,blue,thin] (-0.075,0.74932) --(-0.075,0.52542); 
\draw[majorana,blue,thin] (0.220,0.65) --(0.225,0.52542); 
\draw[scalar2,blue,thin] (0.225,0.65) --(-0.075,0.52542); 
\draw[majorana,blue,thin] (0.075,0.68597) --(0.225,0.52542); 
\draw[scalar2,blue,thin] (0.075,0.68597) --(-0.075,0.52542); 
\draw[majorana,black,thin] (-0.075,0.52542) --(0.0,0.45); 
\draw[majorana,black,thin] (0.225,0.52542) --(0.0,0.45); 
\draw[majorana,black,thin] (-0.35,0.55) --(0.0,0.45); 

\node at (-0.57,0.9) {$\frac{Z}{72 \%}$}; 
\node at (-0.44,0.9) {$\frac{h}{24 \%}$}; 
\node at (0.59,0.9) {$\frac{Z}{69 \%}$}; 
\node at (0.72,0.9) {$\frac{h}{23 \%}$}; 
\node[text=red] at (-0.28,0.82) {$\frac{W}{75 \%}$}; 
\node[text=red] at (-0.135,0.82) {$\frac{W}{25 \%}$}; 
\node[text=red] at (0.435,0.82) {$\frac{W}{72 \%}$}; 
\node[text=red] at (0.28,0.82) {$\frac{W}{24 \%}$}; 
\node[text=blue] at (0.01,0.725) {$\frac{\mu}{60 \%}$}; 
\node[text=blue] at (-0.11,0.72) {$\frac{\mu}{20 \%}, $}; 
\node[text=blue] at (-0.11,0.68) {$\frac{e}{15 \%}$}; 
\node[text=blue] at (0.128,0.65) {$\frac{\nu}{65 \%}$}; 
\node[text=blue] at (0.025,0.65) {$\frac{\nu}{35 \%}$}; 
\node[text=blue] at (0.2,0.6) {$\frac{\mu}{60 \%}$}; 
\node[text=blue] at (0.08,0.55) {$\frac{\mu}{20 \%}, \frac{e}{15 \%}$}; 
\node at (-0.08,0.487) {$\frac{W^*}{100 \%}$}; 
\node at (0.2,0.488) {$\frac{W^*}{100 \%}$}; 
\node at (-0.3,0.5) {$\frac{Z^*}{93 \%}$}; 
\end{tikzpicture}     
    \end{subfigure}
    \begin{subfigure}[c]{\textwidth}
\vspace{2.5cm}
\hspace{-0.5cm}
\begin{tikzpicture}[line width=1.4 pt, scale=10,every node/.style={scale=0.85}]
\draw[linearrow,black,thin] (-0.75,0.47156) --(-0.75,1.3);
\node at (-0.75,1.325) {$Mass ~(GeV)$};

\draw[gray, thick] (-0.775,1.2495) -- (-0.725,1.2495);
\draw[gray, thick] (-0.775,1.2126) -- (-0.725,1.2126);
\draw[gray, thick] (-0.775,1.1800) -- (-0.725,1.1800);
\draw[gray, thick] (-0.775,1.1045) -- (-0.725,1.1045);
\draw[gray, thick] (-0.775,0.70) -- (-0.725,0.7);
\draw[gray, thick] (-0.775,0.6) -- (-0.725,0.6);
\draw[gray, thick] (-0.775,0.6) -- (-0.725,0.6);
\draw[gray, thick] (-0.775,0.52) -- (-0.725,0.52);
\draw[gray, thick] (-0.775,0.47156) -- (-0.725,0.47156);

\node at (-0.85,1.2495) {$1250$};
\node at (-0.85,1.2126) {$1213$};
\node at (-0.85,1.180) {$1200$};
\node at (-0.85,1.1045) {$1105$};
\node at (-0.85,0.7) {$ 500$};
\node at (-0.85,0.6) {$495.5$};
\node at (-0.85,0.52) {$481$};
\node at (-0.85,0.47156) {$479.75$};

\draw[gray, thick] (-0.125,1.2495) -- (-0.025,1.2495); 
\node at (-0.075,1.28) {$\phi_{S}$};

\draw[gray, thick] (0.025,1.2126) -- (0.125,1.2126); 
\node at (0.075,1.25) {$\phi^-$};

\draw[gray, thick] (0.175,1.180) -- (0.275,1.180); 
\node at (0.225,1.22) {$\phi_{P}$};

\draw[gray, thick] (-0.525,1.1045) -- (-0.425,1.1045); 
\node at (-0.475,1.1345) {$\tilde{E}^S_1$};

\draw[gray, thick] (0.625,1.1045) -- (0.725,1.1045); 
\node at (0.675,1.1345) {$\tilde{E}^S_2$};

\draw[gray, thick] (-0.62,0.7) -- (-0.52,0.7); 
\node at (-0.65,0.7) {$\tilde{N}^D_1$};

\draw[gray, thick] (-0.42,0.7) -- (-0.32,0.7); 
\node at (-0.45,0.7) {$\tilde{N}^D_2$};


\draw[gray, thick] (0.375,0.7) -- (0.475,0.7); 
\node at (0.51,0.7) {$\tilde{N}^D_{3}$};

\draw[gray, thick] (0.575,0.7) -- (0.675,0.7); 
\node at (0.71,0.7) {$\tilde{N}^D_{4}$};


\draw[gray, thick] (-0.125,0.6) -- (-0.025,0.6); 
\node at (-0.165,0.6) {$\tilde{E}^D_1$};

\draw[gray, thick] (0.175,0.6) -- (0.275,0.6); 
\node at (0.155,0.6) {$\tilde{E}^D_2$};

\draw[gray, thick] (-0.4225,0.52) -- (-0.3225,0.52); 
\node at (-0.4525,0.52) {$\tilde{N}^S_2$};

\draw[gray, thick] (0.025,0.45) -- (0.125,0.45); 
\node at (0.175,0.45) {$\tilde{N}^S_1$};

\draw[scalar2,blue,thin] (-0.075,1.2495) --(-0.075,0.6); 
\draw[majorana,blue,thin] (-0.075,1.2495) --(0.225,0.6); 
\draw[scalar2,blue,thin] (0.225,1.180) --(-0.075,0.6); 
\draw[majorana,blue,thin] (0.225,1.180) --(0.225,0.6); 
\draw[scalar2,red,thin] (0.075,1.2126) --(0.075,0.45); 
\draw[scalar2,red,thin] (0.075,1.2126) --(-0.57,0.7); 
\draw[scalar2,red,thin] (0.075,1.2126) --(-0.37,0.7); 
\draw[majorana,red,thin] (0.075,1.2126) --(0.425,0.7); 
\draw[majorana,red,thin] (0.075,1.2126) --(0.625,0.7); 
\draw[majorana,black,thin] (-0.475,1.1045) --(-0.57,0.7); 
\draw[majorana,black,thin] (-0.475,1.1045) --(-0.37,0.7); 
\draw[scalar2,black,thin] (-0.475,1.1045) --(-0.075,0.6); 
\draw[majorana,black,thin] (0.675,1.1045) --(0.425,0.7); 
\draw[majorana,black,thin] (0.675,1.1045) --(0.625,0.7); 
\draw[scalar2,black,thin] (0.675,1.1045) --(0.225,0.6); 
\draw[scalar2,black,thin] (0.675,1.1045) --(0.225,0.6); 
\draw[majorana,red,thin] (-0.57,0.7) --(-0.075,0.6); 
\draw[majorana,red,thin] (-0.37,0.7) --(-0.075,0.6); 
\draw[scalar2,red,thin] (-0.57,0.7) --(0.225,0.6); 
\draw[scalar2,red,thin] (-0.37,0.7) --(0.225,0.6); 
\draw[majorana,red,thin] (0.425,0.7) --(0.075,0.45); 
\draw[scalar2,red,thin] (0.425,0.7) --(0.225,0.6); 
\draw[majorana,red,thin] (0.625,0.7) --(0.225,0.6); 
\draw[scalar2,red,thin] (0.625,0.7) --(0.075,0.45); 
\draw[majorana,black,thin] (-0.075,0.6) --(0.075,0.45); 
\draw[majorana,black,thin] (0.225,0.6) --(0.075,0.45); 
\draw[majorana,black,thin] (-0.3725,0.52) --(0.075,0.45); 

\node[text=blue] at (-0.12,0.97) {$\frac{\mu}{6.4 \%}$}; 
\node[text=blue] at (0.01,0.97) {$\frac{\mu}{88 \%}$}; 
\node[text=red] at (0.04,1.12) {$\frac{e}{2.8 \%}$}; 
\node[text=red] at (-0.1,1.12) {$\frac{e}{1.3 \%}$}; 
\node[text=red] at (-0.04,1.12) {$\frac{e}{1.3 \%}$}; 
\node[text=red] at (0.18,1) {$\frac{\mu}{47.7 \%}$}; 
\node[text=red] at (0.27,1) {$\frac{\mu}{45 \%}$}; 
\node[text=blue] at (0.05,0.77) {$\frac{\mu}{6.4 \%}$}; 
\node[text=blue] at (0.26,0.77) {$\frac{\mu}{90 \%}$}; 
\node at (-0.6,0.77) {$\frac{W}{23.3 \%}$}; 
\node at (-0.43,0.77) {$\frac{W}{23.3 \%}$}; 
\node at (-0.31,1.05) {$\frac{Z}{25 \%},~\frac{h}{23.5 \%}$}; 
\node at (0.54,1.05) {$\frac{Z}{24 \%},~\frac{h}{23 \%}$}; 
\node at (0.55,0.84) {$\frac{W}{22.5 \%}$}; 
\node at (0.68,0.84) {$\frac{W}{21.1 \%}$}; 
\node[text=red] at (-0.48,0.65) {$\frac{W^*}{93 \%}$}; 
\node[text=red] at (-0.35,0.645) {$\frac{W^*}{7 \%}$}; 
\node[text=red] at (-0.25,0.63) {$\frac{W^*}{93 \%}$}; 
\node[text=red] at (-0.25,0.71) {$\frac{W^*}{7 \%}$}; 
\node[text=red] at (0.34,0.69) {$\frac{W^*}{2 \%}$}; 
\node[text=red] at (0.25,0.53) {$\frac{Z^*}{98 \%}$}; 
\node[text=red] at (0.45,0.62) {$\frac{W^*}{86 \%}$}; 
\node[text=red] at (0.55,0.63) {$\frac{Z^*}{7 \%}$}; 
\node at (-0.05,0.52) {$\frac{W^*}{100 \%}$}; 
\node at (0.09,0.52) {$\frac{W^*}{100 \%}$}; 
\node at (-0.2,0.46) {$\frac{Z^*}{100 \%}$}; 


\end{tikzpicture} 
    \end{subfigure}
\caption{Decay Cascades at MH1 ($\mN < \mE < \muphi < \mL$) (top) and MH2 ($\mN < \mL < \mE < \muphi$) (bottom). The quoted percentages indicate the corresponding branching ratio and include the conjugate process wherever allowed. Different colours are used only for better readability and do not follow any particular colour scheme. Only dominant (solid lines) and subdominant (dashed lines) decay modes have been shown.}
\label{fig:cascades}
\end{figure}
\vspace{0.25cm}
\subsection{Decays of the $Z_2$-odd particles}
\vspace{0.25cm}
The pair-produced $Z_2$-odd exotics at the LHC each decays into a lighter 
$Z_2$-odd particle and an on-shell or off-shell\footnote{depending on the 
mass splitting between the mother and daughter exotics} SM particle. The decays proceed 
via the gauge couplings or Yukawa couplings (involving the SM Higgs or $Z_2$-odd scalars). The 
daughter $Z_2$-odd particle will decay further until the lightest $Z_2$-odd particle is 
produced. This results in a decay cascade for a given exotic particle, as shown in
Fig.~\ref{fig:cascades} (top) and Fig.~\ref{fig:cascades} (bottom), respectively,
for the two distinct mass hierarchies as had been defined in
Tables~\ref{table:input_para_1} and \ref{table:input_para_2}. The parameters $\mN$ and $d_2$
($\zLN$), however, are chosen from the feasible space in Fig.~\ref{fig:dm_main}. 
Hereafter, we use the notation ${\ND}_{\gamma}$ with $\gamma=1-4$ 
to denote the doublet eigenstates ${\ND}_{X, \kappa}$ and
${\ND}_{Y,\kappa}$ with $\kappa=1,2$.\\

\noindent {\bf {\em Scenario I} (Mass hierarchy 1 in Table~\ref{table:input_para_1}):} The decay 
cascades for this scenario are illustrated in Fig.~\ref{fig:cascades} (top). 
The Yukawa couplings and mass parameters for Mass hierarchy 1 (MH1) yield a 
mass spectrum of the $Z_2$-odd particles where the exotic doublet leptons 
are heavier than the singlets, with the charged component 
of the $Z_2$-odd doublet being the heaviest of them all. 
\begin{enumerate}
    \item While the decays of the charged exotic doublet leptons 
    (${\ED}_{1,2}$) into the neutral components of $Z_2$-odd doublets 
    (${\ND}_{X,Y}$) are 
    kinematically suppressed, the decays into $Z_2$-odd scalars ($\phi_{S,P}$) 
     are suppressed by the smallness of Yukawa coupling, $\yeL$. 
    Therefore, ${\ED}_{1,2}$ predominantly decay into singlet $Z_2$-odd 
    charged leptons (${\ES}_{1,2}$) in association with a $Z$ or Higgs boson. 
    The decays are assisted by the 
    considerable magnitude of Yukawa coupling, $\zRE$, 
    introducing a significant mixing between the charged components 
    of the $Z_2$-odd singlets and doublets.
    \item The neutral doublet $Z_2$-odd leptons (${\ND}_{1-4}$), being lighter 
    than ${\ED}_{1,2}$, cannot decay into ${\ED}_{1,2}$.
    They predominantly decay into the singlet $Z_2$-odd charged leptons 
    (${\ES}_{1,2}$) along with a $W^\pm$-boson, exhibiting an almost 100\% branching 
    ratio for this decay channel. It may be noted that this decay is possible 
    because of the large mixing ($\zRE$) between $\ED$ and $\ES$. 
    Decays into charged $Z_2$-odd scalars are suppressed by the small Yukawa coupling, 
    $\yeL$ while decays into neutral $Z_2$-odd scalars are not possible as there is no 
    coupling involving the $Z_2$-odd doublet and the SM $\nu$. Further, decays into $\NSR$ are 
    also suppressed because of the tiny $\zLN$, despite the phase space being large.
    \item The neutral ($\phi_{S,P}$) and charged ($\phi^{\pm}$) exotic scalars 
    may decay into exotic charged singlets. 
    Decays into ${\NS}_{1,2}$ are highly suppressed as the ratio $\ylN / \ylE \sim {\cal{O}}(10^{-2})$.
    \item For ${\ES}_{1,2}$, the only kinematically allowed decay modes are into 
    ${\NS}_{1}$ and an off-shell $W^{\pm}$. Apart from being a 3-body decay with 
    a relatively small phase space, this is also severely suppressed by the singlet-doublet mixing.
    \item Similarly, the exotic neutral (${\NS}_2$) decays into 
    the lightest one (${\NS}_1$) and a SM fermion-antifermion pair via an off-shell $Z$-boson. 
    The mixing suppression is even more pronounced in this case.
\end{enumerate}
\noindent {\bf {\em Scenario II} (Mass hierarchy 2 in Table~\ref{table:input_para_2}):} The decay 
cascades for this scenario are illustrated in Fig.~\ref{fig:cascades} (bottom). 
This particular scenario is characterised by a mass spectrum where the scalars 
($\phi_{S,P},~\phi^\pm$) are the heaviest among the exotics, followed by the singlet 
charged leptons (${\ES}_{1,2}$), doublet neutral leptons (${\ND}_{1-4}$), doublet 
charged leptons (${\ED}_{1,2}$), and singlet neutral leptons (${\NS}_{1,2}$), in 
that order. Given this spectrum, the relic density of dark matter 
(${\NS}_1$) is satisfied 
through co-annihilations involving the nearly degenerate ${\ED}_{1,2}$. This is in contrast 
to {\em Scenario I} where dark matter relic density is satisfied through co-annihilations 
involving ${\ES}_{1,2}$. The various decays are described below.
\begin{enumerate}
    \item The $Z_2$-odd scalars can decay into singlet and doublet exotic leptons in 
    association with an SM lepton. The neutral exotic scalars ($\phi_{S,P}$) 
    predominantly decay into ${\ED}_{2}$ in association with an electron or a muon, 
    with the decay into muons dominating due to the specific 
    structure of the Yukawa coupling $\yeL$ (refer to Tables~\ref{table:input_para_1} and \ref{table:input_para_2}), 
    as required to obtain the correct anomalous magnetic moments of the electron 
    and muon. For the same reason, the charged exotic scalar ($\phi^\pm$) 
    predominantly decays into ${\ND}_{3,4}$, accompanied by muons. 
    Decays into singlet charged 
    leptons (${\ES}_{1,2}$) are kinematically suppressed while decays into 
    singlet neutral leptons (${\NS}_{1,2}$), although suppressed by the 
    smallness of $y_{lN}$, are still present. 
    \item The singlets ${\ES}_{1,2}$ decay into the doublets
    ${\ED}_{1,2}$ and ${\ND}_{1-4}$ in association with $Z$/Higgs boson and a $W^\pm$-boson,
     respectively. The decays are made possible by virtue of the singlet-doublet 
     mixing induced by the Yukawa coupling, $\zRE$.
    \item The only kinematically allowed decay modes for the neutral doublet leptons, 
    ${\ND}_{1-4}$, are into either ${\ED}_{1,2}$ or ${\NS}_{1,2}$. The decays into 
    ${\ED}_{1,2}$ occur via off-shell $W^{\pm}$ bosons while decays into ${\NS}_{1,2}$ 
    occur via off-shell $Z/h$ bosons\footnote{In addition, 3-body decays of $\ND$ into 
    ${\NS}$ and SM charged leptons via exotic charged scalars are also possible but 
    those are highly suppressed compared to 3-body decays into ${\NS}$ and SM quarks 
    via off-shell $Z/h$ bosons.}. Both the decays depend upon the relative magnitudes 
    of the singlet-doublet components in $\ND$, $\ED$ and $\NS$ and therefore,the Yukawa couplings 
    $\zLN$ and $\zRE$.
    \item The doublet charged leptons (${\ED}_{1,2}$), being the next-to-lightest $Z_2$-odd 
    particles\footnote{The small mass-splitting (approximately a few hundred MeV) between 
    the two singlet neutral leptons (${\NS}_1$ and ${\NS}_2$) can be safely neglected for 
    practical collider analysis. Therefore, we assume ${\NS}_1$ and ${\NS}_2$ are degenerate 
    in the context of collider phenomenology. With this assumption, ${\ED}_{1,2}$ becomes 
    the next-to-lightest $Z_2$-odd particles.}, can only decay into the lightest $Z_2$-odd 
    particles (${\NS}_{1,2}$) in association with an SM fermion-antifermion pair. These 
    decays are tree-level 3-body processes that proceed through a $W^\pm$-boson in the propagator.
\end{enumerate}
\vspace{0.25cm}
\subsection{Signatures at the LHC}
\vspace{0.25cm}
The pair production of the $Z_2$-odd particles at the LHC and their 
subsequent decays into final states comprising of a pair of 
the lightest $Z_2$-odd particles via decay cascades as 
 shown in Fig.~\ref{fig:cascades}
result in signatures characterised by multiple $Z/W$ or Higgs bosons, soft 
leptons/jets, and missing transverse momentum, the last one arising mainly 
from the invisible DM. Similar final 
state topologies have already been explored by the ATLAS and CMS collaborations 
at the LHC, primarily within the framework of R-parity conserving supersymmetric 
(SUSY) scenarios featuring light electroweakinos or sleptons. In Table~\ref{table:collider_searches} (Appendix~\ref{app:LHC_signatures}), 
we have outlined the final state signatures arising 
from the dominant decays of the pair produced $Z_2$-odd particles at the LHC. 
We have also mentioned the relevant LHC 
SUSY searches that investigate similar final state topologies
and therefore, can provide constraints on the masses of the exotic particles 
in our model. In many cases, a direct application or reinterpretation of 
the LHC-derived bounds on the masses of electroweakinos or sleptons from 
SUSY scenarios may not straightforwardly translate to bounds on the masses of the 
$Z_2$-odd particles in our model. 
In such cases, these LHC searches can be reinterpreted 
within the framework of our model through signal and background simulations using a fast detector 
simulator such as Delphes. Performing such simulations to derive specific bounds on the 
parameter space of our model goes beyond the scope of this article. 

The ATLAS search \cite{ATLAS:2019lng} at the 13 TeV LHC with a 
luminosity of $139~\text{fb}^{-1}$ is the only search that can be 
reinterpreted directly for our model. It provides constraints on the allowed 
mass range for both the scenarios. The analysis focuses on electroweakino 
and slepton production in compressed mass spectrum scenarios where an additional 
jet from initial-state radiation enhances the search sensitivity. The constraints 
on electroweakino production are divided into two cases, namely, the wino-bino 
and higgsino scenarios. In the former case, the associated production 
of a chargino ($\tilde{\chi}_1^{\pm}$) and the next-to-lightest neutralino 
($\tilde{\chi}_2^0$) is considered for $m(\tilde{\chi}_1^{\pm}) = m(\tilde{\chi}_2^0)$ 
while in the latter case, the production of pairs $\tilde{\chi}_1^{\pm} \tilde{\chi}_2^0$, $\tilde{\chi}_1^{+} \tilde{\chi}_1^{-}$, and $\tilde{\chi}_2^0 \tilde{\chi}_1^0$ is considered for 
$m(\tilde{\chi}_1^{\pm}) = \frac{1}{2} \left(m(\tilde{\chi}_1^0) + m(\tilde{\chi}_2^0)\right)$. 
Based on the topologies described in Table~\ref{table:collider_searches}, 
 different production channels from \cite{ATLAS:2019lng} are relevant 
to {\em Scenario I} and {\em Scenario II}.

We first discuss the relevance of Ref.~\cite{ATLAS:2019lng} for {\em Scenario II}, 
with $m_{\ND} \sim m_{\ED}$ making the wino-bino scenario 
an appropriate search for deriving the constraint. The associated production 
of a chargino ($\tilde{\chi}_1^{\pm}$) 
and a neutralino ($\tilde{\chi}_2^0$) followed by their 3-body decays into the lightest 
supersymmetric particle (LSP) and a pair of SM fermions via off-shell $W/Z$ bosons is 
quite similar to the associated production and the subsequent 3-body decays of 
${\ED}$ and ${\ND}$ in our case (see Fig. 3 in Table~\ref{table:collider_searches}, 
{\em Scenario II}). By comparing $\sigma(pp \to \ED \ND)$ with the upper limits on 
the production cross-section for winos from \cite{ATLAS:2019lng}, we find that ${\ED}$ 
and ${\ND}$ masses above $160~\text{GeV}$ are allowed\footnote{This supersedes the lower 
bound from LEP results~\cite{L3:2001xsz} which are applicable only to heavy charged 
leptons in {\em Scenario II} of our model.} in {\em Scenario II}.

For the {\em Scenario I,} the final state signature from the 
pair production of $\bar{\ES} \ES$ (see Fig. 2 in Table~\ref{table:collider_searches}, 
{\em Scenario I}) is analogous to the final state topology resulting from slepton 
pair production in the SUSY scenario. However, a direct quantitative 
comparison between the slepton production and the production of $\bar{\ES} \ES$ 
is challenging because sleptons in \cite{ATLAS:2019lng} 
undergo 2-body decays into the LSP ($\tilde{\chi}^0$) and charged leptons ($l$) whereas ${\ES}$ in our 
model decays via a 3-body process into DM (${\NS_1}$), leptons $l$ and neutrinos $\nu_l$. 
While both scenarios result in the same final state, {\em viz.,} 
dilepton + \text{MET}, the kinematics of the final-state leptons differ, making the 
upper bounds on slepton production cross-sections 
derived in \cite{ATLAS:2019lng} with specific kinematic cuts on the final-state leptons 
inapplicable to the pair-production cross-section $\sigma(\bar{\ES} \ES)$ in our model. 
Nevertheless, comparing $\sigma(\bar{\ES} \ES) \times \text{BR}({\ES} \to \text{DM}~ \nu_l~ l)^2$ 
with the upper limits on slepton production for mass splittings up to $5~\text{GeV}$ 
provides a qualitative bound, excluding $m_{\ES}$ up to $100~\text{GeV}$ in our model.
\section{Summary and conclusion}
\label{sec:conclusion}
\vspace{0.25cm}
The BSM scenario in this work is motivated from the observations of neutrino masses, anomalous magnetic moments of electron and muon, and dark matter in the Universe. We explore the potential of vector-like fermions together with an inert scalar doublet in explaining these observations and find that there exists a region of parameter space that satisfies these concerns. We also explain the null observations of dark matter direct detection and lepton flavor violation.

We extend the SM by including two generations of a family of vector-like fermions where the left- and right-handed fields are charged similarly under the gauge symmetry group of the SM. Introduction of a $Z_2$ symmetry ensures the stability of the dark matter candidate while the vector-like nature of leptons renders them a bare mass term in the lagrangian, thus, directly giving them mass at the TeV scale. The neutrino mass is generated at the one-loop level with exotic neutral singlet fermions and 
neutral scalars in the loop where the TeV scale masses of the exotic fermions and a small mass-splitting between the exotic neutral scalar and pseudoscalar together ensure the smallness of the neutrino masses. By virtue of the small mass-splitting $|m_{\phi_S}-m_{\phi_P}|$, it is possible to generate neutrino masses of ${\cal O}\lfb 0.1-0.01\rfb$~eV without making the relevant Yukawa coupling ($\ylN$) exceptionally small. 
For addressing the anomalous magnetic moment of leptons, contributions come from 1-loop diagrams involving $Z_2$-odd neutral fermions ($\NS$, $\ND$) and charged scalars ($\phi^{\pm}$), or $Z_2$-odd charged fermions ($\ES$, $\ED$) and neutral scalars ($\phi_S,\phi_P$). The dominant contributions come from the diagrams with chirality flipping of the fermion in the loop.
The diagram with neutral fermions in the loop is dominant in the region $\lambda_3 \sim {\cal O}\lfb 10^{-8} \rfb$. It can address the anomalies in magnetic moments of electron as well as muon but does not guarantee a simultaneous suppression of $Br(\tau \to e \gamma)$ and $Br(\tau \to \mu \gamma)$. On the other hand, the diagram with charged fermions in the loop becomes dominant for $\lambda_3\sim{\cal O}\lfb 1\rfb$. In this case, it is possible to address the anomalous magnetic magnetic moments of electron as well as muon while also explaining the null observations of cLFV.

We derive analytical expressions for the masses and mixings of the vector-like fermions which are approximated for a {\em simplified scenario}, as described in the text. 
We also evaluate numerical values of various low-energy observables for $10^4$ randomly generated points in the free parameter space of the model and the results are found to be consistent with our calculations. 
In the BSM particle spectrum that we consider, there can be two DM candidates. However, 
the scalar DM does not satisfy relic density observations up to mass $500$ GeV, atleast.
For the case of $Z_2$-odd fermion as a DM candidate, we study two different mass-hierachies. In both the scenarios, the DM satisfies the relic density observations and is a singlet-doublet admixture whose relic abundance is determined either via sizeable self-annihilation cross-sections - requiring a large $d_2$ ($\zLN$), or through coannihilation processes involving heavier $Z_2$-odd particles.
The available parameter space gets further constrained by the bounds on DM-nucleon interaction cross-section from direct detection experiments as well as the cosmological constraint on the decay width of the next-to-lightest $Z_2$-odd particle.

Finally, we comment on the possible collider signatures of the exotic fermions. 
The $Z_2$-odd neutral and charged leptons in the model have collider signatures very similar to SUSY with compressed mass spectra. The possible final state signatures and relevant LHC searches that could constrain the parameter space of this model are presented in Table~\ref{table:collider_searches}. The results of ATLAS search \cite{ATLAS:2019lng} have been used to put
a lower bound of $160$ GeV on the masses of $\ED$ and $\ND$, in scenario II. In scenario I, a qualitative comparison with the sleptons' pair production suggests that only mass $m({\ES})$ upto $100 ~GeV$ can be excluded. An exact reinterpretation of the remaining applicable searches requires signal and background simulations using a fast detector simulator such as Delphes and will be presented in another work.

\section*{Acknowledgments}
V.S. is thankful to Kirtiman Ghosh and Debajyoti Choudhury for their invaluable contributions in the formulation of this manuscript.
V.S. acknowledges the support from research grant no. CRG/2018/004889 of the SERB, India; T.R.Seshadri for providing resources during initial stages of the work and Brajesh Choudhary for providing access to computers bought under the aegis of the Grant No. SR-MF/PS-0212014-DUB (G) of the DST (India).
\vspace{0.25cm}
\appendix

\section{$\beta$-functions upto 2-loop}
\label{app:beta}
\bea
\beta_1(g_1, g_2, g_3, y_t, {\lambda}) ~=~\frac{g_1^3}{(4 \pi)^2} \bigg(b_1^{SM}+b_1^{new} \bigg) ~+~ \frac{g_1^3}{(4 \pi)^4} \bigg(\Sigma_i b_{1i}^{SM} g_i^2 + \Sigma_i b_{1i}^{new} g_i^2 - \frac{17}{10} y_t^2 \bigg) \nn 
\eea
\bea
\beta_2(g_1, g_2, g_3, y_t, {\lambda}) ~=~\frac{g_2^3}{(4 \pi)^2} \bigg(b_2^{SM}+b_2^{new} \bigg) ~+~ \frac{g_2^3}{(4 \pi)^4} \bigg(\Sigma_i b_{2i}^{SM} g_i^2 + \Sigma_i b_{2i}^{new} g_i^2 - \frac{3}{2} y_t^2 \bigg) \nn 
\eea
\bea
\beta_3(g_1, g_2, g_3, y_t, {\lambda}) ~=~\frac{g_3^3}{(4 \pi)^2} \bigg(b_3^{SM}+b_3^{new} \bigg) ~+~ \frac{g_3^3}{(4 \pi)^4} \bigg(\Sigma_i b_{3i}^{SM} g_i^2 + \Sigma_i b_{3i}^{new} g_i^2 - 2 y_t^2 \bigg) \nn 
\eea
\bea
\beta_4(g_1, g_2, g_3, y_t, {\lambda}) ~&=&~\frac{y_t}{(4 \pi)^2} \bigg( b_{4,SM}^{(1)} \bigg) \nn \\
&&+ \frac{y_t}{(4 \pi)^4} \bigg( b_{4,SM}^{(2)} + b_{4,\phi}^{(2)} + b_{4,L^D}^{(2)} + b_{4,E^S}^{(2)} + b_{4,Q^D}^{(2)} + b_{4,D^S}^{(2)} + b_{4,U^S}^{(2)} \bigg) \nn 
\eea
\bea
\beta_5(g_1, g_2, g_3, y_t, {\lambda}) ~&=&~\frac{1}{(4 \pi)^2} \bigg( b_{5,SM}^{(1)} \bigg) \nn \\
&&+ \frac{1}{(4 \pi)^4} \bigg( b_{5,SM}^{(2)} + b_{5,\phi}^{(2)} + b_{5,L^D}^{(2)} + b_{5,E^S}^{(2)} + b_{5,Q^D}^{(2)} + b_{5,D^S}^{(2)} + b_{5,U^S}^{(2)} \bigg) \nn \\
\eea
where the first and second terms in each equation denote the contributions at 1-loop and 2-loops, respectively. For $\beta_1$, $\beta_2$ and $\beta_3$, we have simplified the expressions further and $i=1,2,3$ therein. The individual SM and BSM contributions to these $\beta$-functions are listed below.

\underline{\bf SM}
\bea
&&b_i^{SM}=\bigg\{ \frac{41}{10}, \frac{-19}{6}, -7 \bigg\}, ~~~~b_{ij}^{SM}={\begin{pmatrix} \frac{199}{50} & \frac{27}{10} & \frac{44}{5} \\ \frac{9}{10} & \frac{35}{6} & 12 \\ \frac{11}{10} & \frac{9}{2} & -26 \end{pmatrix}} ~~~\textnormal{where} ~~i,j=1,2,3 \nn 
\eea
\bea
&&b_{4,SM}^{(1)}=\frac{-17}{20} g_1^2 - \frac{9}{4} g_2^2 - 8 g_3^2 + \frac{9}{2} y_t^2, \nn \\
&&b_{4,SM}^{(2)}=\frac{1187}{600} g_1^4 - \frac{23}{4} g_2^4 - 108 g_3^4 - \frac{9}{20} g_1^2 g_2^2 + \frac{19}{15} g_1^2 g_3^2 + 9 g_2^2 g_3^2 + \frac{3}{2} \lambda^2 \nn \\
&&~~~~~~~~~~+ y_t^2 \bigg( \frac{393}{80} g_1^2 + \frac{225}{16} g_2^2 + 36 g_3^2 - 6 ~\lambda - 12 y_t^2 \bigg), \nn \\
&&~~~~ \nn \\
&&~~~~ \nn \\
&&b_{5,SM}^{(1)}=\frac{27}{100} g_1^4 + \frac{9}{4} g_2^4 + \frac{9}{10} g_1^2 g_2^2 - \frac{9}{5} g_1^2 \lambda - 9 g_2^2 \lambda + 12 \lambda^2 + 12 \lambda y_t^2 - 12 y_t^4, \nn \\
&&b_{5,SM}^{(2)}=\frac{-3411}{1000} g_1^6 - \frac{1677}{200} g_1^4 g_2^2 - \frac{289}{40} g_1^2 g_2^4 + \frac{305}{8} g_2^6 + \lambda \bigg( \frac{1887}{200} g_1^4 + \frac{117}{20} g_1^2 g_2^2 - \frac{73}{8} g_2^4 \bigg) \nn \\
&&~~~~~~~~~~+ \lambda^2 \Big( \frac{54}{5} g_1^2 + 54 g_2^2 \Big) - 78 \lambda^3 + y_t^2 \bigg( \frac{-171}{50} g_1^4 + \frac{63}{5}g_1^2 g_2^2 - \frac{9}{2} g_2^4 + \frac{17}{2} g_1^2 \lambda \nn \\
&&~~~~~~~~~~+ \frac{45}{2} g_2^2 \lambda + 80 g_3^2 \lambda - 72 \lambda^2 \bigg) + y_t^4 \bigg( -\frac{16}{5} g_1^2 - 64 g_3^2 - 3 ~\lambda \bigg) + 60 y_t^6. 
\eea
\underline{\bf BSM contribution}
Denoting the number of generations of particle $X$ by $n_X$ and considering $n_{L^D}=n(\LDL)=n(\LDR)$, $n_{E^S}=n(\ESL)=n(\ESR)$, $n_{Q^D}=n(\QDL)=n(\QDR)$, $n_{D^S}=n(\DSL)=n(\DSR)$ and $n_{U^S}=n(\USL)=n(\USR)$,

\underline{At 1-loop}
\bea
&&b_1^{new}=\frac{1}{10} ~n_{\phi}+\frac{2}{5} ~n_{L^D}+\frac{4}{5} ~n_{E^S}+\frac{2}{15} ~n_{Q^D}+\frac{4}{15} ~n_{D^S}+\frac{16}{15} ~n_{U^S}, \nn \\
&&b_2^{new}=\frac{1}{6} ~n_{\phi}+\frac{2}{3} ~n_{L^D}+2 ~n_{Q^D}, \nn \\
&&b_3^{new}=\frac{4}{3} ~n_{Q^D}+\frac{2}{3} ~n_{D^S}+\frac{2}{3} ~n_{U^S}.
\eea

\underline{At 2-loop}
\bea
&&b_{11}^{new}=\frac{18}{100} ~n_{\phi}+\frac{18}{100} ~n_{L^D}+\frac{36}{25} ~n_{E^S}+\frac{2}{300} ~n_{Q^D}+\frac{4}{75} ~n_{D^S}+\frac{64}{75} ~n_{U^S}, \nn \\
&&b_{12}^{new}=\frac{9}{10} ~n_{\phi}+\frac{18}{20} ~n_{L^D}+\frac{6}{20} ~n_{Q^D}, \nn \\
&&b_{13}^{new}=\frac{8}{15} ~n_{Q^D}+\frac{16}{15} ~n_{D^S}+\frac{64}{15} ~n_{U^S}, \nn \\
&&~~~~ \nn \\
&&b_{21}^{new}=\frac{3}{10} ~n_{\phi}+\frac{3}{10} ~n_{L^D}+\frac{1}{10} ~n_{Q^D},, \nn \\
&&b_{22}^{new}=\frac{13}{6} ~n_{\phi}+\frac{49}{6} ~n_{L^D}+\frac{49}{2} ~n_{Q^D}, \nn \\
&&b_{23}^{new}=8~n_{Q^D}, \nn \\
&&~~~~ \nn \\
&&b_{31}^{new}=\frac{1}{15} ~n_{Q^D}+\frac{2}{15} ~n_{D^S}+\frac{8}{15} ~n_{U^S}, \nn \\
&&b_{32}^{new}=3 ~n_{Q^D}, \nn \\
&&b_{33}^{new}=\frac{76}{3} ~n_{Q^D}+\frac{38}{3} ~n_{D^S}+\frac{38}{3} ~n_{U^S}, \nn \\
&&~~~~ \nn \\
&&b_{4,\phi}^{(2)}= \Big(\frac{2}{15} g_1^4 + \frac{1}{2} g_2^4 \Big) ~n_{\phi}, \nn \\
&&b_{4,L^D}^{(2)}= \Big(\frac{29}{75} g_1^4 + \frac{1}{2} g_2^4 \Big) ~n_{L^D}, \nn \\
&&b_{4,E^S}^{(2)}= \frac{58}{75} g_1^4 ~n_{E^S}, \nn \\
&&b_{4,Q^D}^{(2)}= \Big(\frac{54}{225} g_1^4 + \frac{3}{2} g_2^4 + \frac{80}{9} g_3^4 \Big) ~n_{Q^D}, \nn \\
&&b_{4,D^S}^{(2)}= \Big(\frac{58}{225} g_1^4 + \frac{40}{9} g_3^4 \Big) ~n_{D^S}, \nn \\
&&b_{4,U^S}^{(2)}= \Big(\frac{232}{225} g_1^4 + \frac{40}{9} g_3^4 \Big) ~n_{U^S}, \nn 
\eea
\bea
&&~~~~ \nn \\
&&b_{5,\phi}^{(2)}=\bigg( -\frac{63}{500} g_1^6 - \frac{21}{100} g_1^4 g_2^2 - \frac{7}{20} g_1^2 g_2^4 - \frac{7}{4} g_2^6 + \frac{33}{100} g_1^4 \lambda + \frac{11}{4} g_2^4 \lambda \bigg) ~n_{\phi}, \nn \\
&&b_{5,L^D}^{(2)}=\bigg( -\frac{18}{125} g_1^6 - \frac{6}{25} g_1^4 g_2^2 - \frac{2}{5} g_1^2 g_2^4 - 2 g_2^6 + \frac{3}{10} g_1^4 \lambda + \frac{5}{2} g_2^4 \lambda \bigg) ~n_{L^D}*2, \nn \\
&&b_{5,E^S}^{(2)}=\bigg( \frac{-36}{125} g_1^6 - \frac{12}{25} g_1^4 g_2^2 + \frac{3}{5} g_1^4 \lambda \bigg) ~n_{E^S}*2, \nn \\
&&b_{5,Q^D}^{(2)}=\bigg( -\frac{6}{125} g_1^6 - \frac{2}{25} g_1^4 g_2^2 - \frac{6}{5} g_1^2 g_2^4 - 6 g_2^6 + \frac{1}{10} g_1^4 \lambda + \frac{15}{2} g_2^4 \lambda \bigg) ~n_{Q^D}*2, \nn \\
&&b_{5,D^S}^{(2)}=-\frac{1}{125} g_1^4 \bigg( 12 g_1^2 + 20 g_2^2 - 25 \lambda \bigg) ~n_{D^S}*2, \nn \\
&&b_{5,U^S}^{(2)}=\bigg( -\frac{48}{125} g_1^6 - \frac{16}{25} g_1^4 g_2^2 + \frac{4}{5} g_1^4 \lambda \bigg) ~n_{U^S}*2.
\eea
\section{Estimating the scale of exotic quarks}
\label{app:rge_1loop}
Parametrizing $\beta_i^{(1)}(g_i)=\frac{g_i^3}{16 \pi^2} b_i^{(1)}(g_i)$, the RGEs upto 1-loop can be expressed as
\beq
\frac{dg_i}{d \ln Q}=\frac{g_i^3}{16 \pi^2} ~b_i^{(1)}(g_i)
\label{eq:rge_1loop}
\eeq
where $i=1,2,3$. Integrating, we get
\beq
\frac{-1}{2 g_i^2}=\frac{b_i^{(1)}(g_i)}{16 \pi^2} \ln Q + C.
\eeq
For the definite solution, we need an initial condition. Introducing $Z_2$-odd leptons and scalars at scale $Q_1$, the $Z_2$-odd quarks can be introduced stepwise say, $U^S_{(L,R)}$ at $Q_2$, $D^S_{(L,R)}$ at $Q_3$ and $Q^D_{(L,R)}$ at $Q_4$. In the region upto the scale $Q_1$, only SM particles contribute to $\beta$-functions. Thus, we use the initial condition: for $Q=m_Z$, $g_i=g_i(m_Z)$. This gives the solution,
\beq
\alpha_i^{-1} (t) = \alpha_i^{-1} (t_0) - \frac{b_i^{(1)}(g_i)}{2 \pi} (t-t_0),
\eeq
where $t=\ln (Q/GeV)$, $t_0=\ln (m_Z/GeV)$ and $\alpha_i=g_i^2 / (4 \pi)$.

We repeat the procedure for each region $(t_{n-1},t_n)$ to obtain the solution $\alpha_{i,n}^{-1} (t)$, using the boundary condition $\alpha_{i,n}^{-1} (t_{n-1})=\alpha_{i,n-1}^{-1}(t_{n-1})$. This gives,
\bea
t \in (t_0,t_1)&:& \alpha_{i,1}^{-1} (t) = \alpha_i^{-1} (t_0) - \frac{b_i^{(1)}(g_i)}{2 \pi} (t-t_0)    ~~~~~~~~ \nn \\
t \in (t_1,t_2)&:& \alpha_{i,2}^{-1} (t) = \alpha_i^{-1} (t_1) - \frac{b_i^{(1)}(g_i)}{2 \pi} (t-t_1)    ~~~~~~~~ \nn \\
t \in (t_2,t_3)&:& \alpha_{i,3}^{-1} (t) = \alpha_i^{-1} (t_2) - \frac{b_i^{(1)}(g_i)}{2 \pi} (t-t_2)    ~~~~~~~~ \nn \\
t \in (t_3,t_4)&:& \alpha_{i,4}^{-1} (t) = \alpha_i^{-1} (t_3) - \frac{b_i^{(1)}(g_i)}{2 \pi} (t-t_3)    ~~~~~~~~ \nn \\
t > t_4        &:& \alpha_{i,5}^{-1} (t) = \alpha_i^{-1} (t_4) - \frac{b_i^{(1)}(g_i)}{2 \pi} (t-t_4)    ~~~~~~~~ 
\eea
Let the three couplings $g_i$ with $i=1,2,3$ converge at $t=t_G$ (say) {\em i.e.,}
\beq
\alpha_{1}^{-1} (t_G) = \alpha_{2}^{-1} (t_G) = \alpha_{3}^{-1} (t_G).
\label{eqn:unif2}
\eeq
Eq.~\ref{eqn:unif2} implies two independent equations. Thus, fixing $t_1, t_2$ and $t_3$, energy scales $t_4$ and $t_G$ can be determined by solving the two equations.\\
\section{Diagonality of $\yeL\yeL^{\dagger}$}
\label{app:yeLyeLd}
Denoting $\yeL$ and $\ylE$ as,
\bea
\yeL ={\begin{pmatrix} \yeL^{2\times 2} \\ 0 \end{pmatrix}};~~~~~~~
\ylE ={\begin{pmatrix} \ylE^{2\times 2} \\ 0 \end{pmatrix}}, 
\eea
it is easy to show that
\bea
\yeL \yeL^{\dagger}={\begin{pmatrix} \yeL^{2\times 2} ~{\yeL^{2\times 2}}^{\dagger} & 0 \\
                                                            0                       & 0 \end{pmatrix}};~~~~~~
\ylE \ylE^{\dagger}={\begin{pmatrix} \ylE^{2\times 2} ~{\ylE^{2\times 2}}^{\dagger} & 0 \\
                                                            0                       & 0 \end{pmatrix}}. 
\eea
The diagonality of $\ylE \ylE^{\dagger}$ also implies diagonality of $\ylE^{2\times 2} ~~{\ylE^{2\times 2}}^{\dagger}$. From Eq.~\ref{eqn:CLR_2x2_simp},
\bea
\yeL^{2\times 2}&=&{\cal Y}^{-1}~{\rm diag}\lfb -\frac{\Delta a_{e}}{2m_e},~-\frac{\Delta a_{\mu}}{2m_\mu} \rfb \times \lfb{\ylE^{2\times 2}}^{\dagger}\rfb^{-1}\times \zRE^{-1}, \nn \\
\implies \yeL^{2\times 2} \lfb{\yeL^{2\times 2}}\rfb^{\dagger} &=& {\cal Y}^{-2}{\rm diag}\lfb -\frac{\Delta a_{e}}{2m_e},~-\frac{\Delta a_{\mu}}{2m_\mu} \rfb \times \lfb{\ylE^{2\times 2}}^{\dagger}\rfb^{-1}\times \zRE^{-1} \nn \\
&&.\lfb \zRE^{-1}\rfb^{\dagger} \times \lfb\lfb{\ylE^{2\times 2}}^{\dagger}\rfb^{-1}\rfb^{\dagger} \times \lfb{\rm diag}\lfb-\frac{\Delta a_{e}}{2m_e},~-\frac{\Delta a_{\mu}}{2m_\mu}\rfb\rfb^{\dagger} \nn \\
&=& {\cal Y}^{-2}{\rm diag}\lfb -\frac{\Delta a_{e}}{2m_e},~-\frac{\Delta a_{\mu}}{2m_\mu} \rfb \times \ylE^{2\times 2} ~~\lfb{\ylE^{2\times 2}}\rfb^{\dagger} \times {\rm diag}\lfb-\frac{\Delta a_{e}}{2m_e},~-\frac{\Delta a_{\mu}}{2m_\mu}\rfb, \nn \\ 
\eea
where we have used the unitarity of $\zRE$. Thus, being a product of diagonal matrices, $\yeL^{2\times 2} \lfb{\yeL^{2\times 2}}\rfb^{\dagger}$ is ensured to be diagonal.

\section{Final state signal topologies for LHC searches}
\label{app:LHC_signatures}
\begin{table}[h]
    \centering
    \caption{Final state signal topologies for LHC searches}
    \label{table:collider_searches}
    \scalebox{0.9}{
    \begin{tabular}{|m{16 cm}|} 
        \hline \hline \begin{center}
            {\bf {\em Scenario I}}
        \end{center}\\
        \hline \hline
        {\bf 1.} \begin{center}
            \begin{tikzpicture}[line width=1.4 pt, scale=1,every node/.style={scale=1.0}]

\draw[fermion,black,very thick] (-4.5,1) --(-3.5,0);
\draw[fermion,black,very thick] (-4.5,-1) --(-3.5,0);
\filldraw[black] (-3.5,0) circle (2pt);


\draw[fermion,black,thin] (-3.5,0) --(-2.25,0.9);
\draw[fermion,black,thin] (-2.25,0.9) --(-1,1.4);
\draw[vector,red,thin] (-2.25,0.9) --(-1,0.3);
\draw[fermion,black,thin] (-1,1.4) --(0,1.6);
\draw[vector,red,thin] (-1,1.4) --(-0.25,1);
\draw[fermionnoarrow,black,thin] (-0.25,1) --(0.25,0.75);
\draw[fermionnoarrow,black,thin] (-0.25,1) --(0.25,1.25);

\draw[fermion,black,thin] (-3.5,0) --(-2.25,-0.9);
\draw[fermion,black,thin] (-2.25,-0.9) --(-1,-1.4);
\draw[vector,red,thin] (-2.25,-0.9) --(-1,-0.3);
\draw[fermion,black,thin] (-1,-1.4) --(0,-1.6);
\draw[vector,red,thin] (-1,-1.4) --(-0.25,-1);
\draw[fermionnoarrow,black,thin] (-0.25,-1) --(0.25,-0.75);
\draw[fermionnoarrow,black,thin] (-0.25,-1) --(0.25,-1.25);

\node at (-3.25,-0.75) {${\tilde{N}}_{X,Y}^{D}$};
\node at (-3.25,0.75) {$\tilde{N}_{X,Y}^{D}$};

\node at (-1.75,-1.5) {$\bar{\tilde{E}}_{1,2}^{S}$};
\node at (-1.75,1.5) {$\tilde{E}_{1,2}^{S}$};

\node at (0.3,1.8) {$\tilde{N}^S_1$};
\node at (0.3,-1.8) {$\tilde{N}^S_1$};

\node[scale=1] at (-0.6,0.3) {$W^\pm$};
\node[scale=1] at (-0.6,-0.3) {$W^\pm$};

\node[scale=0.6] at (-0.75,1) {$W^*$};
\node[scale=0.6] at (-0.75,-1) {$W^*$};

\node at (-4.75,1.25) {$p$};
\node at (-4.75,-1.25) {$p$};


\end{tikzpicture}
        \end{center} \\
        The Feynman diagram depicting the pair-production and subsequent decay cascades of the doublet neutral leptons ($\tilde{N}_{1,2}^D$) resulting in $WW$ final states at the LHC, accompanied by soft leptons/jets and missing transverse energy ($E_T\!\!\!\!\!\!/$~). Similar final states have been studied in Ref.~\cite{ATLAS:2021yqv,ATLAS:2022hbt} within the context of chargino pair production followed by the decay of the chargino into the lightest SUSY particle in association with a $W^\pm$ resulting in a $WW + E_T\!\!\!\!\!\!/$ ~final state.\\ \hline 
	\end{tabular}}
\end{table}       
\begin{table}[h]
    \centering
    \scalebox{0.9}{
   \begin{tabular}{|m{16 cm}|} 
        \hline \hline 
        {\bf 2.} \begin{center}
            \begin{tikzpicture}[line width=1.4 pt, scale=1,every node/.style={scale=1.0}]

\draw[fermion,black,very thick] (-4.5,1) --(-3.5,0);
\draw[fermion,black,very thick] (-4.5,-1) --(-3.5,0);
\filldraw[black] (-3.5,0) circle (2pt);


\draw[fermion,black,thin] (-3.5,0) --(-2.25,0.9);
\draw[fermion,black,thin] (-2.25,0.9) --(-1,1.4);
\draw[vector,red,thin] (-2.25,0.9) --(-1,0.3);
\draw[fermion,black,thin] (-1,1.4) --(0,1.6);
\draw[vector,red,thin] (-1,1.4) --(-0.25,1);
\draw[fermionnoarrow,black,thin] (-0.25,1) --(0.25,0.75);
\draw[fermionnoarrow,black,thin] (-0.25,1) --(0.25,1.25);

\draw[fermion,black,thin] (-3.5,0) --(-2.25,-0.9);
\draw[fermion,black,thin] (-2.25,-0.9) --(-1,-1.4);
\draw[vector,red,thin] (-2.25,-0.9) --(-1,-0.3);
\draw[fermion,black,thin] (-1,-1.4) --(0,-1.6);
\draw[vector,red,thin] (-1,-1.4) --(-0.25,-1);
\draw[fermionnoarrow,black,thin] (-0.25,-1) --(0.25,-0.75);
\draw[fermionnoarrow,black,thin] (-0.25,-1) --(0.25,-1.25);

\node at (-3.25,-0.75) {$\bar{\tilde{E}}_{1,2}^{D}$};
\node at (-3.25,0.75) {$\tilde{E}_{1,2}^{D}$};

\node at (-1.75,-1.5) {$\bar{\tilde{E}}_{1,2}^{S}$};
\node at (-1.75,1.5) {$\tilde{E}_{1,2}^{S}$};

\node at (0.3,1.8) {$\tilde{N}^S_1$};
\node at (0.3,-1.8) {$\tilde{N}^S_1$};

\node[scale=1] at (-0.6,0.3) {$Z/h$};
\node[scale=1] at (-0.6,-0.3) {$Z/h$};

\node[scale=0.6] at (-0.75,1) {$W^*$};
\node[scale=0.6] at (-0.75,-1) {$W^*$};

\node at (-4.75,1.25) {$p$};
\node at (-4.75,-1.25) {$p$};


\end{tikzpicture}
        \end{center} \\
        The Feynman diagram depicting the pair-production and subsequent decays of the doublet charged leptons (${\ED}_{1,2}$). The decay cascades result in $ZZ/Zh/hh$ final states at the LHC, accompanied by soft leptons/jets and missing transverse energy ($E_T\!\!\!\!\!\!/$~). In the majority of events, the soft leptons/jets arising from the 3-body decays of the final exotic charged leptons (${\ES}_{1,2}$) may not be detected within the coverage of the detector. Similar final states have been studied in Ref.~\cite{ATLAS:2022zwa,ATLAS:2021yqv} within the context of neutralino pair production followed by the decay of the neutralinos into the lightest SUSY particle in association with a $Z$ or Higgs boson, resulting in a $ZZ/Zh/hh + E_T\!\!\!\!\!\!/$ ~final state.  \\\hline \hline
        {\bf 3.} \begin{center}
            \begin{tikzpicture}[line width=1.4 pt, scale=1,every node/.style={scale=1.0}]

\draw[fermion,black,very thick] (-4.5,1) --(-3.5,0);
\draw[fermion,black,very thick] (-4.5,-1) --(-3.5,0);
\filldraw[black] (-3.5,0) circle (2pt);


\draw[fermion,black,thin] (-3.5,0) --(-2.25,0.9);
\draw[fermion,black,thin] (-2.25,0.9) --(-1,1.4);
\draw[vector,red,thin] (-2.25,0.9) --(-1,0.3);
\draw[fermion,black,thin] (-1,1.4) --(0,1.6);
\draw[vector,red,thin] (-1,1.4) --(-0.25,1);
\draw[fermionnoarrow,black,thin] (-0.25,1) --(0.25,0.75);
\draw[fermionnoarrow,black,thin] (-0.25,1) --(0.25,1.25);

\draw[fermion,black,thin] (-3.5,0) --(-2.25,-0.9);
\draw[fermion,black,thin] (-2.25,-0.9) --(-1,-1.4);
\draw[vector,red,thin] (-2.25,-0.9) --(-1,-0.3);
\draw[fermion,black,thin] (-1,-1.4) --(0,-1.6);
\draw[vector,red,thin] (-1,-1.4) --(-0.25,-1);
\draw[fermionnoarrow,black,thin] (-0.25,-1) --(0.25,-0.75);
\draw[fermionnoarrow,black,thin] (-0.25,-1) --(0.25,-1.25);

\node at (-3.25,-0.75) {$\bar{\tilde{E}}_{1,2}^{D}$};
\node at (-3.25,0.75) {$\tilde{N}_{X,Y}^{D}$};

\node at (-1.75,-1.5) {$\bar{\tilde{E}}_{1,2}^{S}$};
\node at (-1.75,1.5) {$\tilde{E}_{1,2}^{S}$};

\node at (0.3,1.8) {$\tilde{N}^S_1$};
\node at (0.3,-1.8) {$\tilde{N}^S_1$};

\node[scale=1] at (-0.6,0.3) {$W^\pm$};
\node[scale=1] at (-0.6,-0.3) {$Z/h$};

\node[scale=0.6] at (-0.75,1) {$W^*$};
\node[scale=0.6] at (-0.75,-1) {$W^*$};

\node at (-4.75,1.25) {$p$};
\node at (-4.75,-1.25) {$p$};


\end{tikzpicture}
        \end{center} \\
       The Feynman diagram depicting the associated production and subsequent decay cascades of the doublet charged leptons (${\ED}_{1,2}$) and doublet neutral lepton (${\ND}_{X,Y}$). The final state comprises of $W^\pm Z/W^\pm h$ accompanied by soft leptons/jets and missing transverse energy ($E_T\!\!\!\!\!\!/$~) from the 3-body decays of the final exotic charged leptons (${\ES}_{1,2}$). Here too, in most of the events, the soft leptons/jets remain undetected at the LHC detectors. Similar final states have been studied in Ref.~\cite{ATLAS:2022zwa,ATLAS:2021yqv,ATLAS:2019wgx,ATLAS:2020qlk,ATLAS:2021moa} within the context of chargino-neutralino associated production followed by the decay of the neutralino (chargino) into the lightest SUSY particle in association with a $Z$ or Higgs boson ($W^\pm$-boson) resulting in a $W^\pm Z/W^\pm h~+~E_T\!\!\!\!\!\!/$ ~final state. \\\hline\hline
     \end{tabular}}
\end{table}       
\begin{table}[h]
    \centering
    \scalebox{0.9}{
    \begin{tabular}{|m{16 cm}|} 
        \hline \hline
        {\bf 4.} \begin{center}
            \begin{tikzpicture}[line width=1.4 pt, scale=1,every node/.style={scale=1.0}]

\draw[fermion,black,very thick] (-4.5,1) --(-3.5,0);
\draw[fermion,black,very thick] (-4.5,-1) --(-3.5,0);
\filldraw[black] (-3.5,0) circle (2pt);


\draw[scalar,blue,thin] (-3.5,0) --(-2.25,0.9);
\draw[fermion,black,thin] (-2.25,0.9) --(-1,1.4);
\draw[fermion,black,thin] (-2.25,0.9) --(-1,0.3);
\draw[fermion,black,thin] (-1,1.4) --(0,1.6);
\draw[vector,red,thin] (-1,1.4) --(-0.25,1);
\draw[fermionnoarrow,black,thin] (-0.25,1) --(0.25,0.75);
\draw[fermionnoarrow,black,thin] (-0.25,1) --(0.25,1.25);

\draw[scalar,blue,thin] (-3.5,0) --(-2.25,-0.9);
\draw[fermion,black,thin] (-2.25,-0.9) --(-1,-1.4);
\draw[fermion,black,thin] (-2.25,-0.9) --(-1,-0.3);
\draw[fermion,black,thin] (-1,-1.4) --(0,-1.6);
\draw[vector,red,thin] (-1,-1.4) --(-0.25,-1);
\draw[fermionnoarrow,black,thin] (-0.25,-1) --(0.25,-0.75);
\draw[fermionnoarrow,black,thin] (-0.25,-1) --(0.25,-1.25);

\node at (-3.25,-0.75) {$\phi_{S,P}$};
\node at (-3.25,0.75) {$\phi_{S,P}$};

\node at (-1.75,-1.5) {$\bar{\tilde{E}}_{1,2}^{S}$};
\node at (-1.75,1.5) {$\tilde{E}_{1,2}^{S}$};

\node at (0.3,1.8) {$\tilde{N}^S_1$};
\node at (0.3,-1.8) {$\tilde{N}^S_1$};

\node[scale=1] at (-0.6,0.3) {$e/\mu$};
\node[scale=1] at (-0.6,-0.3) {$e/\mu$};

\node[scale=0.6] at (-0.75,1) {$W^*$};
\node[scale=0.6] at (-0.75,-1) {$W^*$};

\node at (-4.75,1.25) {$p$};
\node at (-4.75,-1.25) {$p$};


\end{tikzpicture}
        \end{center} \\
       The Feynman diagram depicting the pair production and subsequent decay cascades of the $Z_2$-odd neutral scalars ($\phi_{S,P}$) resulting in two high-$p_T$ lepton final states at the LHC, accompanied by soft leptons/jets and missing transverse energy ($E_T\!\!\!\!\!\!/$~). Similar final states have been studied in Ref.~\cite{ATLAS:2022hbt} within the context of slepton pair production followed by the decay of the slepton into the lightest SUSY particle in association with a hard lepton resulting in a di-lepton $+~E_T\!\!\!\!\!\!/$ ~final state. \\\hline \hline
        {\bf 5.} \begin{center}
            \begin{tikzpicture}[line width=1.4 pt, scale=0.9,every node/.style={scale=1.0}]

\draw[fermion,black,very thick] (-4.5,1) --(-3.5,0);
\draw[fermion,black,very thick] (-4.5,-1) --(-3.5,0);

\draw[fermion,black,thin] (-2.5,-1) --(-3.5,0);
\draw[fermion,black,thin] (-3.5,0) --(-2.5,1);
\filldraw[black] (-3.5,0) circle (2pt);

\draw[fermion,black,thin] (-2.5,1) --(-1.5,1.75);
\draw[vector,red,thin] (-2.5,1) --(-1.5,0.5);
\draw[fermionnoarrow,black,thin] (-0.5,0.25) --(-1.5,0.5);
\draw[fermionnoarrow,black,thin] (-1.5,0.5) --(-0.5,0.75);

\draw[fermion,black,thin] (-2.5,-1) --(-1.5,-1.75);
\draw[vector,red,thin] (-2.5,-1) --(-1.5,-0.5);
\draw[fermionnoarrow,black,thin] (-0.5,-0.75) --(-1.5,-0.5);
\draw[fermionnoarrow,black,thin] (-1.5,-0.5) --(-0.5,-0.25);

\node at (-4.75,1.25) {$p$};
\node at (-4.75,-1.25) {$p$};
\node at (-3.25,-1) {$\bar{\tilde{E}}^{S}_{1,2}$};
\node at (-3.25,1) {$\tilde{E}^{S}_{1,2}$};
\node[scale=0.7] at (-2.25,0.5) {$W^*$};
\node[scale=0.7] at (-2.25,-0.5) {$W^*$};
\node at (-1.25,2) {$\tilde{N}^S_1$};
\node at (-1.25,-2) {$\tilde{N}^S_1$};
\node at (0.2,0.25) {$q ~({\nu}_l)$};
\node at (0.1,0.75) {$q ~(l)$};
\node at (0.1,-0.25) {$q ~(l) $};
\node at (0.2,-0.75) {$q ~({\nu_l}) $};

\end{tikzpicture}
        \end{center} \\
        The Feynman diagram depicting the pair-production and subsequent decay cascades of the singlet charged leptons (${\ES}_{1,2}$) resulting in soft leptons/jets and missing transverse energy ($E_T\!\!\!\!\!\!/$~) final states. The soft-leptons/jets arise from the 3-body decay of the ${\ES}_{1,2}$ into $\tilde{N}_{1}^S$ in association with a pair of SM fermions. Soft di-lepton events in association with $E_T\!\!\!\!\!\!/$ ~have been studied in Ref.~\cite{ATLAS:2019lng} within the context of slepton pair production in quasi-degenerate slepton and lightest neutralino scenario. Although the SUSY signatures do not include the neutrinos as an additional source of $E_T\!\!\!\!\!\!/$~~, the results presented in Ref.~\cite{ATLAS:2019lng} can be re-interpreted in the context of our model using fast-detector simulators. \\
        \hline \hline 
     \end{tabular}}
\end{table}       
\begin{table}[h]
    \centering
    \scalebox{0.9}{
   \begin{tabular}{|m{16 cm}|} 
        \hline \hline \begin{center}
            {\bf {\em Scenario II}}
        \end{center}\\
        \hline \hline
        {\bf 1.} \begin{center}
            \begin{tikzpicture}[line width=1.4 pt, scale=1,every node/.style={scale=1.0}]

\draw[fermion,black,very thick] (-4.5,1) --(-3.5,0);
\draw[fermion,black,very thick] (-4.5,-1) --(-3.5,0);
\filldraw[black] (-3.5,0) circle (2pt);

\draw[fermion,black,thin] (-3.5,0) --(-2.25,0.9); 
\draw[fermion,black,thin] (-2.25,0.9) --(-1,1.4); 
\draw[vector,red,thin] (-2.25,0.9) --(-1.5,0.5);
\draw[fermionnoarrow,black,thin] (-1.5,0.5) --(-1,0.25);
\draw[fermionnoarrow,black,thin] (-1.5,0.5) --(-1,0.75);

\draw[fermion,black,thin] (-3.5,0) --(-2.25,-0.9);
\draw[fermion,black,thin] (-2.25,-0.9) --(-1.25,-1.1);
\draw[vector,red,thin] (-2.25,-0.9) --(-1.5,-0.5);
\draw[fermionnoarrow,black,thin] (-1.5,-0.5) --(-1,-0.25);
\draw[fermionnoarrow,black,thin] (-1.5,-0.5) --(-1,-0.75);

\node at (-3.25,-0.75) {$\bar{\tilde{E}}_{1,2}^{D}$};
\node at (-3.25,0.75) {$\tilde{N}_{X,Y}^{D}$};


\node at (-0.65,1.5) {$\tilde{N}^S_1$};
\node at (-0.95,-1.3) {$\tilde{N}^S_1$};

\node[scale=0.6] at (-2,0.5) {$Z^*$};

\node[scale=0.6] at (-2,-0.5) {$W^*$};

\node at (-4.75,1.25) {$p$};
\node at (-4.75,-1.25) {$p$};


\end{tikzpicture}
        \end{center} \\
       The Feynman diagrams depicting the associated production of the doublet charged lepton (${\ED}_{1,2}$) and doublet neutral lepton (${\ND}_{X,Y}$) and their subsequent decays. The final state comprises soft leptons/jets and missing transverse energy ($E_T\!\!\!\!\!\!/$~). Ref.~\cite{ATLAS:2019lng} has been used to put bound on the masses of $\ED$ and $\ND$ within the context of the wino-bino scenario with $m(\tilde{\chi}_1^{\pm}) = m(\tilde{\chi}_2^0)$. \\ \hline\hline
        {\bf 2.} \begin{center}
            \begin{tikzpicture}[line width=1.4 pt, scale=1,every node/.style={scale=1.0}]

\draw[fermion,black,very thick] (-4.5,1) --(-3.5,0);
\draw[fermion,black,very thick] (-4.5,-1) --(-3.5,0);
\filldraw[black] (-3.5,0) circle (2pt);


\draw[fermion,black,thin] (-3.5,0) --(-2.25,0.9);
\draw[fermion,black,thin] (-2.25,0.9) --(-1,1.4);
\draw[vector,red,thin] (-2.25,0.9) --(-1.5,0.5);
\draw[fermionnoarrow,black,thin] (-1.5,0.5) --(-1,0.25);
\draw[fermionnoarrow,black,thin] (-1.5,0.5) --(-1,0.75);
\draw[fermion,black,thin] (-1,1.4) --(0,1.6);
\draw[vector,red,thin] (-1,1.4) --(-0.25,1);
\draw[fermionnoarrow,black,thin] (-0.25,1) --(0.25,0.75);
\draw[fermionnoarrow,black,thin] (-0.25,1) --(0.25,1.25);

\draw[fermion,black,thin] (-3.5,0) --(-2.25,-0.9);
\draw[fermion,black,thin] (-2.25,-0.9) --(-1,-1.4);
\draw[vector,red,thin] (-2.25,-0.9) --(-1.5,-0.5);
\draw[fermionnoarrow,black,thin] (-1.5,-0.5) --(-1,-0.25);
\draw[fermionnoarrow,black,thin] (-1.5,-0.5) --(-1,-0.75);
\draw[fermion,black,thin] (-1,-1.4) --(0,-1.6);
\draw[vector,red,thin] (-1,-1.4) --(-0.25,-1);
\draw[fermionnoarrow,black,thin] (-0.25,-1) --(0.25,-0.75);
\draw[fermionnoarrow,black,thin] (-0.25,-1) --(0.25,-1.25);

\node at (-3.25,-0.75) {${\tilde{N}}_{X,Y}^{D}$};
\node at (-3.25,0.75) {$\tilde{N}_{X,Y}^{D}$};

\node at (-1.75,-1.5) {$\bar{\tilde{E}}_{1,2}^{D}$};
\node at (-1.75,1.5) {$\tilde{E}_{1,2}^{D}$};

\node at (0.3,1.8) {$\tilde{N}^S_1$};
\node at (0.3,-1.8) {$\tilde{N}^S_1$};

\node[scale=0.6] at (-2,0.5) {$W^*$};
\node[scale=0.6] at (-2,-0.5) {$W^*$};

\node[scale=0.6] at (-0.75,1) {$W^*$};
\node[scale=0.6] at (-0.75,-1) {$W^*$};

\node at (-4.75,1.25) {$p$};
\node at (-4.75,-1.25) {$p$};


\end{tikzpicture}
        \end{center} \\
       The Feynman diagram depicting the pair production and subsequent decay cascades of the $Z_2$-odd neutral leptons (${\ND}_{X,Y}$) resulting in multiple soft leptons/jets and large missing transverse energy ($E_T\!\!\!\!\!\!/$~). \\\hline \hline
     \end{tabular}}
\end{table}      

\FloatBarrier


\bibliographystyle{format}
\bibliography{ref}

\end{document}